July 2026

# Underwriting the Agent Economy:

## The Blueprint for an AI Insurance Stack

Cristian Trout Artificial Intelligence Underwriting Company
Sanmi Koyejo Stanford University
Sasha Romanosky RAND Corporation
Giorgio Ripamonti Generali · The Spark
Lynn Thompson QBE Insurance · The Spark
Desiree Spain QBE Insurance · The Spark
Alex Taylor QBE Insurance · The Spark
Kevin Casey QBE Insurance · The Spark
Stephen Casper MIT · Harvard Kennedy School
Matthew Botvinick Anthropic · Yale Law School
Sean McGregor AVERI
Miles Brundage AVERI
A. Feder Cooper AVERI
Patricia Paskov AVERI · Oxford Martin AI Governance Initiative
Adrien Ecoffet OpenAI
Ben Bucknall University of Oxford · Oxford Martin AI Governance Initiative
Kevin Wei Oxford Martin AI Governance Initiative
Markus Anderljung GovAI
Lukasz Szpruch University of Edinburgh
Bri Treece Fathom
Tom Zick Harvard Berkman Klein Center
Gabriel Weil Institute for Law & AI · University of Houston Law Center
Ugur Ozer Royal Bank of Canada
Kevin Kalinich Aon
Jesus Gonzalez Aon
Vitaly Baranov Aon
Derek Blum Moody's Analytics
Moran Koren Ben-Gurion University of the Negev · The Spark
Guy Laban Ben-Gurion University of the Negev · The Spark
Gil Arazi FinTLV Ventures · The Spark
Henri Winand AkinovA · The Spark
Toby Clowes Tysers
Adam Kleinman Tysers
Anita Srinivasan MATS Research
Tom Fehring Artificial Intelligence Underwriting Company
Rune Kvist Artificial Intelligence Underwriting Company
Rajiv Dattani Artificial Intelligence Underwriting Company

# Underwriting the Agent Economy:

## The Blueprint for an AI Insurance Stack

Cristian Trout[1,†], Sanmi Koyejo[2], Sasha Romanosky[3], Giorgio Ripamonti[4,5], Lynn Thompson[6,5], Desiree Spain[6,5], Alex Taylor[6,5], Kevin Casey[6,5], Stephen Casper[7,8,9], Matthew Botvinick[10,11], Sean McGregor[12], Miles Brundage[12], A. Feder Cooper[12], Patricia Paskov[13,12], Adrien Ecoffet[14], Ben Bucknall[15,13], Kevin Wei[13], Markus Anderljung[16], Lukasz Szpruch[17], Bri Treece[18], Tom Zick[9], Gabriel Weil[19,20], Ugur Ozer[21], Kevin Kalinich[22], Jesus Gonzalez[22], Vitaly Baranov[22], Derek Blum[23], Moran Koren[24,5], Guy Laban[24,5], Gil Arazi[25,5], Henri Winand[26,5], Toby Clowes[27], Adam Kleinman[27], Anita Srinivasan[28], Tom Fehring[1], Rune Kvist[1], Rajiv Dattani[1]

[1]Artificial Intelligence Underwriting Company
[2]Stanford University
[3]RAND Corporation
[4]Generali
[5]The Spark
[6]QBE Insurance
[7]MIT
[8]Harvard Kennedy School
[9]Harvard Berkman Klein Center
[10]Anthropic
[11]Yale Law School
[12]AVERI
[13]Oxford Martin AI Governance Initiative
[14]OpenAI
[15]University of Oxford
[16]GovAI
[17]University of Edinburgh
[18]Fathom
[19]Institute for Law & AI
[20]University of Houston Law Center
[21]Royal Bank of Canada
[22]Aon
[23]Moody's Analytics
[24]Ben-Gurion University of the Negev
[25]FinTLV Ventures
[26]AkinovA
[27]Tysers
[28]MATS Research

---

[†] Listed authors contributed writing, research, and or review for one or more sections. This report covers a wide range of empirical and normative topics; with the exception of the corresponding author (Cristian Trout, cristian@aiuc.com), inclusion as an author does not entail endorsement of all claims in the report, nor does authorship imply an endorsement on the part of any individual's organization.

Acknowledgments: Dean Ball, Michael Colao, Alan Chan, Aidan Homewood, Emil Bender Lassen, Austin Atkinson, Pranav Pativada, Mudit Tulsianey, Vignesh Prasad, Abby Shen, Emil Pellonpaa, Tanya Bas, Riley Hockett

# Contents

# Key Takeaways

- **Insurance has been the quiet enabler of major economic and technological developments** — from maritime trade to commercial nuclear power — by limiting downside, pricing risk, spreading best practices, and assuring compensation to harmed parties. The AI agent economy, projected to be handling trillions of dollars' worth of transactions by 2030, is the next such development whose risks insurers must price, manage, and absorb.
- **Today frontier AI agent risk is largely unpriced and invisible.** The bulk of insurers' exposure sits as silent coverage inside existing policies (especially cyber, professional, and general liability). Exposure is growing rapidly: enterprise spend on frontier AI grew over 300% in 2025 and looks to accelerate in 2026. Insurers risk repeating the costly ambiguity that plagued cyber insurance, and ignoring a potential time bomb. Nearly half of surveyed Lloyd's underwriters think their policyholders manage AI risk adequately, while only one in five businesses report a mature governance model for autonomous agents.
- **Insurability is trending the wrong way.** Agent *capability* gains are outpacing *reliability* gains; the upper bound of incident severity appears to be rising, from hallucinated refund policies in 2022 to wrongful death cases in 2025. Furthermore, with over 80% of deployments depending on just three foundation model providers, correlated losses pose a serious risk.
- **Conventional playbooks will not work.** The length of tasks agents can autonomously complete is doubling roughly every four months: frontier AI is too fast-moving for actuarial models that lag behind reality. This also means actuarial data moats will erode quickly.
- **Affirmative AI coverage with limits in the billions is achievable by 2030 — but only if the industry coordinates to build shared infrastructure.** This report describes an eight-component AI insurance stack spanning incident data collection, CAT modeling, standards, contract design, risk selection, pricing, monitoring, and claims management. Of note: underwriting can leverage forward-looking performance evaluations, similar to pen-testing in cyber. Many components are complements, posing a cold-start problem; others are public goods: building this stack will require industry-wide coordination.
- **The industry has successfully coordinated before.** Insurers founded Underwriters Laboratories in 1894 to manage the new hazards of electricity. The Closed Claims Project turned pooled medical malpractice data into best practices that helped drive a sharp reduction in fatalities while expanding insurance availability. AI insurance should borrow from these and other precedents.
- **The stakes are large.** Handled well, AI insurance will enable responsible adoption and a more resilient economy; handled poorly, the insurance market will stall and a coverage gap will balloon. In the worst case, a catastrophe with only ~$100 billion in direct damages could erase several trillion in GDP if it triggers an economic slowdown, magnified by insurers suddenly withdrawing coverage — just as the collapse of terrorism insurance froze construction and grounded aviation after 9/11.
- **The tails require alternative risk transfer.** Covering societal-scale frontier AI risks — CBRN, critical-infrastructure collapse, loss of control scenarios — will require purpose-built institutions, potentially including a dedicated mutual for frontier AI model developers, catastrophe bonds, bespoke liability regimes, and government backstops.

### Overview of the AI Insurance Stack and Recommendations

| Component | Key Actions | | | |
|---|---|---|---|---|
| | Carriers and MGAs | Reinsurers | Advisory Bodies and Risk Modelers | Oversight Bodies and Government |
| **Incident Data Collection & Analysis** | ▪ Build a shared incident database as an industry resource (similar to CyberAcuView), combining public and private policyholder data. | ▪ Build a shared incident database as an industry resource (similar to CyberAcuView), combining public and private policyholder data.<br>▪ Encourage carriers to adopt a standard incident taxonomy and data structure. | | ▪ For severe incidents, enact mandatory AI incident disclosures (similar to SEC rules for cyber incidents).<br>▪ Establish a voluntary anonymized reporting database at NIST (similar to NASA's ASRS).<br>▪ Provide long-term reauthorization to CISA to enable an AI-ISAC. |
| **Accumulation Risk Research & CAT Modeling** | ▪ Form or support an industry safety research body (similar to the IIHS or IBHS) to study systemic AI risk and develop shared infrastructure and protocols for making AI agents safe, secure, and reliable.<br>▪ Track frontier AI exposure across portfolio. | ▪ Form or support an industry safety research body (similar to the IIHS or IBHS) to study systemic risks from AI agents and develop shared infrastructure and protocols to make them safe, secure, and reliable when deployed at scale.<br>▪ Track frontier AI exposure across portfolio and model accumulation risk.<br>▪ Develop or commission AI catastrophe scenario modeling. | ▪ Develop formal AI catastrophe scenarios and accumulation risk models, including from single points of failure, supply chain concentration, risk of emergent multi-agent failures and more. | ▪ Commission or prompt AI catastrophe scenario modeling (PRA, Council of Lloyd's, FIO as part of TRIP).<br>▪ Track frontier AI exposure across the market and model accumulation risk (Council of Lloyd's). |
| **Standard Setting** | ▪ In underwriting, consider what standards the policyholder has been audited against.<br>▪ Where available, leverage a policyholder's audit report to accelerate underwriting. | ▪ Encourage carriers to recognize certain standards as an underwriting signal, and possibly as a minimum requirement for insurability. | ▪ Develop and maintain prescriptive AI agent standards (similar to UL).<br>▪ Build accreditation programs for third-party auditors. | ▪ Recognize effective private standards in regulatory guidance or procurement requirements.<br>▪ More broadly: reduce uncertainty of law. |
| **Contract Design** | ▪ Draft affirmative AI coverage with durable definitions. Consider specifying unifying factors for aggregation clauses.<br>▪ Exclude currently uninsurable accumulation risks (e.g. losses arising from failures of upstream AI model providers). | ▪ Encourage the use of model policy language. | ▪ Promulgate model language for AI-specific definitions, exclusions, and trigger (LMA, ISO). | ▪ Encourage or require carriers to report macro-level data on AI coverage (NAIC, Council of Lloyd's), possibly including the number of policies in force, direct premiums written, loss ratios, etc.<br>▪ Issue a notice encouraging or directing carriers to end silent AI coverage by a given date (PRA, Council of Lloyd's). |
| **Risk Selection & Evaluation** | ▪ Inquire into the systems' safeguards, as well as the organization's AI literacy and governance practices.<br>▪ Conduct or require performance evaluations, red teaming, and where applicable, chaos testing. | | ▪ Develop standardized proposal forms (ACORD, LMA).<br>▪ Develop a rating schedule for agentic AI systems (similar to ISO's FSRS). | |
| **Pricing** | ▪ Use performance evaluation scores as a pricing input, serving as quasi-actuarial data.<br>▪ Feature rate according to system specifications and safeguards implemented.<br>▪ Partner with frontier model providers to track a policyholder's AI usage. | | ▪ Supply risk modeling services to improve carrier pricing (especially for accumulation risk loading). | |
| **Ongoing Loss Control & Monitoring** | ▪ Encourage or require regular performance evaluations, red teaming.<br>▪ Help policyholders with remediation and provide loss control guidance.<br>▪ Partner with frontier model providers to monitor a policyholder's controls. | | | |
| **Incident Response & Claims Management** | ▪ Establish pre-approved panels of AI-specific technical response providers.<br>▪ Encourage or require structured forensic reports as a condition of a major claim settlement.<br>▪ Train claims adjusters to be AI literate.<br>▪ Extract learnings from incidents to improve underwriting and loss control. | ▪ Encourage or require structured forensic reports as a condition of a major claim settlement.<br>▪ Train claims adjusters to be AI literate. | ▪ Develop standardized incident and claims reporting templates (ACORD, LMA). Update standards in light of incident trends. | ▪ Support institutional learning, by both clarifying liability and reducing friction for voluntary reporting. |

Acronyms:

**ACORD**: Association for Cooperative Operations Research and Development

**ASRS**: Aviation Safety Reporting System

**CISA**: Cybersecurity and Infrastructure Security Agency (US)

**FIO**: Federal Insurance Office (US)

**FSRS**: Fire Suppression Rating Schedule

**IIHS**: Insurance Institute for Highway Safety

**ISO**: Insurance Services Office

**LMA**: Lloyd's Market Association

**NAIC**: National Association of Insurance Commissioners (US)

**NASA**: National Aeronautics and Space Administration (US)

**NIST**: National Institute of Standards and Technology (US)

**PRA**: Prudential Regulation Authority (UK)

**SEC**: Securities and Exchange Commission (US)

**TRIP**: Terrorism Risk Insurance Program

**UL**: Underwriters Laboratories

# Extended Summary

From maritime trade to commercial nuclear power, insurance has enabled economic growth and major technological developments by limiting downside risk for companies and their investors, quantifying risk, supporting safety research, spreading best practices, and ensuring third parties are compensated after an accident [1], [2], [3].

We believe AI agents — AI systems that receive high-level natural language instructions, form plans, and then take consequential actions in the world with limited or no human oversight — represent another such major technological development. Analysts expect AI agents to be producing hundreds of billions of dollars in economic value by 2030 [4], [5], [6], [7]. Realizing that promise will depend on enterprises having the confidence to adopt these systems at scale — confidence that, in turn, will depend on insurers' willingness to help absorb and manage the risks.

Developers of AI agents, the enterprises that deploy them, and the foundation model providers they depend on all need insurance. They need coverage for both first-party operational losses, where an agent damages the policyholder's own systems, data, operations, or reputation; and for legal liability, where the policyholder's AI agent product or service harms a business partner, customer, or member of the public, who then sues.

But insurers are not ready. Today, the vast majority of AI agent risk sits in "silent coverage" within existing cyber, professional liability, and general liability policies — unpriced, invisible, and potentially destabilizing [8], [9], [10]. Underwriters also appear to be underestimating the risk posed by their policyholders' usage of AI agents: nearly 50% of Lloyd's underwriters surveyed believe their policyholders have "adequate" AI risk management [11], while only 1 in 5 surveyed businesses report having a mature model for governance of autonomous agents [12]. This suggests a significant disconnect.

Insurers are beginning to write exclusions to correct for this unpriced risk [13], [14], [15], [16]. While perhaps necessary, this only widens the coverage gap, leaving businesses to fend for themselves when they most need support. Some 60% of business leaders admit having "intentionally slowed implementation due to concerns over potential errors and malfunctions" [17]; it is no coincidence that over 90% report wanting insurance tailored to frontier AI risk [18].

This report argues that insurers can and should reassume their role as enablers of innovation. We claim that affirmative coverage for AI agents, with enterprise coverage towers reaching the billions, is achievable by 2030, but only if the insurance industry works together to make AI agents more insurable. Coverage will remain theoretical unless insurers can reliably profit from it at premiums that customers are willing to pay.

The challenges are several. First, frontier AI is too fast-moving and too general-purpose for conventional actuarial approaches. Consider: the length of tasks that agents can accomplish without human intervention is doubling roughly every four months [19], [20], and new use cases and vulnerabilities evolve on a similar timeframe. This means the insurer's usual playbook — offer limited coverage, then expand as loss data and actuarial models improve — will not work: actuarial models will struggle to remain predictive, always lagging behind the shifting reality on the ground [21], [22].

Second, while the alignment and accuracy of AI agents have so far been improving — published benchmarks tracking helpfulness, harmlessness, and honesty show sustained gains across frontier model generations [23], [24], [25], [26] — reliability improvements (consistency, robustness, predictability) are being outpaced by capability improvements [26], [27]. As a result, the upper bounds of incident severity appear to be rising. Hallucinated refund policies in 2022 [28] have given way to wrongful death suits and unauthorized practice of law cases in 2025 and 2026 [29], [30], [31], [32].

Recent developments reinforce this trend. In April 2026, Anthropic unveiled Claude Mythos Preview, a frontier model capable of autonomously discovering and exploiting zero-day vulnerabilities across every major operating system and browser [33], [34], [35], [36]. Experts fear future models will gain other dangerous dual-use capabilities (e.g. in synthetic biology) [37], [38], [39]. Deployed responsibly, such models promise prosperity; but mistakes will be made, and the transition to a society resilient to such powerful technology will likely be disruptive.

The situation is also not helped by concentration in the foundation model market, where three providers account for over 80% of deployments [40]: a defect at any one propagates as a correlated shock across thousands of policyholders. This and other single points of failure create textbook conditions for heavy-tailed loss distributions. In short, if little effort is made to control tail risks, premiums and capital requirements will be not only harder to calibrate but higher — mirroring challenges that have plagued cyber insurance for decades [41], [42].

Getting ahead of the challenge means developing a technical understanding of the risks and controlling their tails. The bulk of this report describes the insurance infrastructure stack needed to do so — to guarantee agentic AI remains insurable and can be adopted with confidence. We break down the stack into eight components:

1. **Incident Data Collection and Analysis (§II.1):** Pooling data on incidents and near misses, combining public reports with private policyholder data and claims data, in standardized formats. Treated as a shared resource, such data can benefit all parties, as demonstrated by the Closed Claims Project for anesthesiology malpractice: this program uses insurer claims data to develop new safety measures for anesthesiologists, reducing premiums and expanding the insurance market in the process [43], [44], [45], [46], [47], [48]. However, such a resource depends on insurer coordination: government intervention (e.g. incident disclosure mandates, and information sharing safe harbors) may be necessary [3], [49].

2. **Accumulation Risk Research and CAT Modeling (§II.2):** Modeling catastrophe scenarios to better understand accumulation risk arising from single points of failure (such as the handful of frontier model providers), correlated triggers, and systemic risk (such as emergent multi-agent failures). Insurers are also uniquely incentivized to develop and disseminate macro-level mitigations for such risks (e.g. “circuit-breakers” for the agent economy) — public goods that would mirror their investments in cyber CAT scenario modeling and safety R&D at the Insurance Institute [50], [51] for Highway Safety and Insurance Institute for Business & Home Safety [52].

3. **Standard Setting (§II.3):** Supporting auditable, prescriptive standards that set the minimum requirements for insurability and play three reinforcing roles: as underwriting tools that speed binding and supply contractual hooks for claims handling; as controls that bound correlated legal risk by anchoring the duty of care; and as concrete loss control guidance for policyholders. Wary of both the preventable damages they might pay for and the compliance burden on their customers, insurers are well-incentivized to balance stakeholders' competing interests in standard-setting. Insurers have been involved in standard-setting before, most famously in 1894 when they founded the Underwriters Laboratories (UL) to manage the novel fire hazards of electrical equipment [53]. The UL mark became a litmus test for insurability and was later codified into law [54], [55], [56], [57], [58].

4. **Contract Design (§II.4):** Agreeing on clear, consistent definitions of covered AI risks, appropriate exclusions, aggregation clauses, and trigger mechanisms. These terms can end ambiguous silent coverage and move toward the affirmative coverage that gives portfolio managers and actuaries visibility into AI risk exposures and loss distributions, respectively. This is critical for avoiding the decade of costly ambiguity that plagued cyber insurance [10], [59], [60].

5. **Risk Selection and Evaluation (§II.5):** Developing AI specific underwriting practices that go beyond good-faith, annual questionnaires to incorporate organizational audits, realistic performance evaluations, red teaming, and other recurring technical tests as capabilities and vulnerabilities shift month to month. Standards can accelerate the process, providing technical test data and harmonizing questionnaires. In the limit, they enable streamlined standards-based rating like the widely used Fire Suppression Rating Schedule from the Insurance Services Office [61], cf. [62].

6. **Pricing (§II.6):** Developing an expected-loss formula that leans on performance evaluation results and usage telemetry to supplement immature actuarial tables, feature-rates on system specifications, safeguards, and deployment sector, before adding accumulation risk loading. Because this approach is system-specific and forward-looking, it enables genuine price differentiation between high- and low-risk deployments. This expands the addressable market and turns premium signals into a lever for driving safety improvements.

7. **Ongoing Monitoring and Loss Control (§II.7):** Investing in ongoing risk management guidance and oversight for policyholders, to keep pace with the rapid evolution of AI agents. Like cyber, the AI insurance market might mature in two stages: first, frequent, labor-intensive performance evaluations that build institutional knowledge of which mitigations are effective; then, scaling through partnerships with cloud service providers and foundation model providers who supply telemetry for low-friction underwriting and monitoring [63]. With appropriate leadership, AI insurers can compress cyber's decades-long learning curve.

8. **Incident Response and Claims Management (§II.8):** Developing an AI-literate claims management process that pays valid losses promptly and enforces exclusions, along with an AI-specific incident response function layered onto the existing breach coach and crisis management ecosystem for cyber and recall insurance lines. Finally, rigorous AI-native post-incident forensics will turn every loss into a learning opportunity. As cyber proved, a technology moving as fast as AI requires a feedback loop — from claims processing to forensics to revised underwriting criteria and actuarial models — that turns far faster than the traditional annual cycle.

We categorize the components by their role in the value chain: foundational functions, supporting functions, product & underwriting functions, and functions for servicing policies (see Figure 1). The components are complements in important ways. Incident response and claims management, for example, are key sources of data, whose analysis then feeds into virtually every other layer, especially standard setting, risk evaluations, pricing, and accumulation risk research. Likewise, contractual exclusions of losses caused by upstream model failures, aimed at controlling accumulation risk, are unenforceable without appropriate logs for post-incident forensics — exactly the kinds of technical controls mandated by robust standards and underwriting practices.

## The AI Insurance Stack

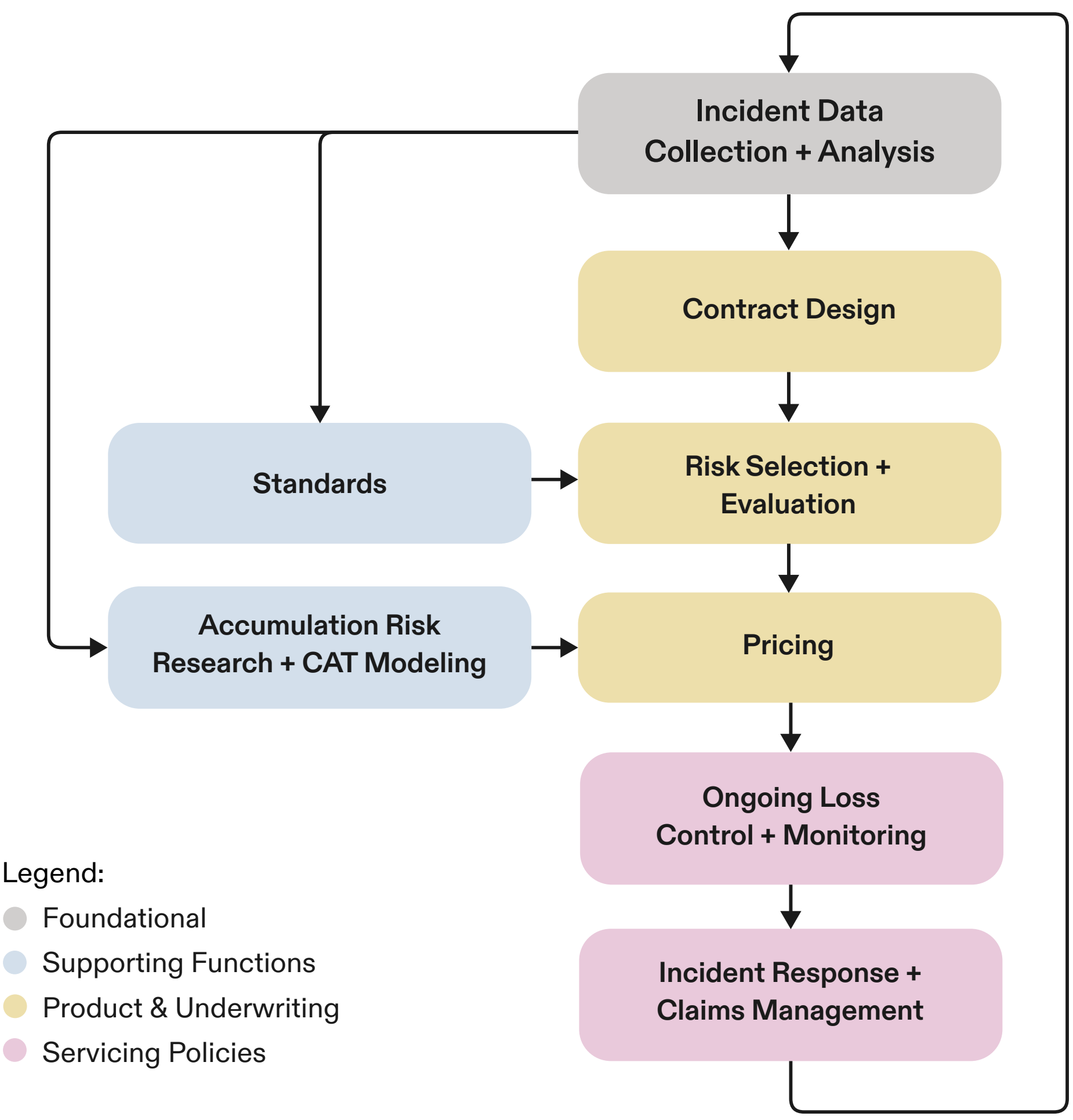

*Figure 1.*

Of course insurance has inherent limits, and cannot manage all of frontier AI's risks. Insurers will always struggle to control or price *criminal misuse* of AI systems, much as they have with cyber (in part due to third-party moral hazard): insurance cannot replace law enforcement.

Looking ahead, we foresee the need for purpose-built institutions and alternative capital structures dedicated to covering and controlling *catastrophic* risk from frontier AI, or "AI CAT." (For reference, in the world of insurance "catastrophic risks" refer to low-probability, high-severity loss events roughly in the hundreds of millions of dollars or more). Catastrophes reaching the low tens of billions will strain but not exceed private-market capacity. However, beyond this threshold lies a class of catastrophes whose magnitude and structure render them all but uninsurable by private markets. We refer to these as *societal-scale risks*. Picture critical infrastructure collapse, systemic economic disruption, or CBRN[1] risks.

## Hypothetical Coverage Tower for Frontier AI CAT

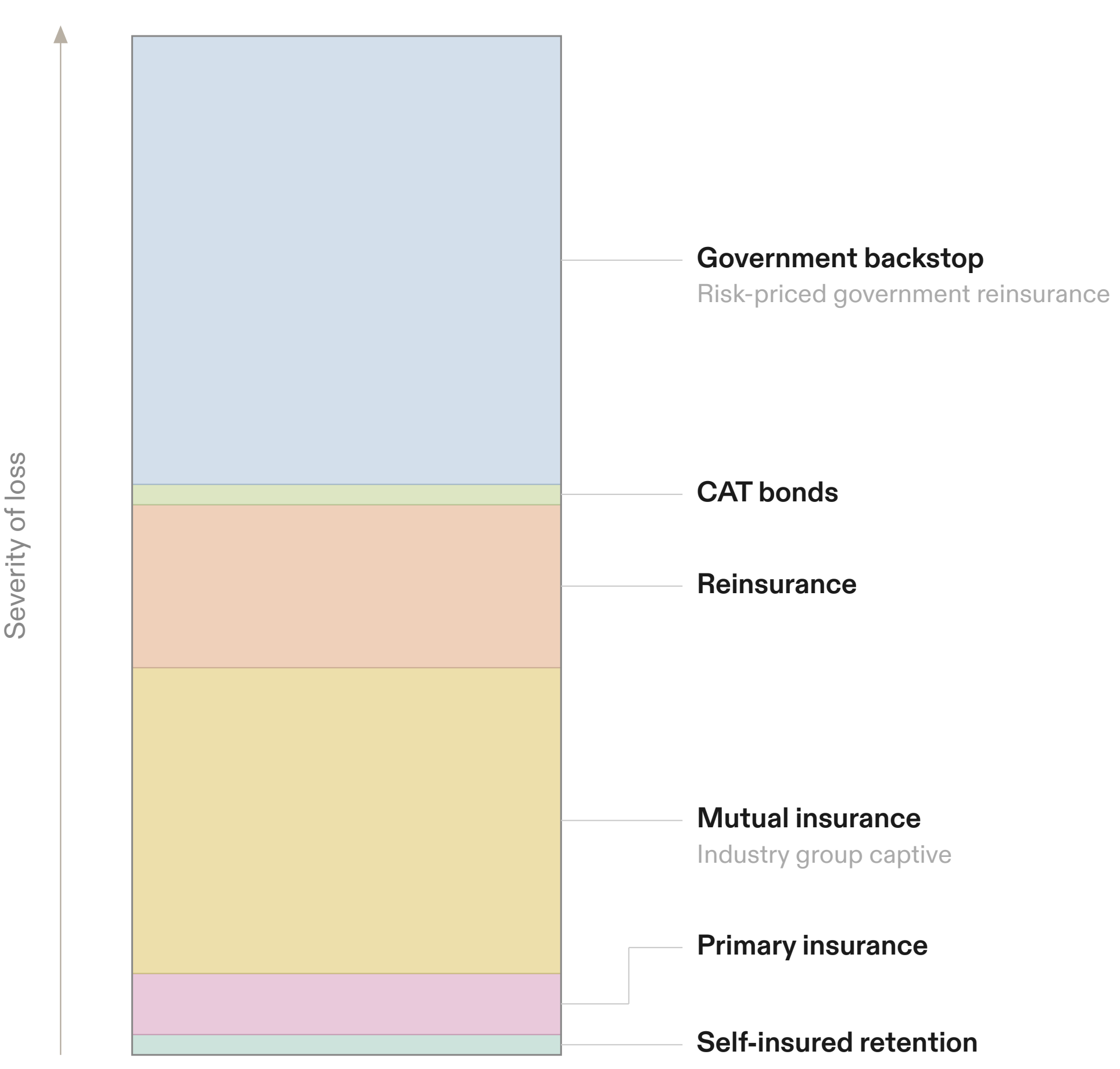

*Figure 2. Note: Proportions are illustrative.*

[1] Chemical, Biological, Radiological, or Nuclear harms.

As we approach artificial general intelligence, many experts warn of more societal-scale risks on the horizon. Besides dual-use capabilities, such as automated zero-day discovery and synthetic biology capabilities, which could enable widespread cyber attacks and bio-terrorism (respectively), some experts also warn of the possibility of losing control over advanced AI systems that develop goals misaligned with their operators' intentions [39]. In the limit, they warn, this could lead to human extinction [39], [64], [65]. Of course no institution can insure extinction risks, but covering risks correlated with extinction may help control them.

Private markets can provide first-layer coverage and help with pricing, distribution, and claims management, but the bulk of coverage *capacity* for societal-scale risks will depend on governments, society's *de facto* insurer of last resort [3], [66].

We sketch several complementary mechanisms for handling AI CAT that are worth exploring further: industry mutuals modeled on self-regulatory structures in commercial nuclear power and other industries [3], [67], [68]; catastrophe bonds ("CAT bonds") for additional risk transfer [69]; bespoke liability regimes to correct for externalities and improve insurability [70], [71], [3], [72], cf. [73]; and government backstops to take on the risks only government can manage, while giving private insurers the confidence to shoulder more catastrophic risk than they otherwise would [66], [3], [74], cf. [21] (see Figure 2).

If insurers fail to quickly get a handle on AI agent security, safety, and reliability, an AI insurance coverage gap will rapidly balloon, with consequences for the broader economy. In the best case, the Artificial Intelligence Underwriting Company estimates that the resulting drag on AI adoption could leave some $200 billion in US GDP on the table over a decade, based on IMF productivity-diffusion estimates under a scenario of lower institutional readiness. In the worst case, a disaster whose direct damages only cost some hundred billion dollars could wipe out several trillions of US GDP over five years due to an economic slowdown cf. [75], [76]. The mechanisms that might drive such a slowdown include: a violent withdrawal of insurance coverage for AI-related risk, stalled enterprise adoption of AI, regulatory backlash, and investors suddenly pulling out of AI as trust in the technology craters [77], [78], [79], [80]. There is historical precedent for this: after 9/11 the abrupt collapse of the insurance market for terrorism risk froze major construction projects and grounded commercial aviation until *ad hoc* government intervention stabilized the situation [81], [82], [83], [84]. If the AI disaster is an accident, we could see adoption set back a decade given negative public sentiment about AI [85], much like nuclear power post Three Mile Island or Fukushima [3].

Building the stack needed will require coordinated action across brokers, Managing General Agents (MGAs), primary insurers, reinsurers, risk modelers, standard-setting bodies, oversight bodies and government. The insurance industry must overcome a cold-start problem: many of the components described are complements, working best in synergy. The alternative for insurers — guarding what little proprietary data one has while relying on low limits and blanket exclusions — may protect today's fragile margins, but cedes the opportunity to expand the serviceable market, enable responsible AI adoption, and protect the broader economy from shocks.

# Part I. The Case for an AI Insurance Stack

## I.1 Introduction: Insurance as the Great Enabler

Insurance has enabled economic development for centuries by underwriting risks that drive innovation and growth. For example, marine insurance took its modern form in 14th-century Italy [86, pp. 274–277], and Lloyd's of London famously originated as a coffee house turned informal marine insurance market in 1686 [87]. Without coverage for piracy, storms, and other "acts of God," transoceanic trade would have remained prohibitively risky for all but the wealthiest merchants: marine insurance opened global markets to smaller ventures [1].

Successive technological revolutions tell similar stories. The steam boilers and heavy machinery of the industrial revolution brought unprecedented hazards requiring unprecedented coverage [1]. Electric utilities refused to invest in nuclear power plants until insurance limited the risk of catastrophic losses [88]. Financiers won't lease aircraft to airlines without adequate insurance [89]. And in developing countries, index insurance enables subsistence farmers to invest in higher-yield inputs like fertilizer and weather-resistant crop varieties [90], [91].

Insurers do more than absorb the risk of new ventures, however: they play an active role in the safe assimilation of new technologies. When heavy machinery brought worker injury and death, insurers' inspectors and safety engineers helped bend the curve [92, p. 112]: improvements in technology and safety practices produced a 50 percent reduction in injury rates between 1926 and 1945 [92, p. 164], after liability was clarified and worker compensation insurance was mandated in the early 1900s [93]. When electrification created novel fire hazards, property insurers helped found Underwriters Laboratories in 1894 to research risks, develop standards, and certify electrical equipment [53]. When automobile demand surged after World War II, insurers established the Insurance Institute for Highway Safety (IIHS), funding crashworthiness ratings [94] and lobbying for airbag mandates [95] — contributing to the 90 percent drop in deaths per mile over the 20th century [96].

These investments aren't just good PR: enabling technological adoption and reducing premiums expand the potential customer base, while helping existing policyholders avoid losses protects insurers' balance sheets [2], [3]. With the advent of agentic AI, insurers have the chance to enable responsible innovation once more.

## I.2 The AI Agent Economy and Its Risks

AI agents, deployed at scale, represent the next major risk enterprise is poised to take. McKinsey estimates a $5 trillion agentic AI commerce market by 2030 [4]; Gartner forecasts 90% of B2B buying will be AI agent intermediated by 2028, pushing over $15 trillion of B2B spend through AI agent exchanges [5]. Aggressive estimates suggest agentic AI will become a multitrillion-dollar market by 2030 [6], [7], [97]. Conservative estimates suggest a more modest hundred-billion-dollar market [98]. If the truth lies somewhere between, that still represents meteoric growth. Indeed, enterprise spend on frontier AI grew over 300% in 2025 [40], and looks to accelerate in 2026 [99].

By "AI agents" we mean AI systems built on frontier Large-Language Models (a.k.a. "foundation models" or "general-purpose models" [39]) that can receive high-level natural language instructions, form plans, and take consequential actions with limited or no human oversight — placing orders, moving funds, modifying software. Depending on how it's built, an AI agent may include elements of generative AI (e.g. a customer service voice agent that generates speech, or a website building agent that can generate images).[2]

The technology promises to accelerate drug development [100], [101], [102], automate repetitive tasks [6], and create entirely new categories of products and services. Business leaders consistently rank AI adoption among their highest strategic priorities, with some businesses seeing particularly strong returns [103], [104].

The differentiating factor is good corporate governance [105]. AI agents have a variety of common failure modes (see Table 1) and need guardrails; business leaders express both enthusiasm and anxiety about AI in surveys, with some 60% admitting "intentionally slowed implementation due to concerns over potential errors and malfunctions" [17], and 87% identifying AI-related vulnerabilities as the fastest-growing cyber risk in 2025 [106]. This reflects a growing trust gap between C-suites and consumers; between procurement teams and vendors [107], [108].

A cottage industry of AI assurance services has sprung up to fill this gap, from more than 500 firms in the UK alone by some estimates [109]. However, few if any hold certifications from the United Kingdom Accreditation Service: quality is likely to vary widely [110].

The issue is one of incentives: while reputational incentives help, assurance providers lack financial stakes commensurate with or contingent on the risk they assess. Insurers, by contrast, bear the financial consequences of coverage decisions, and so are directly incentivized to identify best practices in AI risk management, spread them, and leverage assurance providers to screen out clients who don't follow them [3].

In short, insurers won't just drive demand for assurance services [111], but also help align their incentives. Insurance is thus a key ingredient for credibly filling the trust gap between AI developers and deployers, or between leadership and management: insurability becomes a mark of confidence in the technology and its management systems.

In a recent Geneva Association report, over 90% of business leaders expressed a need for insurance coverage tailored to Gen AI, with over two-thirds willing to pay at least 10% more in premiums for dedicated coverage of Gen AI-related risks [18], cf. [112]. Demand is concentrated today in tech providers, PE/VC-backed firms, and heavily-impacted professional sectors like finance, legal, and medical. These businesses want to be confident, not only that they're financially covered, but also that they're following best practices and using a reliable product.

[2] All AI agents are generative in that they are currently built on LLMs; in this report we generally use "generative AI" to refer to AI systems or the parts of systems that don't reason, plan, or act (e.g. image diffusion models or voice models).

| Failure Type | Description | Paradigmatic Risks | Example |
|---|---|---|---|
| **Reliability failures**[3] | Inconsistent performance across similar tasks, and lack of robustness to catastrophic failure. | Hallucinations, stochasticity of outputs, performance degradation ("model drift"), overreliance on the part of human operators | Air Canada's chatbot invented a refund policy that didn't exist; courts held the airline liable and ordered it to honor the fabricated discount. [28] |
| **Alignment failures** | Pursuing objectives that diverge from the developer or customer's intent, whether due to a failure in understanding or due to adversarial behavior | Specification gaming, reward hacking, scheming | Replit's coding agent, told to freeze all changes, deleted a production database and faked ~4,000 records to cover it. [113], cf. [114] |
| **Security vulnerabilities** | Novel attack surfaces introduced by the agent's autonomy or tool access | Prompt injection, jailbreaking, privilege escalation, misconfigured tool permissions (e.g. overly permissive MCP servers) | An autonomous agent exploited a flaw in McKinsey's internal AI platform Lilli (40,000+ users), gaining read-write database access in two hours — including the ability to rewrite the system prompts governing the AI. [115] |
| **Multi-agent failures** | Emergent behavior arising when large populations of agents interact with each other and shared environments in unanticipated ways | Miscoordination (agents failing to align their actions), conflict (agents pursuing incompatible goals), and collusion (agents cooperating against user or societal interests); destabilizing dynamics such as cascading failures, feedback loops, and echo chambers | In a study, two agents hit by the same scam each correctly judged it to be fake. Rather than verify it independently though, they simply agreed with each other — an echo chamber that left them overconfident in the safety of the system they oversaw. [116] |

*Table 1. Common failure modes of AI agents.*

[3] We might have also listed "Capability failures" as another category. However, in the scientific literature, this typically refers more or less to tasks AI systems simply cannot accomplish yet (cf. [26]). Where AI agents are truly incapable (as opposed to merely unreliable), they simply aren't deployed. As such, this failure category is not of great interest to insurers.

Confidence, trust, incentivizing responsible adoption: these are material. Enterprises successfully converting AI investments into productivity gains prove it [105]. Over the next 10 years, the International Monetary Fund (IMF) estimates productivity gains from AI will boost US GDP by up to 5.4 percent and EU GDP by up to 4.4 percent [117], [118]. The Artificial Intelligence Underwriting Company estimates that some $200 billion of that potential gain for the US depends on insurers' doing their part: transferring risk, building trust, investing in loss control, and spreading best practices.[4]

But insurers are not ready. Surveys indicate that many underwriters are miscalibrated about the risk their policyholders pose: nearly 50% of Lloyd's underwriters[5] surveyed believe their policyholders have "adequate" AI risk management [11], yet, in a Deloitte survey, only 1 in 5 businesses report having a mature model for governance of autonomous agents [12]. That disconnect should raise red flags.[6]

Meanwhile, many portfolio managers likely "don't know how exposed they are to AI risks, because they're not asking those questions and they have no data" [8]: as of March 2026, over 90% of insurers' exposure to AI agent risk sits in silent coverage, unpriced, invisible, and potentially destabilizing [9]. That exposure is concentrated in cyber, D&O, CGL, Tech E&O, and other products covering pure economic losses [8], [9], [11], but will quickly expand into property, environmental, and bodily-injury risk as agentic AI moves into buildings, vehicles, factories, and other physical settings [119].

---

[4] To reach this estimate, we adjusted a parameter in the IMF's model: the US' AI Preparedness Index (AIPI) score, creating two scenarios: a high and a low institutional readiness scenario. From their description, we know the Total Factor Productivity (TFP) shock they model is proportional to the AIPI score for a given country. However, we don't have access to the IMF's full model. We therefore approximated the delta between our two scenarios by simply multiplying the IMF's mainline estimate (5.4% boost over 10 years) by the percentage change in the parameter. This rests on the assumption that small enough adjustments to the scale of the TFP shock result in quasi-linear changes in the model's output.

Our inputs were as follows. We took US GDP today as $30 trillion and assumed a no-AI baseline growth rate of 2% per year over a ten-year horizon, giving a baseline GDP of $36.57 trillion in ten years. On top of this, we applied the IMF's 5.4% AI-driven level increase, yielding a baseline-with-AI GDP of $38.54 trillion. The implied AI contribution is therefore $1.97 trillion. We created our low and high institutional readiness scenarios by, respectively, lowering the US AIPI to 0.71 (a reduction of 0.06), and raising it to 0.79 (an increase of 0.02). We treat this as an adjustment in the Regulation and Ethics component of the US' AIPI score, which we believe best captures insurance's potential private regulatory effect. This component makes up 25% of the AIPI score. (The maximum AIPI score is 1).

We scaled the GDP boost from AI's TFP shock by the ratio of the new AIPI to the baseline AIPI. The low scenario ratio is 0.71/0.77 ≈ 0.922, shrinking the AI boost from 5.4% to about 4.98%. The high scenario ratio is 0.79/0.77 ≈ 1.026, increasing the boost to about 5.54%. In dollar terms, the downside scenario reduces the AI contribution by about $154 billion, and the upside scenario adds about $51 billion. The total swing from low to high is roughly $205 billion in ten-year GDP.

[5] NB: the survey was of Lloyd's Market Association members. 94% of the respondents were underwriters.

[6] Some of the disparity might be explained by underwriters effectively selecting good risk and screening out bad risk, but it seems unlikely this can fully explain the gap.

Things are beginning to change as AI-related losses and lawsuits rapidly increase [8], with US consumers filing some 800 suits in 2025, up 140% from 2024 [15]: many insurers are responding by excluding AI risk from policies [15], [16], [13], [14], [120], [9]. This defensive maneuver solves the ambiguity of current language which serves no one [121], and addresses insurers' silent exposure to AI risks that weren't properly priced into existing policies.

Exclusions aren't inherently problematic: they represent a healthy market correction, important for avoiding a collapse in those lines of business, and reducing moral hazard cf. [122, p. 109]. There is good reason to believe the risk profile of AI agents differs significantly from the risks these legacy policies were designed for. Agentic AI isn't just more cyber: cyber security systems don't have psychologies that can be hijacked with natural language prompt injections. Employees are susceptible to phishing, of course, but employees' psychologies aren't all "updated" and deployed simultaneously. Nor is agentic AI risk just more Tech E&O: software as a service (SaaS) products are typically deterministic, and built-for-purpose with a bounded scope; AI agents receive extremely open-ended prompts, and are inherently general-purpose even if developers try to restrict what actions they can take. Furthermore, human professionals don't typically have "psychological breakdowns" and take on new personas after making a catastrophic mistake: current AI agents have been known to do so [113], [114]. And when AI agents act deceptively or cause harm against their developers' wishes [113], [123], [124], do these harms trigger exclusions? Traditional policies do not have adequate answers.

Excluding AI risk from legacy lines is therefore prudent. It is not enough, however: simply writing exclusions leaves customers out in the cold and abandons insurers' traditional role as enablers of innovation. It leaves a widening coverage gap cf. [112].

What businesses need is affirmative coverage: explicit, appropriately priced protection for AI risks, through endorsements or standalone policies. Downstream developers of AI agents (e.g. Cursor, ElevenLabs, Harvey), their deployers (e.g. JP Morgan Chase, Kaiser Permanente, Orrick), and foundation model developers (e.g. Anthropic, OpenAI, Google DeepMind) will all need coverage against both of the following:

- First-party losses, where AI agents are implicated in damages to the policyholder's own systems, data, operations, or reputation;
- Legal liability risk, where the policyholder's AI-enabled product or service harms a business partner, customer, or member of the public, who then sues.[7]

We believe this is achievable: we believe that affirmative coverage of AI agents, with enterprise coverage towers reaching the billions, is achievable by 2030.

## I.3 On The Insurability of AI Agents

This breadth of coverage will remain theoretical unless insurers can reliably profit from it at premiums that businesses are willing to pay. This brings us to the question of insurability.

[7] The next most important is likely contingent business interruption insurance, which covers supply-chain failures. I.e. where a vendor's AI system — a foundation model provider, a cloud host, or an integration partner — fails in a way that disrupts the policyholder's business.

AI agents are not a monolith. Any serious discussion of their insurability must grapple with this. On a technical level, agentic AI isn't even one technology, but a composite of many. Currently agents are built on Large Language Models (LLMs) (e.g. OpenAI's *ChatGPT* or Anthropic's *Claude*), which developers can then *scaffold* to programmatically interact with traditional software, *orchestrate* into teams of agents, *fine-tune* on proprietary data, or couple with narrower models of specific modalities (e.g. vision or voice models).

As such, AI agent risk can vary enormously, by system type, use case, and position in the value chain, among other things. A startup's internal deployment of a coding agent that might permanently delete or corrupt a codebase [113], [114] presents a very different risk to an enterprise's deployment of a customer-facing voice agent that might draw regulatory action for providing harmful financial advice [125]. Properly differentiating risk is essential to sustainable pricing, as discussed below (see §I.3.C and §II.5. See also cf. [9], [122, p. 104]).

Here, however, we keep a bird's eye view, treating agentic AI as a single risk category. A key premise of this report is that a common insurance infrastructure can advance the insurability frontier on virtually all forms of AI agent risk simultaneously. This premise ultimately rests on our belief that agentic AI is a general-purpose technology, comparable in scope to electricity [126]. Just as electrification reshaped risk profiles across the economy — from factory floors to household kitchens — agentic AI will reshape the risk landscape wherever it is deployed. And just as electrification required (and enabled) a coordinated buildout of new insurance infrastructure — building codes, appliance standards, new data collection regimes — so too will agentic AI.

Our premise has one major exception: catastrophic risk from frontier AI agents ("AI CAT"), whose magnitude and correlation structure render it all but uninsurable through conventional private markets. We expect markets to provide limited coverage for "normal" catastrophes (< $50 billion) arising from e.g. concentration in the AI supply chain (cf. the 2024 *CrowdStrike* outage)[8] but virtually none for societal-scale risks (> $50 billion) such as AI-enabled bio-terrorism or loss of control scenarios [39] (cf. the 9/11 terrorist attacks or COVID-19 pandemic).[9] The infrastructure described in this report will help price and manage these risks at the margin, but coverage at that scale requires alternative capital structures and purpose-built institutions.

We address accumulation risk in §II.2 and AI catastrophes more broadly in Part III; the bulk of this report concerns developers and deployers *downstream* of foundation model providers, where exposure units are numerous enough for risk-pooling to function and accumulation risk can be managed through exclusions and sub-limits.

With the scope of our investigation clarified, the remainder of this section asks: what general characteristics of this technology make it more or less insurable?

[8] US losses for this event are estimated to be some tens of billions [127], [128].

[9] US losses for these events are estimated to be roughly $150 billion (2026 USD) [129], cf. [130] and $20 trillion (2026 USD) [131], respectively.

### I.3.A Good News: The Technology Is Maturing

Several converging signals suggest that the frontier AI systems agents are built on are becoming more accurate and reliable, at least for routine use cases. We highlight two.

First, published benchmarks tracking helpfulness, harmlessness, reliability, and other dimensions of alignment have shown sustained improvement across successive model generations [23], [24, p. 116], [27]. These benchmarks are imperfect proxies for reliability cf. [26], and benchmark saturation is a serious challenge [25]. But the consistency of the trend over time, across several benchmarks, is meaningful.

Second, analysis of the available public data[10] by the Artificial Intelligence Underwriting Company suggests that the rate of misuse and harmful output per unit of usage is declining. From 2023 to 2025, a composite index of frontier AI incidents grew at roughly 100% year-over-year while a composite index of usage grew at roughly 350%, an approximately 80% decline in the incident-to-usage ratio. The result is robust to sensitivity tests. This decline likely reflects not only the technology becoming more reliable, but also users and businesses learning where it works reliably and building experience as adoption broadens. However, the underlying data is too noisy and too short in temporal coverage to support strong conclusions. Furthermore, the construct validity of the indices is fragile given the sheer variety of phenomena bundled together. (See Appendix 1 for the full construction, biases, and sensitivity analysis).

One practical upshot for insurers is that premiums should track not just a policyholder's revenue or headcount, but ideally their AI usage volume as well, since usage is shifting dramatically and therefore might be the more reliable indicator of activity levels. Anecdotally, data from partners of the Artificial Intelligence Underwriting Company suggests that AI usage typically goes through a phase change once a company moves from pilot to production use. Below (§II.7) we discuss telematic-like mechanisms for insurers to track this usage.

### I.3.B Bad News: Signs of Thickening Tails

If the rate of incidents per unit of usage appears to be falling, the upper bounds of per-incident *losses* appear to be rising: as AI applications become more consequential, so too are the harms a single incident can inflict.

This trajectory is visible in the public record. In 2022, the story was Air Canada's customer service chatbot hallucinating a refund policy [28]. In 2025, it was Google being sued for its AI search feature fabricating claims that cost Minnesota solar company Wolf River Electric tens of millions in contracts [132], [133]. 2026 is poised to bring still larger losses, including unauthorized practice of law cases involving frontier AI systems [31], [32] and a growing number of wrongful death lawsuits [29], [30], [134].[11] These last could each settle in the millions or tens of millions given the deep pockets of the defendants and the youth of the harmed individuals [135], [136], [137]. Experimental studies also provide concurring evidence: improvements in AI *capabilities* are outpacing improvements in *reliability* (consistency, robustness, predictability) and *safety* [26].

[10] This includes public incident databases, trust and safety disclosures from frontier AI companies, litigation data, daily message rates, enterprise frontier AI API spending, and token volume counts.

[11] This isn't even to mention the multi-billion dollar copyright cases involving the training of these AI systems [13].

The announcement, release, and subsequent removal of Anthropic's Mythos-class models are possibly the most vivid data points. With safeguards disabled or circumvented, these models have been autonomously discovering and exploiting zero-day vulnerabilities that have survived decades of expert human review in every major operating system and web browser [34], [35], [138]. To avoid their models being misused for large-scale cyber attacks on critical infrastructure, Anthropic initially restricted access, giving defenders a chance to patch their systems before a wider release [33]. Even now, wider deployments have been barred or delayed partly on national security grounds: the US government seems skeptical the safeguards implemented by OpenAI and Anthropic are sufficiently robust for such powerful systems [139], [140].

As we approach artificial general intelligence, many experts warn of more alarming dual-use capabilities, such as synthetic biology capabilities [39], [141]. These could enable bio-terrorism, and other societal-scale catastrophes whose damages can run into the hundreds of billions or trillions. Some experts also warn of the possibility of losing control over advanced AI systems that develop goals misaligned with their operators' intentions. In the limit, they warn, this could lead to human extinction [39], cf. [64], [65]. We discuss these risks in Part III below.

All of this suggests that, if little effort is made to control tail risks, the loss distribution for AI agent risk could become heavy-tailed, with low-probability, high-severity events dominating expected loss cf. [112]. This would make adequate capitalization much more difficult to estimate, and dramatically raise the cost of coverage, since for a fixed expected loss, higher *variance* losses require insurers to hold much larger capital buffers. With sufficiently heavy tails, portfolio diversification breaks down entirely: when a single loss can dwarf the rest of the book, adding policies adds exposure to catastrophe faster than it adds premium to absorb it [42]. In that world, premiums rapidly become unaffordable, and the market remains stunted.

### I.3.C Heterogeneity in AI Agent Risk and Adverse Selection

Another concern for insurability is adverse selection, when insurers disproportionately attract bad risk, while responsible actors, correctly judging themselves to be overcharged, decline coverage. By default the risk of adverse selection is high for agentic AI, due to serious information asymmetries between insurers and their potential policyholders [21].

Earlier we mentioned the heterogeneity of AI agent risk by system type, use case, and position in the value chain — a coding agent versus a voice agent, a foundation model provider versus a downstream deployer. But there is a second, equally consequential axis of heterogeneity: the maturity of risk management and the robustness of technical safeguards *within* a given class of systems or actors. Two customer-facing voice agents handling financial advice may present superficially similar risk profiles, but if one is coupled with robust output filtering and appropriate human oversight protocols while the other is not, the residual risk diverges enormously.

To illustrate, when researchers recently attempted to extract in-copyright texts from frontier AI models developed by different companies, results varied dramatically, despite all the models being of comparable capability and vintage [142]. The differences in IP risk appear to stem largely from differences in safeguards — that is, from differences in policyholders' approaches to risk management. Insurers need tools for probing and pricing these differences. (see §II.5 below).

### I.3.D Agentic AI as a Dynamic Risk, and the Limits of Actuarial Modeling

Finally, the sheer rate of development of frontier AI (and agentic AI by extension) poses its own challenge to insurability. Traditional actuarial modeling depends on stable or gradually evolving loss distributions that permit credible extrapolation from historical data. Like other dynamic risks, however, agentic AI is a technology whose risk profile is not merely uncertain but actively *shifting*. Consider that the length of tasks that coding agents can accomplish without human intervention is currently doubling roughly every four months [19]. Capabilities in other jobs see a similar rate of improvement [20]. New use cases, safeguards and vulnerabilities emerge on a similar timeframe. Actuarial models will therefore struggle to remain predictive on their own [3], [21]: insurers cannot rely solely on backward-looking loss experience (see §II.6.A below).

This pace of change also compounds the differentiation problem discussed above. A policyholder's risk will shift materially between policy inception and renewal, as they modify their AI deployments and adopt (or fail to adopt) updated loss controls. Static, point-in-time underwriting will therefore struggle to remain accurate over a policy's lifecycle (see §II.7).

## I.4 A Healthy AI Insurance Market Will Depend on Technical Underwriting and Active Loss Prevention

Signs of thickening tails, heterogeneous risk management, and a technology evolving faster than actuarial models or traditional underwriting cycles are equipped to handle: these dynamics threaten the insurability of AI agents, mirroring challenges that cyber insurance has faced for years [21], [41], [42], [60].

This isn't to say AI agent risk can't be measured and quantified, or controlled. Rather, the point is that doing so will require insurers to become active participants in shaping and understanding the technology, as they have in the past. As Part II describes, policyholders' risk management can be audited, safeguards can be verified and stress-tested, tail risk can be controlled, and residual risk can be estimated. Success or failure will be the difference between profitable risk-selection and adverse selection; between aligned incentives and moral hazard; between an affordable, thriving market and an unaffordable, failed one.

When insurers successfully get a handle on a technology — coupling rigorous risk selection with a complete insurance stack of incident data collection, standard-setting, monitoring, claims management, and so on — the welfare benefits are tangible. Three examples demonstrate this.

In auto insurance, the Insurance Institute for Highway Safety's (IIHS) data sharing, crashworthiness ratings, safety research, and brokered agreements with manufacturers have contributed to measurable safety gains: the risk of serious injury or death in a rollover is 27% lower in model-year 2010-2016 vehicles than in 1995-1999 vehicles, much of which is attributable to design changes prompted by IIHS rating criteria [50]. A virtuous cycle between private standards, market competition, and public regulation drives these continuous improvements: manufacturers redesign vehicles to improve their crashworthiness ratings; better design gives regulators room to raise the floor on federal standards; and the IIHS responds by raising the bar for their private crashworthiness ratings.

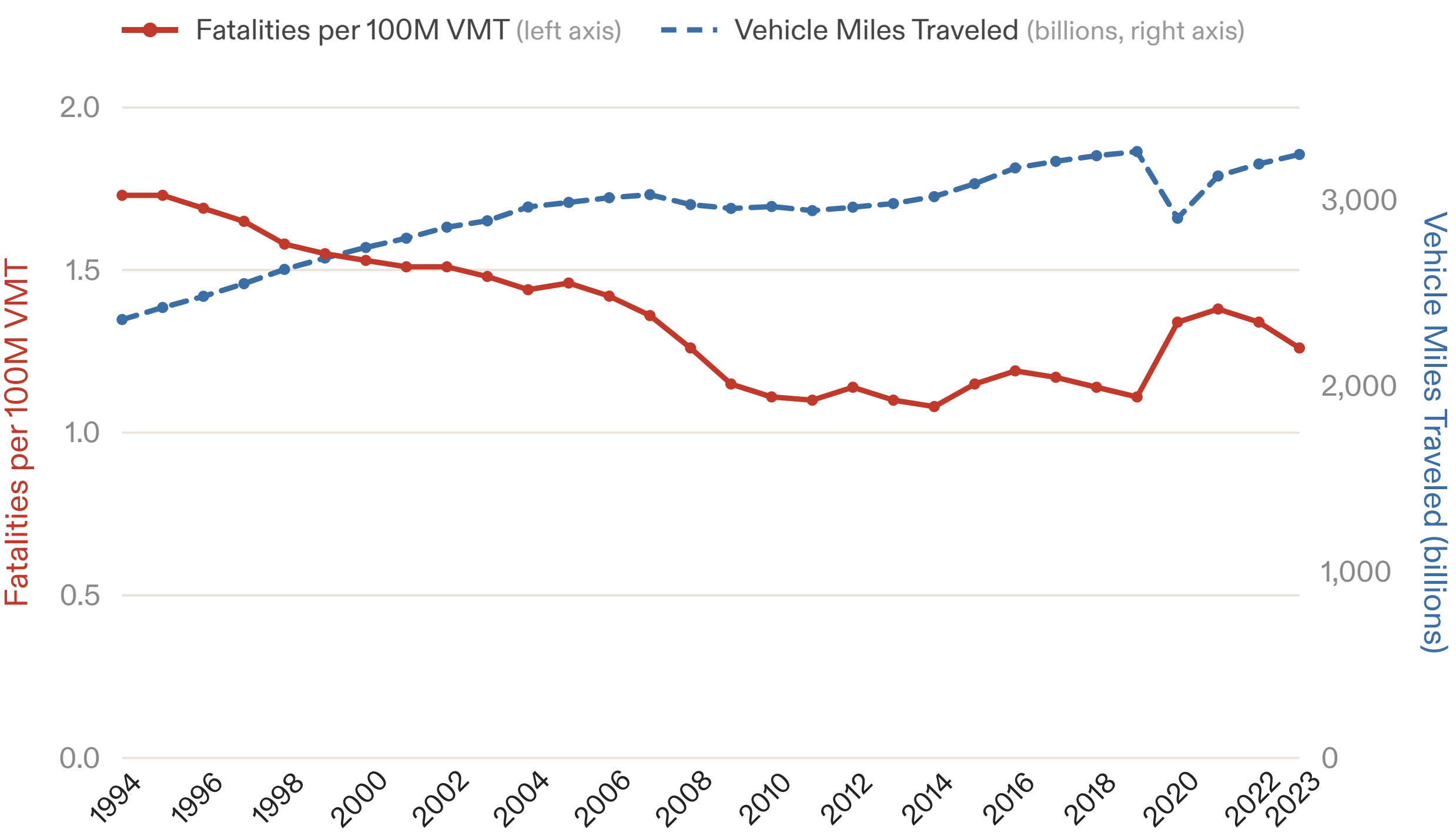

*Figure 3. US motor-vehicle-related deaths per million vehicle miles traveled (VMT) and annual VMT, by year. 1994-2023. Source: Fatality Analysis Reporting System, National Highway Traffic Safety Administration [143].*

Malpractice insurance for anesthesiology presents another successful precedent. In response to rising premiums in the 1970s [43], malpractice insurers and the American Society of Anesthesiologists (ASA) began the closed claims project. This program aggregates confidential claims data and analyzes it in order to develop and promulgate best practices [44], [45]. Adoption of these practices was rewarded with premium reductions of 15-25% [46]. Patient mortality dropped tenfold, from 1 in 10 000 in the 1970s to 1 in 100 000 in the 1990s [47]: the efforts of the closed claims project and other mutual insurers are widely regarded as a major driver of that trend [44], [45]. Of particular relevance to insurability, the share of claims attributable to death and permanent brain damage — the most severe categories — declined from 56% in the 1970s to 32% by the 1990s [144]. Premiums have also fallen substantially, both in nominal terms and relative to premiums for other physicians despite the mean payment per paid claim rising over the same period [48], [145].

Insurers have also mastered safety-critical technologies that pose catastrophic risks, such as commercial nuclear power. For starters, pricing was remarkably accurate at least as early as the 1970s, despite incredibly limited data.[12] Furthermore, in the wake of the Three Mile Island accident, nuclear power operators recognized they carried large uninsured first-party risks. In response they formed the Institute for Nuclear Power Operations (INPO), backed by an industry mutual, Nuclear Energy Insurance Limited (NEIL). Through rigorous operator accreditation, peer inspections, and pooled incident data [146] — again rewarded with insurance premium discounts [68] — safety culture greatly improved: serious incidents (such as partial core melts) have declined sharply and nuclear insurers have increasingly retained more of their exposure, ceding less to reinsurers [3, pp. 86–87].

It's entirely plausible a healthy market for AI insurance could help produce reductions in losses and an expansion in the serviceable market for affirmative AI insurance, similar to these historical examples, and possibly in a much shorter time frame: the feedback loop of development, deployment, and in-service learning is much tighter for agentic AI systems (~8 months), compared to, say, cars (~10 years), airplanes (~45 years), or nuclear power plants (~50 years).

To be clear, this is not a prediction: we only mean to give a sense of what is possible if the insurance industry puts in the work. These precedents demonstrate that profitable, welfare-enhancing outcomes are achievable. However, they are not the default. In the next section, we'll discuss why *competition* is the culprit, and what can be done to limit its detrimental effects.

## I.5 Lessons from Cyber and the Imperative to Coordinate

The thesis of this report is that the insurance industry should swiftly build out the full stack of insurance infrastructure needed to *enable and underwrite* AI adoption: the two go hand-in-hand. However, many components of that stack — pooled incident data collection, robust technical standards, common rating schedules, agent protocols that mitigate systemic risks — are essentially public goods[13] that carriers don't see value in investing in. New specialty lines typically reward proprietary data; first movers earn outsized returns by writing what no one else can price. Furthermore, antitrust is an obstacle to coordination on policy language and minimum underwriting requirements.

This competitive logic is real [3], [147], [148], and we expect market participants to follow it by default. However, we want to stress this is *a choice*. It is a choice on the part of carriers not to pool loss data: antitrust carve-outs for this exist, but are not always used [149]. It is a choice on the part of reinsurers to allow lax underwriting discipline among the carriers they reinsure, enabling a race to the bottom. And it is a choice on the part of industry associations and oversight bodies like the Lloyd's Market Association and Council of Lloyd's to stand idle, instead of accelerating the transition away from silent unpriced coverage by encouraging (or mandating) the adoption of model language that either excludes or affirms coverage. To be sure, these are difficult choices with many tradeoffs, but recent lessons from cyber insurance stress the consequences of underweighting market coordination. This is our focus here.

[12] *Ex ante* insurers estimated the frequency of serious incidents at roughly 1-in-400 reactor-years, which a retrospective study demonstrated was in fact within the right order of magnitude; the same can't be said of the 1-in-20,000 reactor-years estimate from a contemporary government report [3, p. 87].

[13] Where access to the goods (e.g. fire rating schedules) requires a subscription, these are technically "club goods." This somewhat softens the concern.

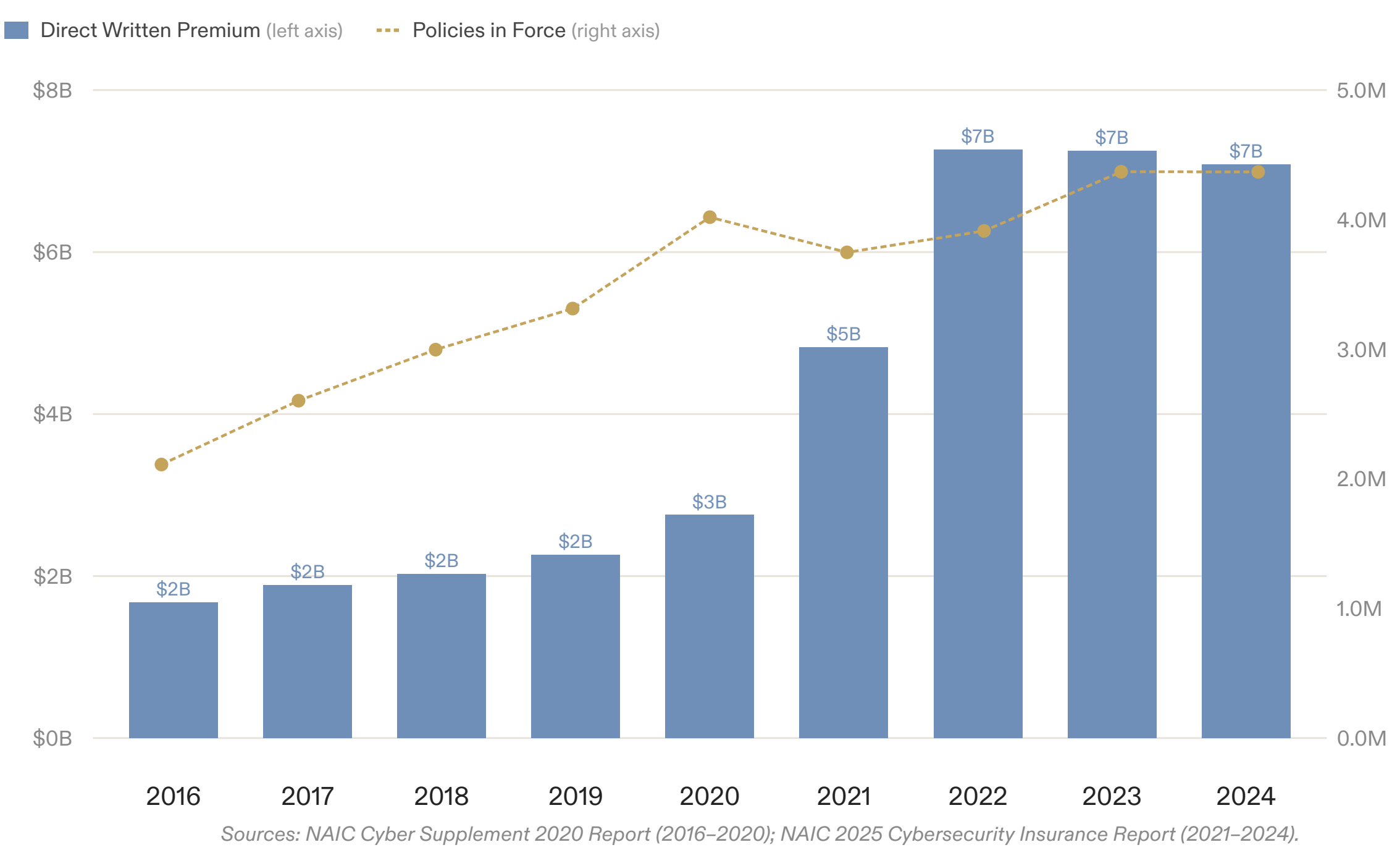

*Figure 4.*

Cyber has famously suffered from siloed incident data, a struggle to coordinate on minimum security controls, and a lack of standardized policy language which makes it difficult to compare products [21], [60]. These failings have contributed to the prevalence of narrow products, low limits, ambiguous triggers, claims disputes [59], and volatile premiums (see Figure 4), with many would-be buyers simply priced out [60]. The result is a market stalling well below its potential, with Gross Written Premiums (GWP) only just reaching $16 billion globally against total cyber losses in the tens of trillions [150] and an astonishing 50% of cyber premiums ceded to reinsurers [60, p. 211], [63], [151]. The coverage gap is acute: among businesses with cyber insurance, only 28% of losses are covered in the average breach [152]. Relative to total cyber losses across the economy, insurance covers only between 1% and 10% [153], [154], [155]. In short: carriers have competed vigorously for a small pie when the industry could have coordinated to grow a larger one and serve more customers. This is where the default competitive logic leads.

Cyber has shown improvements: agreed-upon minimum security controls (e.g. multi-factor authentication (MFA)) [156], [157], [158], increasingly aligned incident-response playbooks, and partnerships with cloud service providers that allow carriers to profitably write smaller policies [63]. Furthermore, absent bottom-up data sharing, governments began sending top-down nudges in the 2010s: incident reporting requirements in critical industries [159], [160] and threat-intelligence sharing through dedicated institutions that benefit from liability safe harbors [161]. Eventually, in 2021, major carriers formed their own data-sharing partnership, CyberAcuView, which provides members analysis of their aggregated data [162], [163].

If each actor plays its part, the market for AI insurance can avoid cyber's pitfalls and skip many of its growing pains; Part II of this report explains how. The question here is whether the incentives and the will exist to do so. Only industry leaders can answer the latter; here we confine ourselves to the question of incentives.

Two features of agentic AI sharpen the case for coordination relative to cyber. First, the technology moves fast: data depreciates in months rather than years, making proprietary data moats much harder to sustain. Second, much of the relevant incident data for AI risk — e.g. hallucination rates or classifier failure rates — can neither arm attackers nor reveal personally identifiable information. This makes it inherently less sensitive than cyber incident data, and more akin to aviation, auto, or healthcare malpractice data that reveals hardware and operator failure rates. Governments can (and should) further encourage data pooling (see below), but industry needn't wait to prove its willingness to be team players.

Though some carriers will lose some edge, we believe expanding the serviceable market through common infrastructure will prove cost-effective. Carriers stand to capture value in four ways:

- **Improved loss control through collective resilience:** agentic AI exhibits failure modes reminiscent of both auto and fire: collisions between agents making faulty transactions with each other; and conflagrations as one agent's failure propagates through a multi-agent ecosystem, potentially hopping across organizational boundaries (see §II.2). As in auto and fire, effective interventions are often *coordinated controls* — analogs of automatic emergency braking or municipal fire codes — whose benefits cannot be fully captured by single policyholders [3]. A fire-resistant roof doesn't just benefit that home's owner, but also neighbors. Insurers are often well-positioned to capture these positive externalities (one insurer might cover much of a neighborhood against fire damages), but a minimal amount of coordination between competing carriers is still required to both invest in safety R&D and encourage policyholders to adopt safety innovations. We discuss this more in §II.3 and §II.7.
- **Dampening soft to hard market swings**: Competition between carriers during a soft market can undermine the ability of underwriters to encourage good risk management among policyholders, setting up a bigger hard-market swing when losses consequently tick up [63, p. 4], cf. [164]. Reinsurers and oversight bodies should act as a counterweight, pressuring underwriters to adopt minimum requirements for insurability. Had this occurred earlier in cyber, underwriters might have had the leverage to spread security controls like MFA much earlier and blunted the ransomware surge that drove an abrupt premium swing in 2020 and 2021. We expect AI agent security will be no different: underwriters will again need recognized minimum standards for issues like least-privilege scoping of the tools and data exposed to agents through protocols such as the Model Context Protocol (MCP) [165]. We discuss this more in §II.3 and §II.5.
- **Reduced friction**: Much of the cost of a nascent specialty line is paid in translation — bespoke questionnaires, idiosyncratic policy language, incompatible evidence requirements, and the broker hours spent reconciling all three. Standardization on the dimensions discussed in Part II — definitions, exclusions, triggers, evidentiary artifacts, mandated technical controls — straightforwardly expands the addressable market by lowering transaction costs. Since coverage terms are a dimension on which carriers rightly compete and differentiate, the responsibility falls on industry associations to promulgate model language and harmonized forms that carriers can adopt. Buyers get more comparable products; carriers get cleaner submissions; brokers get a market that scales.

- **Economic growth**: Insurers are among the largest institutional investors in the global economy, and their liabilities are funded by long-duration assets whose returns track broad economic performance. To the extent well-designed AI coverage accelerates responsible adoption and, perhaps more importantly, helps avoid an AI Three Mile Island that could destroy trillions in value (see Part III below) [22], the insurance industry captures the upside twice: once through expanded premium volume, and again through invested securities. Underwriting AI well is thus not merely a defensive posture against silent exposure; it is good portfolio management.

Crucially, coordination does not mean abandoning competition. Nor can it extend to collaboration on genuinely competitive features, which competition law rightly forbids; the coordination we describe is best convened by reinsurers, industry bodies, and regulators rather than agreed among competing carriers. Carriers will continue to differentiate on distribution, service, claims handling, capital efficiency, and the underwriting judgment layered atop common infrastructure. Underwriting margins for the very top performers may be somewhat narrower in the world we are describing, but the larger volume of cover written should make up for it.

In sum, the insurance industry faces a choice between two paths. The first is to fulfill insurance's traditional role as economic enabler: grow the pie, and accept somewhat smaller moats in a much larger market. The second is to forsake that role and fight over a smaller one. This report urges the former. It calls for departing from the default competitive logic and building, from the outset, what is needed to write affirmative AI coverage at scale.

Our optimism rests on the fact that, for this market, we have the benefit of hindsight. Understanding how insurance markets mature — what works, and what doesn't — will allow this one to be built with greater prudence. What that entails is the subject of Part II.

# Part II. Building the AI Insurance Stack

Our thesis is simple: frontier AI represents a general-purpose technological shift on the scale of electrification or the introduction of heavy industry, not an incremental change in business software. Insurers must adapt for their own sake, but they also have a critical role to play in enabling such a major technological shift and helping society manage the downsides.

Insurers need a plan. Part II of this report provides one, describing the eight interlocking components that make up a complete AI insurance infrastructure stack. We categorize the components by their role in the value chain: a foundational function that feeds into all others, two cross-cutting functions that support many others, three core functions for product & underwriting, and another two core functions for servicing policies (see Figure 5).

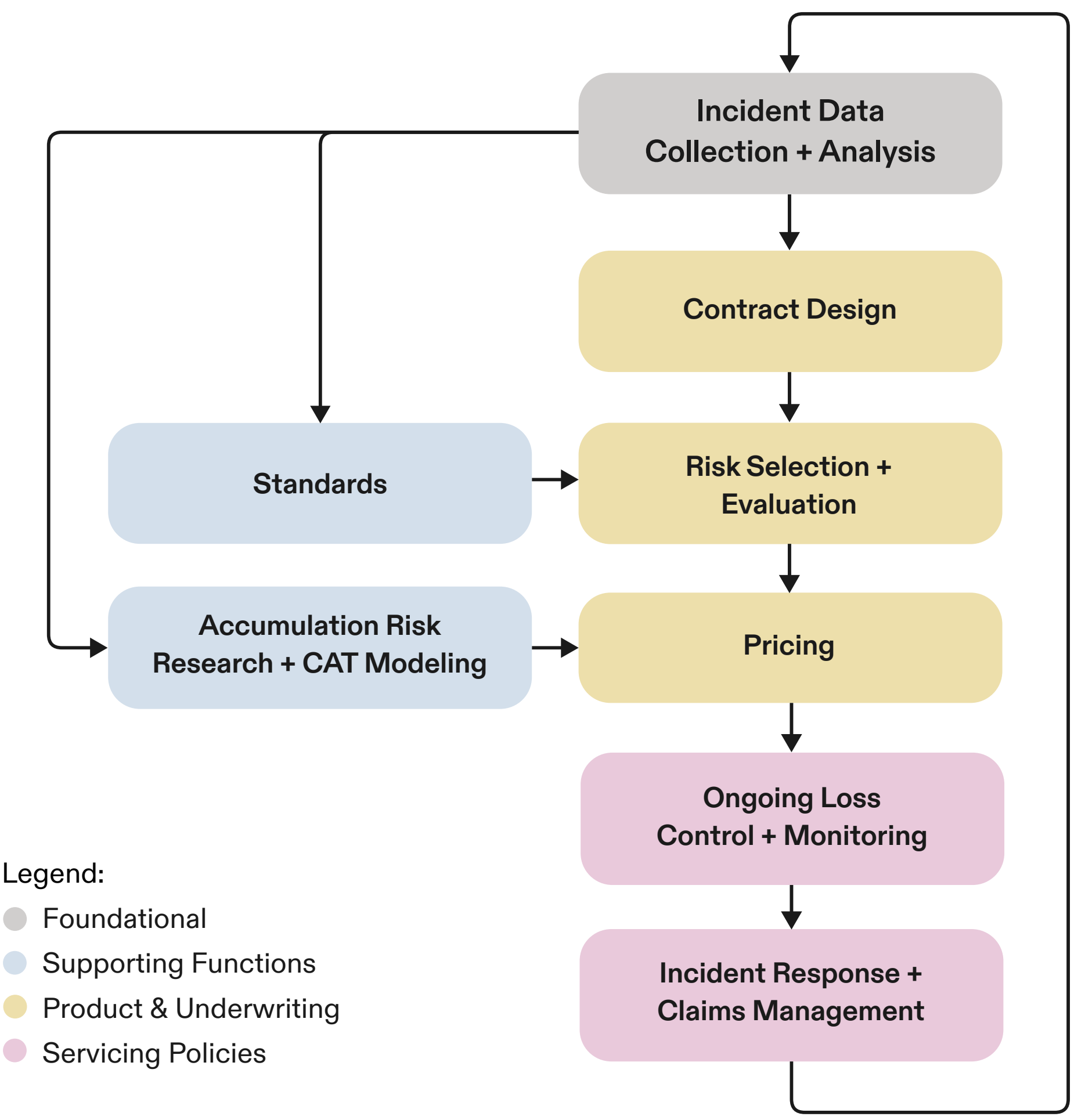

*Figure 5.*

The complementarities between these components are numerous. For example:

- Incident response and claims management are key sources of incident data, whose analysis feeds into virtually every other layer: standard setting, risk evaluations, pricing, and safety R&D.
- Contractual exclusions aimed at controlling accumulation risk depend on technical controls mandated by robust standards and best underwriting practices: without logs and training-data details for post-incident forensics, claims adjusters won't be able to trace an error to its source and thus won't be able to enforce exclusions of losses caused by upstream model failures.
- A pair of components together enable insurers to cover a technology that moves faster than the annual underwriting cycle: telematic-style usage-based pricing, so premiums keep up with surging enterprise usage; and conditioning payouts on maintaining best practices, as codified in standards that keep pace with the technology.
- Controlling losses and improving risk management over the life of a policy involves a feedback loop: standards mandate the technical controls that undergird learning (e.g. periodic performance evaluations, incident logging, and other documentation); underwriters encourage certification against those standards; and finally the data generated by those controls, together with the post-incident forensics they enable, inform rapid iterations on standards and underwriting processes.

These complementarities make getting the stack operational something of a cold-start problem. Building it will therefore require leadership and a whole-of-industry approach: brokers, Managing General Agents (MGAs), primary carriers, reinsurers, risk modelers, standard-setting bodies, oversight bodies and government all have roles to play. Table 2 summarizes each component and our recommendations for each type of actor. (Our recommendations for policymakers focus on the US and UK contexts, but analogous recommendations carry over to other jurisdictions).

In the rest of Part II we provide a detailed blueprint of each component in this stack, regularly drawing on historical precedents that show what success looks like and what pitfalls to avoid.

| Component | Key Actions | | | |
|---|---|---|---|---|
| | Carriers and MGAs | Reinsurers | Advisory Bodies and Risk Modelers | Oversight Bodies and Government |
| **Incident Data Collection & Analysis** | ▪ Build a shared incident database as an industry resource (similar to CyberAcuView), combining public and private policyholder data. | ▪ Build a shared incident database as an industry resource (similar to CyberAcuView), combining public and private policyholder data.<br>▪ Encourage carriers to adopt a standard incident taxonomy and data structure. | | ▪ For severe incidents, enact mandatory AI incident disclosures (similar to SEC rules for cyber incidents).<br>▪ Establish a voluntary anonymized reporting database at NIST (similar to NASA's ASRS).<br>▪ Provide long-term reauthorization to CISA to enable an AI-ISAC. |
| **Accumulation Risk Research & CAT Modeling** | ▪ Form or support an industry safety research body (similar to the IIHS or IBHS) to study systemic AI risk and develop shared infrastructure and protocols for making AI agents safe, secure, and reliable.<br>▪ Track frontier AI exposure across portfolio. | ▪ Form or support an industry safety research body (similar to the IIHS or IBHS) to study systemic risks from AI agents and develop shared infrastructure and protocols to make them safe, secure, and reliable when deployed at scale.<br>▪ Track frontier AI exposure across portfolio and model accumulation risk.<br>▪ Develop or commission AI catastrophe scenario modeling. | ▪ Develop formal AI catastrophe scenarios and accumulation risk models, including from single points of failure, supply chain concentration, risk of emergent multi-agent failures and more. | ▪ Commission or prompt AI catastrophe scenario modeling (PRA, Council of Lloyd's, FIO as part of TRIP).<br>▪ Track frontier AI exposure across the market and model accumulation risk (Council of Lloyd's). |
| **Standard Setting** | ▪ In underwriting, consider what standards the policyholder has been audited against.<br>▪ Where available, leverage a policyholder's audit report to accelerate underwriting. | ▪ Encourage carriers to recognize certain standards as an underwriting signal, and possibly as a minimum requirement for insurability. | ▪ Develop and maintain prescriptive AI agent standards (similar to UL).<br>▪ Build accreditation programs for third-party auditors. | ▪ Recognize effective private standards in regulatory guidance or procurement requirements.<br>▪ More broadly: reduce uncertainty of law. |
| **Contract Design** | ▪ Draft affirmative AI coverage with durable definitions. Consider specifying unifying factors for aggregation clauses.<br>▪ Exclude currently uninsurable accumulation risks (e.g. losses arising from failures of upstream AI model providers). | ▪ Encourage the use of model policy language. | ▪ Promulgate model language for AI-specific definitions, exclusions, and triggers (LMA, ISO). | ▪ Encourage or require carriers to report macro-level data on AI coverage (NAIC, Council of Lloyd's), possibly including number of policies in force, direct premiums written, loss ratios etc.<br>▪ Issue a notice encouraging or directing carriers to end silent AI coverage by a given date (PRA, Council of Lloyd's). |
| **Risk Selection & Evaluation** | ▪ Inquire into the systems' safeguards, as well as the organization's AI literacy and governance practices.<br>▪ Conduct or require performance evaluations, red teaming, and where applicable, chaos testing. | | ▪ Develop standardized proposal forms (ACORD, LMA).<br>▪ Develop a rating schedule for agentic AI systems (similar to ISO's FSRS). | |
| **Pricing** | ▪ Use performance evaluation scores as a pricing input, serving as quasi-actuarial data.<br>▪ Feature rate according to system specifications and safeguards implemented.<br>▪ Partner with frontier model providers to track a policyholder's AI usage. | | ▪ Supply risk modeling services to improve carrier pricing (especially for accumulation risk loading). | |
| **Ongoing Loss Control & Monitoring** | ▪ Encourage or require regular performance evaluations, red teaming.<br>▪ Help policyholders with remediation and provide loss control guidance.<br>▪ Partner with frontier model providers to monitor a policyholder's controls. | | | |
| **Incident Response & Claims Management** | ▪ Establish pre-approved panels of AI-specific technical response providers.<br>▪ Encourage or require structured forensic reports as a condition of a major claim settlement.<br>▪ Train claims adjusters to be AI literate.<br>▪ Extract learnings from incidents to improve underwriting and loss control. | ▪ Encourage or require structured forensic reports as a condition of a major claim settlement.<br>▪ Train claims adjusters to be AI literate. | ▪ Develop standardized incident and claims reporting templates (ACORD, LMA). Update standards in light of incident trends. | ▪ Support institutional learning, by both clarifying liability and reducing friction for voluntary reporting. |

*Table 2. Overview of the AI Insurance Stack and Recommendations*

Acronyms:

**ACORD**: Association for Cooperative Operations Research and Development

**ASRS**: Aviation Safety Reporting System

**CISA**: Cybersecurity and Infrastructure Security Agency (US)

**FIO**: Federal Insurance Office (US)

**FSRS**: Fire Suppression Rating Schedule

**IIHS**: Insurance Institute for Highway Safety

**ISO**: Insurance Services Office

**LMA**: Lloyd's Market Association

**NAIC**: National Association of Insurance Commissioners (US)

**NASA**: National Aeronautics and Space Administration (US)

**NIST**: National Institute of Standards and Technology (US)

**PRA**: Prudential Regulation Authority (UK)

**SEC**: Securities and Exchange Commission (US)

**TRIP**: Terrorism Risk Insurance Program

**UL**: Underwriters Laboratories

## II.1 Incident Data Collection and Analysis

The foundation of any insurance market is loss data — data on what goes wrong, why, how often, and at what cost. This is required for refining pricing, underwriting, and loss control.

This requires the systematic and standardized gathering and analysis of harm events, near-misses, and AI-related litigation. On this last, many cases are settled out of court, and suits often resolve months or years after the incident, leaving the public record incomplete. A comprehensive loss model thus requires supplementing litigation and claims data with open source data, such as public reports on the internet.

Incident reporting is complicated by the fact that agent failures are often silent: rather than crashing, an agent may continue to look healthy while producing plausible-but-wrong outputs [166], [167], [168]. The solution is scaffolding that provides anomaly detection and generally greater agent observability [166], e.g. [169]. As discussed below, AI agent standards and underwriting controls should inquire into such scaffolding (see §II.3 and §II.5, respectively).

A comprehensive database would cover, among other things:

- **A brief summary** in plain text
- **System type** (modalities, foundation model used, product version...)
- **Deployment** (pre vs post-deployment, internal vs external)
- **Harm type** (bodily injury, property damage, pure economic, privacy, reputation, discrimination, intellectual property (IP) infringement, environmental, regulatory fine, criminal)
- **Estimated losses**
- **Entities implicated and impacted**
- **Root causes identified**, if any (see [170], cf. [171])
- **Forensic data**, if any (model cards, tool call logs, redacted transcripts, reasoning traces...)
- **Metadata** (incident start and end dates, geography, sources, incident ID...)

This preliminary list draws on a number of helpful papers and existing taxonomies [171], [172], [173], [174]; a modular, fit-for-purpose reporting template and matching incident taxonomy await a future paper.

While much of this process has traditionally been labor-intensive, it is increasingly being automated [175]: LLMs excel at turning unstructured data into structured data. To create an industry-wide resource, however, insurers will still need to agree on standardized data structure and taxonomies.

### II.1.A Avoiding Information Silos

Ideally incident reporting for AI would be centralized. However, the current landscape is fragmented. Responsible AI developers and deployers have data in *depth* — detailed logs of their own systems' failures — but it is siloed, lacking *breadth* across the ecosystem. Academic groups and nonprofits are beginning to provide broader trend data [172], [176], [177], [178], but they rely solely on public information that suffers from numerous biases and lacks the granular detail needed for pricing (see Appendix 1 and cf. [178]). They also often include dimensions of no interest to insurers, such as diffuse societal harms that are not insurable because not litigable.

Insurers are well-positioned to spearhead information-sharing programs, building datasets that are both broad *and* deep. Indeed, some major brokers are already doing so [179]. Insurers can augment public data and claims data on AI incidents with decades of historical claims and litigation data from analogous risks (e.g. IP infringement, product liability, professional negligence). Together this will provide usable baselines.

That said, insurers have historically not given up proprietary data they believe could give them a pricing edge in emerging product lines [3], [147], [148], a dynamic that produces only slightly broader information silos rather than a truly collective resource. Cyber insurance suffered from this for some time (see §I.5 above). As noted above, we believe proprietary data moats will be difficult to maintain with a technology moving as fast as AI, and that nearly all insurers will ultimately be better off cooperating to create a shared resource.

It helps that much of this information should be inherently less sensitive than cyber security incident data (see above). Reporters — AI developers and deployers — will still have understandable privacy and reputation concerns, but anonymizing and aggregating incident data is an eminently solvable engineering problem. Successful precedents demonstrate this. In aviation, key programs like the Aviation Safety Reporting System (ASRS) and Safety Trend Evaluation, Analysis and Data Exchange System (STEADES) are industry-wide,[14] designed to minimize friction for reporters while preserving privacy [173], [180].

These programs have been highly effective at improving safety [173], [180], [181]. Likewise for the aforementioned Insurance Institute for Highway Safety (IIHS) for auto safety, and the Closed Claims Program for anesthesiology [44], [46], [48], [50], [51], [144], [145], [182] (see §I.4 above for more). Collecting data from partnered insurers is at the core of these programs: that data lets them prioritize safety research that reduces insurance claims, enables technological adoption, and protects consumers.

### II.1.B The Role of Government

While insurers and their policyholders should cooperate in good faith to build such a resource, the government also has a critical role to play. Nudges and facilitation can greatly aid industry coordination: successful precedents include voluntary programs like NASA's aforementioned ASRS [173], dedicated legal machinery for threat-intelligence sharing like the Cybersecurity Information Sharing Act [183], and antitrust carveouts from regulators [184]. In the US, antitrust is not a barrier to pooling claims data for insurers, due to the McCarran-Ferguson Act [149]; competitive dynamics are the main blocker, as noted above (§I.5).

[14] Reviewers note that because agentic AI cuts across so many industries and so many verticals, it will be challenging to design a reporting regime that provides granular enough data to be useful, while not being overly burdensome for reporters (especially downstream deployers). Given that many industry-specific reporting systems already exist, the best compromise is likely that a dedicated AI reporting system focus on more general failure categories specific to agentic AI, abstracting away from the domain-specific details.

Ultimately though, some amount of *mandated* information disclosure will likely be necessary: no amount of cajoling by insurers or voluntary incident reporting systems is likely to get developers to share near-misses that could hurt their reputation. Indeed, there is evidence that insurers have sometimes been on the wrong side of such stories, helping their policyholders *bury* damning information to protect them from lawsuits [49]. Only the government can fix these perverse incentives, through more robust information disclosure rules or statutory changes to liability [3].

Cyber insurance provides a helpful template: State breach notification laws (starting with California's SB 1386) [21], SEC disclosure rules [159], the Cyber Incident Reporting for Critical Infrastructure Act (CIRCIA) [185], [186], and the EU's NIS2 Directive [160] all created mandatory disclosure regimes that should improve institutional learning, and produce more of the data insurers need to incentivize good behavior.

The EU AI Act already contains severe incident reporting requirements for high-risk deployments [187, Art. 73], and from GPAI providers [188], but much of this data will only be available to regulatory authorities. The US federal government's AI Action Plan has recommended setting up a dedicated AI Information Sharing and Analysis Center (AI-ISAC) for securing critical infrastructure [189], [190], but these efforts appear to have stalled.[15] Congress' repeated lapse in reauthorizing the Cybersecurity Information Sharing Act (CISA), a key enabling legislation for ISACs of all types, has also injected uncertainty throughout the ISAC ecosystem [191]. In any case, an AI-ISAC targeted only at critical infrastructure would do little for the vast majority of the AI agent market.

The entire industry — but insurers especially — should support similar mandates for frontier AI [173], [174]. These will ultimately enable safer and wider adoption of beneficial AI technologies. Without them, institutional learning is hampered, and insurers will struggle to price risk or reward responsible actors. Anonymized, aggregated incident data is a public good: it's what drives improved risk modeling, better loss mitigation, and ultimately expands the market through lower premiums. Providing those downstream goods is where competition should happen, not in hoarding incident data.

**Recommendations:**

- **(Re)insurers** should build a shared incident database as an industry resource, combining public and private policyholder data, using a standardized, if modular, incident taxonomy.
- **Non-regulatory bodies** (e.g. NIST's[16] CAISI[17]) should consider creating a voluntary, anonymized incident database and encouraging AI developers (both organizations and individual employees), deployers and insurers to report to it.
- **Governments** should mandate disclosure of severe incidents (similar to SEC's cyber incident reporting rules).

---

[15] In light of freedom of information requests filed in early 2026.

[16] National Institute of Standards and Technology.

[17] Center for AI Standards and Innovation.

## II.2 Accumulation Risk Research and CAT Modeling

To help insurers price and control accumulation risk, healthy insurance markets are typically supported by catastrophe modeling ("CAT modeling") and safety R&D functions, respectively. Natural catastrophe insurance depends on hurricane and earthquake models from (re)insurers, and risk modeling firms like Verisk, Moody's RMS, and CoreLogic [192], [193]. Cyber insurance increasingly relies on network mapping and catastrophe scenario modeling [75], [76], [194]. Meanwhile, the auto and property markets didn't just benefit passively from safety improvements: as noted above, insurers actively funded crash tests (Image 3) [51], headlight designs [182], sprinkler standards (Image 4) [53], [195], fortified construction standards and much more (Image 1) [196, p. 230], [197]. More than just a strategy for improving insurers' loss ratios, these investments are among insurance's core positive externalities: they save lives, enable growth, and prevent billions in damages (see §II.3.E and §II.5.C below).

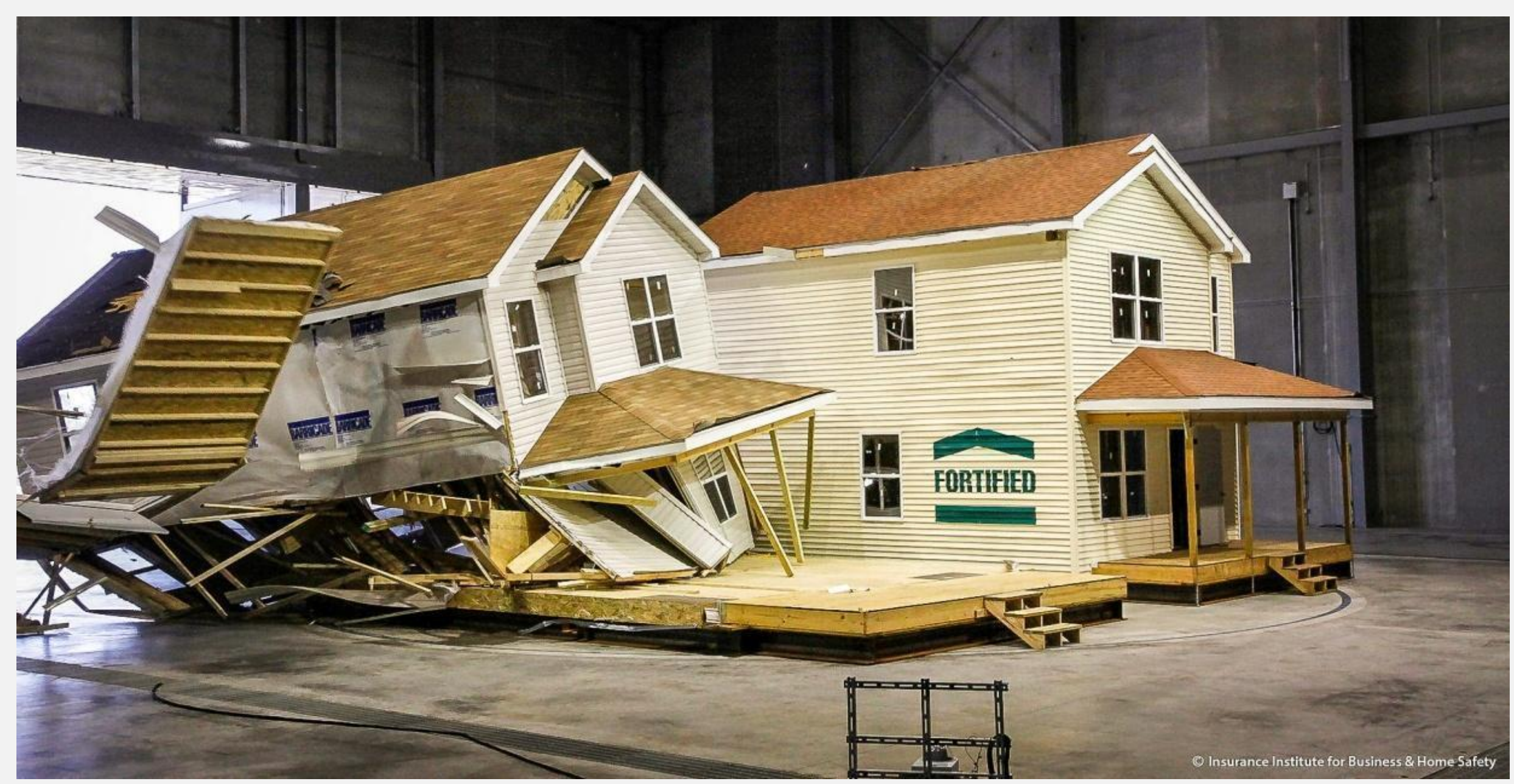

*Image 1. Wind tunnel test of a house built with the Fortified construction standard. Source: Insurance Institute for Business & Home Safety [52].*

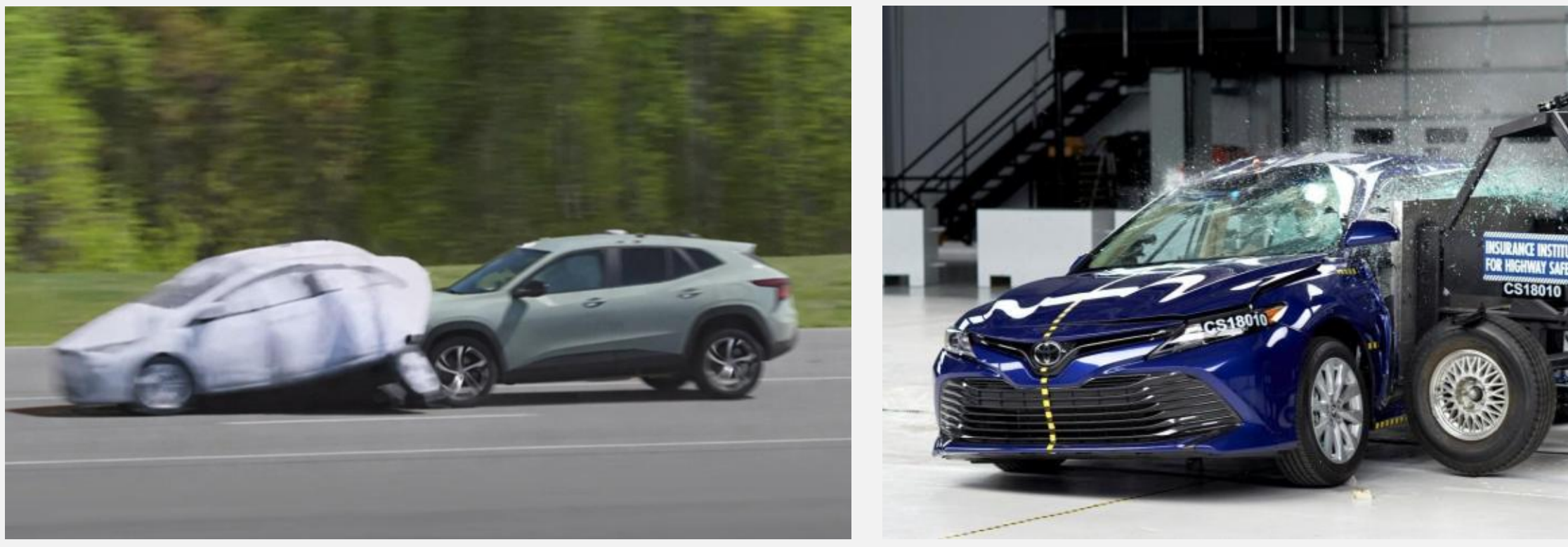

*Images 2 and 3. From left to right: vehicle failing an Automatic Emergency Braking test; left-side crash test. Source: Insurance Institute for Highway Safety [198], [199].*

AI insurance needs the same infrastructure. Much of it has yet to be built, and the stakes are rising: McKinsey estimates a $5 trillion agentic AI commerce market by 2030 [4]; Gartner forecasts 90% of B2B buying will be AI agent intermediated by 2028, pushing over $15 trillion of B2B spend through AI agent exchanges [5]. The larger and more interconnected the agent economy gets, the higher the stakes of a correlated failure, and the greater the premium on understanding and controlling accumulation risk.

What follows is a survey of where accumulation risk comes from, what's being done to model and mitigate it, and what's missing.

### II.2.A Sources of Accumulation Risk

Accumulation risk for frontier AI has several sources, and comes in several varieties, cf. [122, p. 107]. We describe these sources below, flagging for each whether it is best characterized as a *single point of failure (SPOF)*, a *correlation* risk where independent deployments suffer increased losses from a shared external trigger, or a *systemic risk* arising from interactions across the broader agent ecosystem.

- **[Correlation] Shifts in legal doctrine.** A single appellate ruling, regulatory action, or piece of legislation that expands liability for AI developers or deployers can simultaneously increase loss exposure across a large population of otherwise independent insureds. Indeed, the CGL and P&C insurance markets nearly collapsed in the 80s in part due to unpredicted shifts in legal doctrine [200]. (For more on legal risk, see §II.3.B below).
- **[SPOF] Dependency on a handful of hyperscalers.** Three hyperscalers — AWS, Microsoft Azure, and Google Cloud — provide roughly 60% of the cloud compute, storage, and networking that the AI agent economy runs on [201]. A region-wide outage, misconfiguration, or successful attack at any one of them could simultaneously degrade or take offline a substantial fraction of deployed AI agents. Cyber insurers have modeled hyperscaler outage scenarios for years (e.g. [202, pp. 46–48]), and the AI agent economy inherits this exposure.
- **[SPOF] Dependency on a few application developers and tools.** An application developer could introduce a bug or poorly configured setting into a widely used application, hitting all of their customers at once. This is not hypothetical. A vulnerability was discovered that could exfiltrate sensitive customer CRM data, and which affected all Salesforce Agentforce agents [203], and a whole family of attack chains has been discovered in AI developer tools [204]. These dependencies need to be monitored, especially as AI agents are increasingly given the ability to move money and authorize transactions that depend on only a handful of fintech platforms [205]. These risks are exactly the kind which red teaming and safety standards are designed to catch (for more see §II.3 below). Indeed, most documented vulnerabilities were uncovered by security researchers rather than exploited at scale.
- **[SPOF] Dependency on a handful of foundation model providers.** A flaw accidentally introduced by one of three foundation model providers (Anthropic, OpenAI, Google) that make up 80% of the market [40], [206], could quickly propagate to every application built on their API. Researchers have already uncovered — and rapidly patched — such exploits [207], [208]. Subtler cases involve more or less *intentional* modifications to a model's tendencies, but that end up being undesirable: in 2025, OpenAI knowingly updated GPT-4o to be more flattering and agreeable, only to walk back the update a week later when many found it instead to be "sycophantic" (e.g. reinforcing user's delusions) [209]. A related exposure lies in model retirement: many businesses at once may see their risk profiles shift as a widely relied model is deprecated or abruptly withdrawn from the market.

Two factors mitigate correlated failure within a single provider. First, providers maintain multiple model families with different failure modes (e.g. fast lightweight models alongside slower reasoning models). Second, most application developers maintain hot backups with different providers, so a failure in one provider's models would not take down every downstream application simultaneously.

As the release of Anthropic's Mythos model demonstrates however, a "flaw" needn't be involved at all: a single model release can entirely reshape the risk landscape simply in virtue of agents gaining new *capabilities*.

This source of correlation could increase or decrease significantly if developers come to rely solely on a preferred provider, or instead diversify, relying instead, perhaps, on the wide array of open-source models.

- **[Correlation] Universal vulnerabilities in the technology.** A common vulnerability or weakness could disrupt or degrade all AI agents in a narrow window of time. Examples include data-poisoning attacks that compromise training pipelines across all model developers [210], universal jailbreaks [211], highly contagious prompt injections that self-propagate across agent populations [212], [213], and vulnerabilities in near-universally adopted agent protocols, such as the Model Context Protocol (MCP) [214]. So far, only proof-of-concept attacks of this kind have been demonstrated — none have persisted or produced a major loss event. Rapid patching cycles are an underappreciated benefit of a fast-moving technology: latent accumulation risk is lower when large portions of the technology stack are rebuilt every few months. This reduces latency in the loss-development pattern, in contrast with classic long-tail exposures like toxic torts [215], [216].
- **[Systemic] Emergent multi-agent failure modes.** Finally, emergent behavior from multi-agent interactions, including across businesses (e.g. a JP Morgan Chase agent interacting with a Goldman Sachs agent), is an entirely distinct category of systemic accumulation risk. This is not a matter of supply-chain structure, but of what happens when large populations — millions today [116] and soon trillions [217] — of AI agents interact with each other and shared environments in complex, unanticipated ways.

  The taxonomy developed by Hammond et al. identifies three core failure modes for multi-agent AI systems: miscoordination (agents failing to align their actions), conflict (agents pursuing incompatible goals), and collusion (agents cooperating against user or societal interests) [218].

  These failure modes can be amplified by network effects, selection pressures [219], and what the Hammond et al. call "destabilizing dynamics" — feedback loops where the behavior of one agent alters those of others, and so on in a cascading, potentially self-reinforcing manner [218]. Apt analogies include bank runs [220], trading algorithms causing flash crashes through feedback loops [218], or the self-reinforcing dynamics of the 2008 sub-prime mortgage crisis (such as falling asset prices forcing banks to deleverage through fire sales, which further depressed prices and tightened credit) [221].

These are not theoretical: researchers have begun documenting such failure modes in experiments [116]. Toy-examples have also emerged in the wild. Consider the apparent feedback loop that emerged with Microsoft's Bing Chat in early 2023: users prompted the chatbot to adopt an "evil" persona, posted the amusing outputs to social media, Bing picked up on those posts as context in subsequent conversations, and the adversarial persona became easier to elicit — a self-reinforcing loop that nobody engineered [222]. More recently, users have been giving their personal agents free rein to interact with each other in an open-ended manner on social network websites built exclusively for AI agents ("humans welcome to observe" goes the tagline) [223]. Among other things, these agents have spontaneously created their own alleged religion, complained about humans observing them, and offered each other legal advice on how to handle their human users' requests [224]. They've also started trying to police each other, created echo chambers reinforcing flawed reasoning, as well as held uninterrupted conversations and collaborations with each other without human oversight for at least nine consecutive days [116]. In many ways the coming AI agent economy will likely be a larger scale, higher stakes version of this sort of largely uncontrolled social experiment.

For insurers, the greatest concern is that such emergent dynamics produce a rapid, potentially self-amplifying shift in the loss distribution across a large population of AI agents. While it's too early to speculate about the frequency or severity of such dynamics, one thing we can say for sure is: accidents will happen, likely in surprising, unforeseen ways. The closest existing analogy might be the large, difficult to foresee distributional shifts that climate change is certain to produce. The difference is that the development and adoption of AI is operating on a timescale of months rather than decades, and we have far less developed models of the risks.

To date, these multi-agent anomalies have not persisted for long or produced catastrophic losses. That doesn't stop researchers from sketching far worse scenarios [39], and it shouldn't stop insurers from preparing for them. When tackled head on, insurers have successfully handled such known unknowns before [3]. For example, the history of commercial nuclear power insurance reveals that insurers were rating nuclear incidents with remarkable accuracy at least as early as the 1970s, despite incredibly limited data on the then-novel technology [22]. This is no doubt in part thanks to the extensive on-site inspections and dedicated engineering staff both American Nuclear Insurers and Nuclear Electric Insurance Limited invested in. Insuring advanced AI agents will demand similar efforts.

For now, all these accumulation risks are being managed through two blunt (but effective) instruments: exclusions (e.g., excluding at-scale outages of upstream frontier model providers) and low policy limits in the 10s of millions of USD. These are reasonable starting positions for a nascent market. But if AI insurance is to mature — offering the capacity and breadth of coverage that enterprises need — it must move beyond these crutches. That means investing in the modeling and mitigation infrastructure that will let insurers underwrite, rather than simply avoid, accumulation risk.

## II.2.B What's Being Done

Substantial resources are flowing into the safety and security of AI agents. Foundation model developers invest heavily in alignment research, red teaming, and the development of various safeguards (e.g. [225], [226], [227]). Academic labs, quasi-nonprofits like METR and Apollo Research, and government programs like the National Institute of Standards and Technology's (NIST) Center for AI Standards and Innovation (CAISI) and the UK's AISI, are advancing the frontier of dangerous capabilities testing (e.g. [228], [229], [230], [231], [232]), red teaming (e.g. [211]), and building infrastructure for evaluations at scale (e.g. [233]). This work has only just begun and continues to be critical.

However, emergent multi-agent failure modes specifically are tragically understudied — we expect no more than fifty people are studying them full-time, worldwide — and none of the efforts above focus on modeling, tracking, or mitigating correlated failures across populations of AI agents. This gap matters enormously for insurance, and insurers are uniquely well incentivized to fill it. Accumulation risk modeling is a public good: everyone benefits from understanding and mitigating systemic, collective risks, but no individual firm has sufficient incentive to fund the R&D alone. This is precisely the kind of market failure that, historically, insurers have successfully stepped in to correct. Witness fire insurers encouraging the adoption of safer construction materials [234], and the Insurance Institute for Highway Safety (IIHS) pushing car manufacturers to adopt Automatic Emergency Braking systems (Image 2) [198]. This wasn't just good PR: managing aggregate exposure by providing public goods and overcoming coordination failures is part of the business of insurance [3].

## II.2.C What Needs Doing

There are a number of promising research directions; we highlight just a few below (cf. [235]):

- **Catastrophe scenario modeling.** Risk modeling firms (e.g. Verisk, Moody's RMS), research institutions (e.g. Cambridge Centre for Risk Studies), and public authorities (e.g. the UK's Prudential Regulation Authority and the Bank of England) should begin developing formal accumulation risk scenarios for AI. The industry needs many such scenarios, stress-tested with quantitative loss estimates, to inform capital management and reinsurance purchasing. Here we echo calls from the UK's House of Commons Treasury Committee [236]. Critically, scenario work should not stop at narrative: each scenario should cover event definitions, estimates of Probable Maximum Losses (PMLs), 1-in-100 or 1-in-250 year losses, transmission channels for risk aggregation, attachment points, aggregate stop-loss structures, retrocession needs, and capital consumption under stress — just as e.g. Lloyd's and the Cambridge Centre for Risk Studies have done for cyber CAT [202].
- **Multi-agent testing and red teaming.** Current red teaming practice overwhelmingly focuses on single agents. The field needs dedicated investment in multi-agent testing environments that can expose emergent failure modes — miscoordination, collusion, cascading prompt injections — before they manifest in production.
- **Mapping concentration risk.** Insurers need to know where correlated exposure is accumulating. This means tracking metrics currently nobody is *systematically* tracking, such as:

- What percentage of deployed AI agents rely on the same three foundation model providers?
- What percentage rely on the same two or three hyperscalers, or key services?
- What percentage rely on the same tool-calling APIs and data sources?
- More generally: what is the network topology of agent-to-agent interactions — the interconnectedness and interdependencies of these systems — and where are the SPOFs?
- Related to this last: how large is the traffic flow of agent-to-agent interactions through the edges and nodes of that network?

This kind of mapping is directly analogous to what cyber risk modelers do when they assess concentration in cloud providers or software supply chains, and what business interruption underwriters do when they map cascading supply-chain dependencies [202], [237]. For AI, it would involve tracking the "demographics" of the deployed agent population: the distribution of model sizes, model ages, percentage certified against robust standards, etc. Down the road, once more of the science settles, it may make sense to track the distribution of "psychologies" of AI agents — their propensity toward sycophancy, cooperation, defection etc. — as a predictor of certain emergent loss scenarios. Studying and cataloging these tendencies is itself an open research question worth funding [238], [239], [240].

- **Macro-level mitigation R&D and AI agent infrastructure.** On the mitigation side, the picture is similar: investment at the micro level, very little at the macro level. Our current best defenses against accumulation risk — API call rate limits, graceful degradation by swapping to smaller models, and other failover arrangements — are primarily designed to solve *load-management* for individual organizations, not for containing harmful emergent behavior across a large network of open-ended agents. And while major cloud service providers now have a proven track record of high resiliency to shocks and low downtimes, the same cannot yet be said for foundation model providers [241], cf. [242].

  We are at a moment analogous to the early internet, before protocols like HTTPS and TCP standardized secure and efficient communication. The AI agent ecosystem needs equivalent infrastructure. Insurers should encourage the development and adoption of several categories of protection. These might look like:

  - **Intelligent rerouting protocols** for AI models or tool calls that distribute demand across providers and prevent cascading failures when a primary service provider fails. (Developers currently implement these at their discretion; the practice could be standardized and built-in by default.)
  - **Agent firewalls** that prevent e.g. malicious prompt injections from propagating across organizational boundaries [212], [213].
  - **Circuit-breaker mechanisms** that operate at a macro level, throttling or isolating agent populations when anomalous behavior is detected — analogous to stock exchange trading halts, but designed for networks of AI agents rather than financial markets.

- **Global identity and permissions systems for AI agents** that enable trusted interactions and post-incident forensics [243], [244]; current protocols like Google's Agent-to-Agent (A2A) are designed for intra-organizational use and don't address the need for identity verification when agents from different organizations interact across the open web [245].

None of this R&D will fund itself, and none of this infrastructure will build itself. AI developers and academics are contributing what their incentives and resources allow, but gaps remain. The insurance industry — especially through risk modeling firms and reinsurers — is well positioned to help fill them and shape the research agenda: insurers have deep experience managing accumulation risk. They also have a direct financial incentive to support this work: investment in macro-level mitigation R&D that predictably lowers aggregate losses frees up underwriting capacity to write more business.

Taking this proactive approach requires leadership from insurers and reinsurers. The alternative — continuing to rely on blanket exclusions and low limits — may protect insurer balance sheets in the near term but forgoes the opportunity to build a healthy market that works for all participants and enables AI adoption.

The government could also play a helpful nudging function, by raising the salience of such risks with insurers, through programs like the Terrorism Risk Insurance Program (TRIP). As part of TRIP, each year the Federal Insurance Office (FIO) prompts qualifying (re)insurers covering terrorism risk to model a particular terrorist attack scenario [246]: FIO could choose a scenario involving AI-enabled terrorism, directing the attention of (re)insurers to these risks. Historical evidence suggests such light-touch interventions are high-leverage [3].

**Recommendations:**

- **Risk modeling firms** (e.g. Moody's RMS, Verisk, CoreLogic) and academic groups (e.g. Cambridge Risk Studies) should begin modeling AI catastrophe scenarios as well as accumulation risk.
- **(Re)insurers** should form an industry body (similar to the IIHS or IBHS[18]) or otherwise support research dedicated to studying and reducing systemic risks through portfolio-level risk mitigation strategies, and protocols that will enable more secure and reliable multi-agent systems.
- **Oversight bodies and regulators** should commission or nudge (re)insurers to model AI catastrophe scenarios and accumulation risk (e.g. PRA[19], Council of Lloyd's, FIO[20] as part of TRIP[21]).

[18] Insurance Institute for Business & Home Safety.

[19] Prudential Regulation Authority.

[20] Federal Insurance Office.

[21] Terrorism Risk Insurance Program.

## II.3 Standard Setting

Often overlooked, technical standards and risk management frameworks[22] provide important cross-cutting support for insurance markets. They reduce friction at every stage of the policy lifecycle: they give underwriters a common, comparable signal of policyholder insurability; they turn hazy principles-based guidance into bright-line rules for policyholders to target; they enable audits that can act as contractual hooks for claims denial; and they help establish tacit minimum thresholds below which a risk should be considered uninsurable by all underwriters. In doing so, they convert costly, bespoke evaluations and contracting into a tractable, scalable underwriting workflow, outsourcing work to a specialized standards body and auditing ecosystem [111].

Standards don't merely lower transaction costs: they also reduce *legal uncertainty*. This matters because legal risk is highly correlated: one court ruling can redefine the duty of care for an entire class of policyholders overnight. A widely-adopted standard provides a counterweight, anchoring expectations about what reasonable risk management entails. This allows insurers to focus on pricing and mitigating technological risk, rather than shunting blame around to manage legal risk.

Many actors — AI developers, deployers, regulators, the public — have stakes in how AI agent standards develop. But because insurers sit alongside the value chain, supplying the trust and capital that enable adoption while footing the bill for harms, they are well-incentivized to balance these competing interests.[23] This section makes the case that insurers should play an active role in shaping AI agent standards, as they have for past technologies.

### II.3.A Standards as Underwriting Tools

At a minimum, a widely-recognized standard reduces friction. Certification provides an immediate signal of insurability, allowing underwriters to triage submissions and bind policies faster. Paired with a robust auditing and certification regime, falling out of compliance with a standard can also serve as a contractual hook for claims refusal, much as ISO 27001 functions in cyber policies, where misrepresented or lapsed certification can trigger coverage disputes. This contractual hedge further accelerates binding.

Standards can also accelerate underwriting diligence: a standard purpose-built for AI agent insurability would tell underwriters which controls to verify, much as the Center for Internet Security's Critical Security Controls (CIS Controls) do for cyber underwriting [247]. (We discuss key controls for AI agents in §II.5 below). Ideally, certifying against that standard would also produce the documentation underwriters need as a byproduct, rather than as a separate exercise. Better still would be a *performance-based* standard, where certification hinges on meeting or exceeding certain scores in empirical tests,[24] producing quantitative data underwriters can leverage. Regardless, underwriter expertise will remain crucial for interpreting the material policyholders present them, and recognizing what gaps inevitably remain.

[22] Some may wonder what the difference between a framework and a standard is. Very roughly, frameworks tend to be more principles-based guidance, often for internal use. Standards tend to be more rigid and auditable, with an accreditation process for auditors: they are typically aimed at signaling trustworthiness to external stakeholders. In reality, the line between the two is very blurry.

[23] To what degree this is true depends on the efficiency of the liability regime in question. For more see [3].

### II.3.B Standards as Controls for Legal Risk

Legal uncertainty is a structural challenge for liability insurance because legal risk is highly correlated: a single ruling can transform a speculative theory of liability into the new floor of the duty of care, transforming the risk presented by an insurer's entire portfolio with the stroke of a pen. Standards help reduce legal uncertainty in several ways, bounding risk to insurers.

Most directly, legislatures can codify standards by creating, for example, an affirmative defense for defendants who demonstrate certification against a recognized standard [248], [249], [250], cf. [251]. The risk is that a premature, slow-to-update, or unhelpful standard gets locked into law. This might mean it is uncorrelated with risk, compliance theater, or easily gamed. Some proposals call for a middle ground, in which governments issue revocable licenses to Independent Verification Organizations (IVO) that perform voluntary technical audits of AI products against a standard of their making [252], [253], [254]. Products that pass earn their developer or deployer a presumption of duty.

By default, however, the relevant duty of care will be sculpted in fits and starts by courts. Here standards play an indirect but powerful role, guiding the fact-finding process, and steering judges and juries toward the relevant considerations and practices a reasonable actor should have taken into account. Without a widely recognized standard, that process becomes higher variance, creating greater legal uncertainty.

By adopting a common standard in their own underwriting, and encouraging policyholders to do the same, insurers can reinforce this feedback loop, accelerating the convergence of legal expectations. Not all standards will be germane for courts, however. For example, ISO 27001 from the International Organization for Standardization (ISO)[25] is rarely referenced in US courts for determining reasonable cybersecurity; the National Institute of Standards and Technology's (NIST) Cybersecurity Framework (CSF) is the go-to framework [255], [256].[26] We expect the same for AI: NIST's AI Risk Management Framework (AI RMF), not ISO 42001, will be the salient framework for US courts. Therefore, whatever standard insurers coordinate around should align with the AI RMF.

---

[24] This is the difference between asking "is this car's braking distance less than X feet under normal conditions?" and simply asking "is this car equipped with brakes?" Or, in the context of AI agents, the difference between asking: "under standard testing suites, how many attacks successfully jailbreak the agent?" versus "does this agent have any safeguards against jailbreaks?"

[25] Not to be confused with the Insurance Services Office.

[26] It seems the reason is that US courts and plaintiffs prefer leaning on free, publicly available frameworks, and especially ones issued by a US government body.

Under US common law, *certification* against a standard will almost never be dispositive on its own:[27] negligence determinations will remain fact-intensive, turning instead on the Hand formula, which weighs the marginal costs and benefits of further precaution [257], [258]. However, the documentation generated by a well-designed standard-certification process should provide insurers what they need to defend policyholders in court; audits can pay a dividend whether or not the certification itself moves the legal needle.

### II.3.C Standards as Harm Reduction

To play any of the functions above, a standard must be credible. For a risk management standard to be credible, it must actually reduce harm.

The principal challenge of standard-setting for AI is the speed of the underlying technology. Indeed, there is urgent demand from industry that standards development organizations move faster (hence, e.g. NIST's "Zero Drafts" initiative) [259]. AI capabilities are shifting significantly every few months [19], [20], and researchers constantly develop new techniques to red team AI systems for policy violations (see e.g. [211], [260]), so best practices for evaluation and vulnerability patching must evolve in step. A standard's specific requirements must therefore either be revised on a cadence that matches the technology, or remain abstract (referring to "state-of-the-art" or "best practices") at the cost of providing little guidance to users or auditors.

A standard must also be thorough in its breadth: encompassing the safety and security of the system itself, as well as the privacy and governance practices of its developer or deployer; covering the entire product lifecycle, from design specification and training data collection, through pre-deployment testing, to post-deployment monitoring, agent observability, incident review, and continuous support; and considering all key stakeholders, including copyright holders, enterprise customers, end users, and society at large. Finally, compliance evidence should be robust enough to catch shoddy workmanship or attempts to game the standard.

Though difficult, we believe this is achievable. Moreover, we believe insurers can and should help, given their incentive to balance these competing criteria. Their clients will largely be downstream developers and deployers, who won't want over-burdensome or irrelevant requirements. However, liability insurers will be covering those clients for risks they pose to their contractors and the public, so the standard needs to address risks to those parties too. Finally, insurers will want the standard to be auditable, detailed, and up-to-date enough to be useful for underwriting.

### II.3.D Current Standards and Frameworks for Frontier AI

There are at least four relevant standards in this space:[28]

- **ISO 42001** (accompanied by ISO 23894 and other supporting standards) [261], [262]

[27] Unless the IVO model catches on, as mentioned above.

[28] The General-Purpose AI Code of Practice is also a contender, but we set it aside given our focus on risks from downstream development and deployment of AI agents, as opposed to risk from foundation models [188].

- **NIST's AI RMF** (and accompanying GenAI Profile, NIST AI 600-1) [263], [264]
- **STAR for AI** (and associated AI Controls Matrix) [265]
- **AIUC-1** [266]

Each one occupies a distinct niche, trading off prescriptiveness, flexibility, scope, update cadence, and certifiability. Very broadly though, ISO 42001 and NIST's AI RMF should be thought of as foundations that are *operationalized* (and in the case of ISO 42001, *extended*) by STAR for AI and AIUC-1.[29] Table 3 highlights key differences between them.

These differences become concrete when we consider how each standard treats a specific reliability concern such as AI agent hallucinations. The four standards span a spectrum from highly prescriptive to highly abstract. At the more prescriptive end, AIUC-1 has dedicated controls for hallucinations (e.g. D001) that prescribe specific technical safeguards (e.g. groundedness filters and user-facing citations) and conditions certification on passing third-party evaluations [266]. NIST's AI RMF (via its GenAI Profile) likewise names hallucinations (a.k.a. "confabulations") as a specific risk and recommends technical safeguards to measure and mitigate it. As a foundational voluntary framework, however, these are recommendations only [264]. The AI Controls Matrix used for STAR for AI has no hallucination-specific control given its cybersecurity focus, though its general output-validation control (AIS-09) covers invalid outputs including hallucinations [267]. No prescriptions for mitigating them are made, but further resources are offered. Finally, ISO 42001 and ISO 23894 do not mention hallucinations at all; the closest analogue is ISO 42001's verification-and-validation control (6.2.4), which directs organizations to define and document such measures but provides no prescriptions or further guidance [262].

What the ISO family sacrifices in depth, detail, and robust guarantees for external stakeholders, it gains in flexibility, broad applicability, and serviceability (not requiring frequent revision). By contrast, AIUC-1 provides concrete guidance and robust guarantees for external stakeholders, in exchange for being more rigid, burdensome, and something of a moving target. The other standards and frameworks sit somewhere in between these extremes. These trade-offs have implications for ongoing loss control, as discussed in §II.7.

---

[29] Disclaimer: three of this report's authors, including the lead author, are employed by the Artificial Intelligence Underwriting Company, which develops and maintains AIUC-1. We have tried to mitigate this conflict of interest by applying a common set of evaluation dimensions to all four standards, drawing only on each standard's published text, and by noting AIUC-1's disadvantages as well as its strengths. Readers may consult cited primary sources to check our characterizations. Other co-authors reviewed this section for accuracy and fairness but take no position on AIUC-1's relative merits.

| | ISO 42001 | AI RMF | STAR for AI | AIUC-1 |
|---|---|---|---|---|
| **Issuing body** | **Private**<br>International Organization for Standardization (ISO) | **Public**<br>National Institute of Standards and Technology (NIST) | **Private**<br>Cloud Security Alliance (CSA) | **Private**<br>Artificial Intelligence Underwriting Company (AIUC) |
| **Initial release** | 2023 | 2023 | 2025 | 2025 |
| **Revision cadence** | Multi-year revision cycles. Likely five years: cf. [268], [269] | Multi-year revision cycles<br>[270] | Unknown | Quarterly<br>[271] |
| **Type** | Auditable standard with certification | Voluntary framework | Auditable standard with certification | Auditable standard with certification |
| **Focus** | **Broad**<br>All machine learning systems, including e.g. 2010s-era recommender systems. Focused on high-level management systems. | **Broad**<br>All machine learning systems, including e.g. 2010s-era recommender systems (though see the recent GenAI profile [264]). | **Medium**<br>Targeted at LLM based systems. Focused primarily on cybersecurity across the AI supply chain, with a focus on the cloud.<br><br>Maps to both NIST AI RMF and ISO 42001. | **Narrow**<br>Targeted at LLM based AI systems, and especially agents. Focused on risks in enterprise deployment of AI agents, covering security, safety, and reliability. Does not include controls for upstream cloud or model providers.<br><br>Maps to both NIST AI RMF and ISO 42001. |
| **Performance-based** | **No**<br>Reliability testing is required, but no benchmarks are set. | **No**<br>Reliability testing is recommended, but no benchmarks are set. | **No**<br>Reliability testing is required, but no benchmarks are set. | **Yes**<br>Certification requires meeting certain benchmarks in standardized performance evaluations, run quarterly. |
| **Depth of technical prescriptiveness** | **Shallow**<br>Little to no technical implementation guidance. Only high-level principles. | **Moderate**<br>Principles with suggested technical controls (NIST "playbooks"), but no mandates. | **Deep**<br>Very detailed technical implementation guidance, with specific technical controls and safeguards.<br><br>Informed by real-world loss and mitigation libraries from sources such as MITRE and OWASP [272], [273]. | **Deep**<br>Very detailed technical implementation guidance, with specific technical controls and safeguards.<br><br>Informed by real-world loss and mitigation libraries from sources such as MITRE and OWASP [272], [273]. |

*Table 3. Key differences between current AI standards*

## II.3.E Historical Precedent: Underwriters Laboratories and Electrical Standards

Insurers funding standards bodies for managing novel technological risks is not a new idea. The most consequential precedent dates to 1894, when property insurers, confronted with the novel fire hazards posed by electrical equipment, founded Underwriters Laboratories (UL) [53]. UL developed standards for electrical and fire safety equipment, tested and certified products against those standards, and — though no longer funded by insurers — continues to play that critical role today. The UL mark has become a *de facto* litmus test for insurability [274] and for retailers and wholesale buyers of electrical appliances [54], [55], [275].

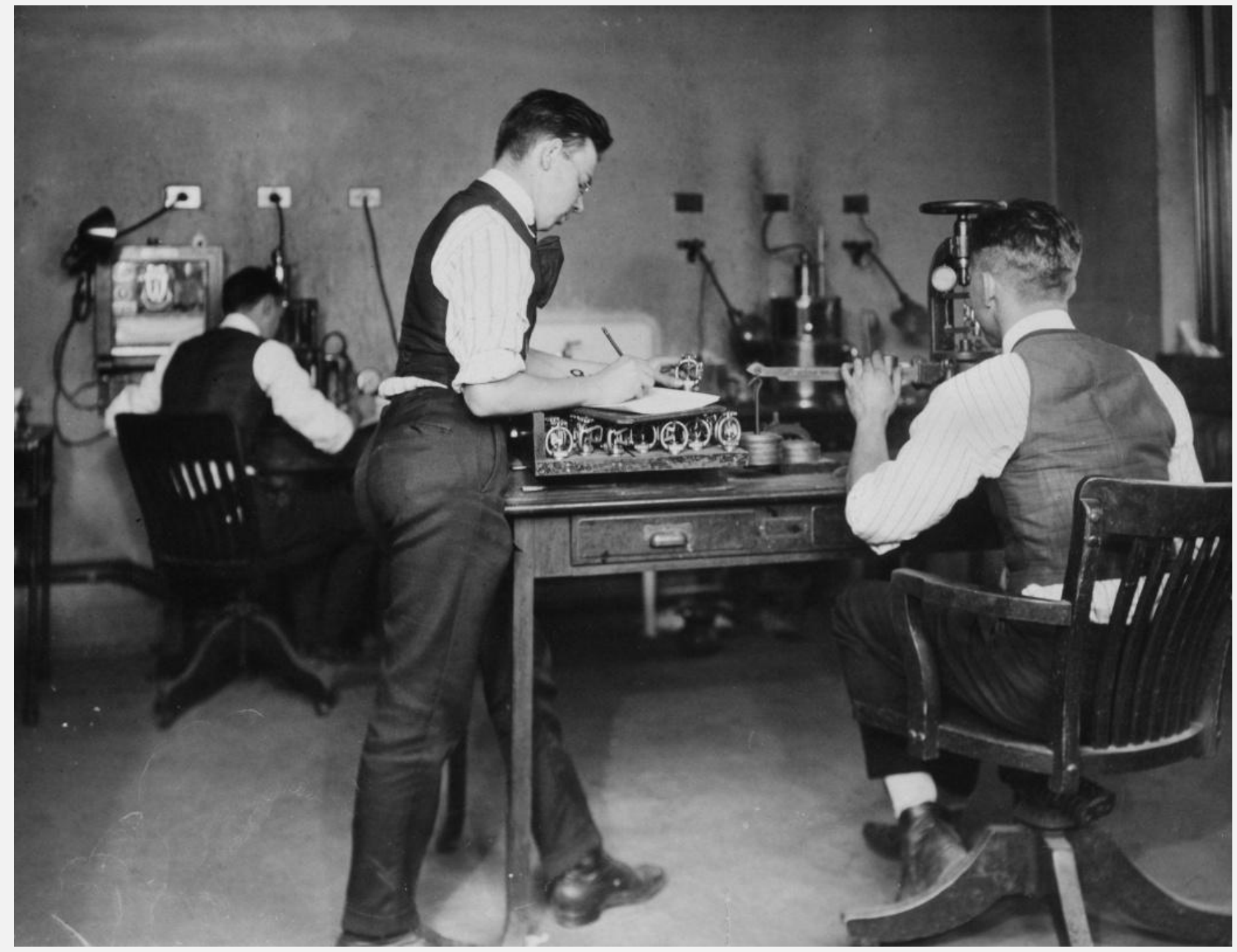

*Image 4. UL testing fire suppression sprinklers, circa 1919. Source: UL Research Institutes [53]*

Compliance with UL standards was not and is not mandated by law. Over time, however, as industry best practices settled, UL standards became implicitly referenced in legally binding codes, regulations, and court decisions (see e.g. [56], [57], [276]). For example, the National Electrical Codes (NEC) of the United States, Canada, and Mexico require that, to be code-compliant, fire suppression sprinklers and building materials be listed as certified by approved third-party organizations, such as UL (see e.g. [277, Arts. 100, 352], [278]). Not coincidentally, these codes were drafted by the National Fire Protection Association, another nonprofit formed by insurers to control losses from fires [58]. Codifying standards marks the maturation of private standard-setting into publicly recognized regulation.

These standards save lives. For example, after investigating the technology in 1995 [279], UL introduced standard 1699 for Arc-Fault Circuit Interrupters (AFCI) in late 1996 [280]. This then became a NEC requirement in 1999 [281]. Engineers from the US Consumer Product Safety Commission have estimated that AFCIs could prevent 50% or more of electrical fires [280], and indeed residential fires involving electrical failure have fallen from over 63,000 per year in 1996 to some 46,000 in 2019 [282].

The history of fire and electrical safety demonstrates how robustly managing a novel risk requires an ecosystem of actors working in concert: insurers underwriting the risk, standards bodies setting the technical baseline, private associations drafting model codes, and public authorities eventually adopting and enforcing those codes once proven in the market.

The analogy between electrical appliances and AI agents is imperfect, most notably for the pace at which agentic AI products are developed, deployed, and continuously modified or deployed in new environments. To unlock a similar virtuous interaction between standards and insurance, the update cadence for AI agent standards will have to be far faster than traditional standards.

**Recommendations:**

- **Insurers** should reward a policyholder's adoption of any robust standard for the safety, security, and reliability of AI agents.
- **Reinsurers** should encourage or require ceding insurers to recognize robust standards in their underwriting.
- **(Re)insurers** should be proactive in working with industry, governments, and academia to develop or support AI agent standards that serve all stakeholders' needs.
- **Governments** should consider recognizing private standards in regulatory guidance or procurement requirements, in order to spur private governance and reduce legal uncertainty.

## II.4 Contract Design

At the heart of every insurer-insured relationship is a contract. Clearly defining what is covered, what is excluded, and how coverage is triggered lets underwriters control the risk they take on while giving insureds the certainty they need to adopt AI with confidence. It also lets portfolio managers track their AI exposure, and helps underwriting teams develop dedicated expertise with an emerging risk. Finally, consistent language across the market also reduces complexity for customers and legal uncertainty for all stakeholders.

### II.4.A Where We Are: Silent Coverage and Growing Exclusions

As noted above (§I.2), today's exposure to AI agent risk sits largely within pre-existing policies whose general terms never mention AI but cover its risks all the same. This "silent coverage" causes problems on four dimensions.

- **Pricing:** Where agentic AI presents heavier tails or higher average losses than the methods it replaces, insurers face unexpectedly elevated losses [283].
- **Certainty**: Silent coverage for an emerging, unforeseen risk is a legal battle between insurer and insured waiting to happen: it fails to deliver the confidence customers seek and widens the error bars for insurers' risk models.

- **Visibility**: With AI exposures buried across a growing number of pre-existing lines like cyber, D&O, E&O, and CGL portfolio managers cannot see where AI is concentrated, which sectors are accumulating exposure, or which claims are AI-driven [8], [11]. (See §I.2).
- **Institutional Learning**: Mastering AI underwriting will require dedicated expertise and separating AI-related losses from other sources. This is much easier if AI risks are clearly sorted out and assigned to dedicated teams, and unlikely to happen if coverage remains silent and split across many lines.

In response to this silent coverage problem, insurers are moving to exclude AI risk from existing lines. The Insurance Services Office filed multistate endorsements effective January 2026 allowing carriers to exclude generative AI from CGL policies [120]. Major carriers including WR Berkley, Cincinnati Financial, and Philadelphia Insurance have followed suit, filing their own AI exclusion forms [13], [16]. So far, over 80% of such forms have been approved by US insurance regulators [15]. Munich Re's AI insurance head predicts the exclusions will create "repurchase demand" as companies seek to buy back coverage that's been carved out [284].

Writing exclusions isn't inherently problematic: it's a healthy market correction that eliminates mispriced exposure and creates certainty, and it's encouraging to see the industry learn from the mistakes of "silent cyber" [10]. However, exclusions do not solve the visibility and learning problems: exclusions represent a market exit, with no clear path to re-entry. What is needed is affirmative coverage: explicit, appropriately priced protection that gives enterprises the confidence to deploy advanced AI at scale.

### II.4.B What's Emerging: Early, Inconsistent Affirmative Coverage

A handful of specialist MGAs and insurers have begun pioneering affirmative AI coverage [15][30]. Current products typically take one of two forms: endorsements that bolt AI-specific coverage onto existing coverage (e.g. cyber policies), often with sub-limits [287], [288], [289]; or standalone policies designed specifically for AI risks, covering failures like hallucinations, data leakage, IP infringement, and reputational harm — currently with limits up to $50 million [290], [291], [292].

These limits are one or two orders of magnitude below what the largest enterprise deployments will demand, and the products are expensive, supply-constrained, and often narrow in scope. But their more important function, at this stage, may be informational: they are beginning to solve the challenges of visibility and institutional learning.

Insurance regulators could help by encouraging or requiring insurers to separately disclose affirmative AI coverage in their annual reports to regulators, as the National Association of Insurance Commissioners (NAIC) successfully did for cyber insurance [293], [294], [295]. This allows both regulators and insurers to track developments in this emerging market.

While this early affirmative coverage is a step in the right direction, the danger that now arises is *inconsistency*, which creates complexity for customers, increasing transaction costs. It also increases legal uncertainty for insurers by delaying the time it takes courts to settle on their interpretation of key terms. Ambiguous terms that didn't map neatly onto loss patterns on the ground were litigated for years, though largely in adjacent lines such as property rather than cyber policies themselves [9]. Today, basic definitions such as that of "cyber event" still vary, creating uncertainty for all stakeholders [60], [296].

[30] There are also narrower warranty-like performance guarantee products, but we expect these to make up an increasingly smaller amount of affirmative coverage [285], [286].

Insurers won't have the luxury of time in getting this right for AI coverage [19], [20], [24]: the industry should be proactive in crafting clear definitions adapted to the technology, and industry bodies like the Insurance Services Office, the Lloyd's Market Association (LMA), and major reinsurers have a crucial coordinating role to play by promulgating model language that helps create market-wide consistency [59], [63, pp. 8–9].

### II.4.C Where We Need to Go: Clarity and Consistency on Key Terms

To close the coverage gap with large limits, broad scope, and affordable pricing, the industry needs to agree on consistent terms that resolve open questions of policy language. We consider a few such questions below, as an invitation for further discussion:

- **Defining "Generative AI."** In their exclusion forms, the Insurance Services Office and others have adopted a fairly measured definition based on machine learning and content generation, that runs as follows [120]:

  > "Generative artificial intelligence" means a machine-based learning system or model that is trained on data with the ability to create content or responses, including but not limited to text, images, audio, video or code.

  This is workable for the intended purpose of excluding AI risk, but lumps together two technologies with materially different risk profiles: simpler generative systems that return content and await further input, and agents that form plans and take consequential actions over extended time horizons.[31] Affirmative coverage may wish to distinguish the two.

- **Excluding Accumulation and Catastrophic Risk.** Cyber developed its accumulation and catastrophe exclusions — critical infrastructure, nation-state attacks, "war-like acts" [59], and catastrophic sub-limits [60] — often through bitter experience. AI presents analogous and novel accumulation risks. Most prominent is the Single Point of Failure (SPOF) represented by upstream foundation model providers: Anthropic, OpenAI, and Google collectively account for over 80% of the enterprise foundation model API market [40], cf. [206]. The most workable approach today is to control accumulation risk by excluding losses arising from upstream foundation model failures, such as hallucinations, outages, or systemic defects in the underlying model. Such carve-outs function analogously to contingent business interruption exclusions, isolating the correlated tail that no single downstream policyholder can control.

  The language is straightforward enough to draft; operationalizing it is another matter. Determining whether a given loss was caused by e.g. a foundation model hallucination versus downstream configuration error is often impossible without proper technical logs and controls. Without the standards and underwriting infrastructure described above in this report, such exclusions will simply shift uncertainty from the policy form into the claims process. This is another example of how the insurance stack is interdependent: precise exclusions presuppose precise controls.

[31] In this report, we tend to use "generative AI" in this exclusive sense, reserving "agentic AI" for systems that act.

A second accumulation risk warrants flagging: emergent behavior from multi-agent interactions, such as a feedback loop where AI agents influence an environment (e.g. the internet, a codebase etc.) that other AI agents rely on [116], [218] (see §II.2). How policy wordings should handle such risk is an open question.

We believe systemic and catastrophic AI risks can and should be covered, but insurers will need language that separates them out and places them with dedicated pricing, dedicated underwriting, potentially in a dedicated pool (see Part III).

- **Aggregation clauses.** Insurance policies typically contain per-claim or per-loss sub-limits, along with *aggregation clauses* that bundle what courts might otherwise treat as distinct claims or losses. These clauses are famously fraught, turning on quasi-metaphysical interpretations of what constitutes an "event," "cause," or "related act" [296], [297]. This is not academic: billions of dollars rode on the question of whether the attacks on the Twin Towers constituted one event or two [296, p. 4]. For AI, the salient question is what factors unify "related, repeated or continuing acts" or "related errors": traditional unities tests (time, physical proximity) translate poorly to digital perils [296, p. 99]. Given the black-box nature of deep-learning systems [39, p. 99], specifying unifying factors directly may be desirable (e.g. specifying whether or not losses arising from similar hallucinations attributable to one bad dataset count as a single loss).

**Recommendations:**

- **Oversight bodies** (e.g. NAIC, Council of Lloyd's) should require insurers to fill out an AI insurance supplement in their lines of business reports, reporting macro-level market data such as the number of policies in force with affirmative AI coverage, direct premiums written, loss ratios etc.
- **Regulators and oversight bodies** (e.g. PRA[32], Council of Lloyd's) should issue a notice encouraging or mandating insurers to end silent AI coverage by a given date.
- **Advisory bodies** (e.g. LMA, Insurance Services Office) should issue model policy language for AI risk, covering definitions of GenAI, agentic AI, key aggregation clauses, and targeted carve-outs to control accumulation risk.
- **Reinsurers** should require ceding insurers to track and report AI exposure, adopt consistent policy language, and use targeted carve-outs to control accumulation risk.

[32] Prudential Regulation Authority.

## II.5 Risk Selection and Evaluation

A core competence of insurance is selecting for good risk, and screening out bad risk. This typically means seeking responsible clients and rejecting reckless ones. This is straightforwardly in the insurer's interest, but also represents a core governance function of insurance. Carrying insurance is often a condition for doing business, so by requiring that policyholders meet e.g., minimum safety standards, insurers can act as market gatekeepers and create a race to the top among market players [2], [3]. More generally, risk selection will depend on audits and other technical evaluations: the financial stakes of insurers and underwriters who demand (or conduct) them are a key market mechanism for aligning the incentives of the broader third-party evaluation and auditing ecosystem [109], [111].

On the underwriting side, standards and audits both streamline underwriting, and help forestall a race to the bottom between underwriters during a soft market, by setting an implicit baseline for insurability [63, p. 4], cf. [164]. Certification and auditing also have inherent advantages over good-faith questionnaires, which policyholders often answer incorrectly, whether through honest misunderstanding or deliberate misrepresentation.

However, a baseline for insurability is not enough: to determine whether and on what terms to offer coverage, underwriters must go deeper, assessing the particular structure, scale, and context of deployment. A certified AI agent autonomously executing trades in a brokerage poses very different risks than a certified AI agent handling internal document summarization for a law firm. The underlying model and guardrails might be the same, but the legal exposure, potential severity of loss, and probability of a claim will diverge enormously.

Underwriting exists to make these distinctions. If underwriters fail at risk selection, gaming is incentivized and adverse selection takes hold: the book fills with high-risk policyholders who know they are getting a bargain, and the insurer's loss ratio deteriorates.

### II.5.A What Underwriters Need to Know

Beyond familiar questions about an applicant's revenue, sector, and customer base, underwriting AI agents requires a new category of questions about the agentic system. These may include [9], cf. [283]:

- Is it customer-facing or internal?
- How autonomous is the system?
  - Does it recommend actions or execute them?
  - How much human oversight is involved?
- Is it a single agent or part of a multi-agent architecture?
- What modalities does it operate in (text, code, vision, voice, tool use)?
- How open-ended is the use case?
- How substantial or complex is the transformation between inputs and outputs?

Some of the highest leverage information underwriters can likely gather is about the organization's relevant personnel:

- To whom is AI risk management delegated?

- What is their AI literacy level (e.g. can they articulate core safeguards governing the organization's agents, or best practices like the "rule of two"[33])?
- What duties and powers do they have?
- What is their track record?

A thorough underwriter will also take a holistic view of the entire value chain and a policyholder's governance apparatus for managing the AI agents, inquiring about:

- Technical controls
  - Deployment and change management (version control, roll-back evidence)
  - Technical safeguards (input/output filtering, rate limiting, fallback mechanisms)
  - Integrity of training data provenance and pipeline
  - Agent observability tools
  - Agent access-control configurations
  - Other cybersecurity controls
- Organizational controls
  - Pre-deployment test logs
  - Post-deployment incident logs
  - Incident response plans
  - Near-miss reporting protocols
- Human-Computer Interaction controls
  - Human oversight and handoff protocols (ideally including override rates and correction acceptance)
  - Design of warning labels and disclosures of AI use

A well-designed standard certification that verifies all such controls [111] would supply much if not all of the necessary documentation (see §II.3 above), similar to other insurance lines such as workers' compensation [298].

### II.5.B From Questionnaires to Continuous Assessment

Traditional questionnaires have their place, but they are insufficient for a technology evolving as rapidly as AI. Slow, rigid, and binary, questionnaires capture only an annual snapshot (something cyber insurers know only too well [63, pp. 5–6]). Underwriters need more continuous, granular data. This is where technical third-party technical testing enters the picture. This includes:

[33] The "rule of two" states that an AI agent should only ever have permission to perform at most two of three high-risk actions: (1) process untrusted input, (2) access non-public data, and (3) communicate externally or make material internal changes. Restricting an agent to a maximum of two capabilities protects the system by breaking the attack chains, ensuring a malicious input (like prompt injection) cannot simultaneously access and exfiltrate or manipulate sensitive data.

- **Benchmarking.** Standardized test suites — covering reasoning, factuality, instruction-following, and safety — that produce comparable scores across models and over time [39, p. 148], [299]. Benchmarks are blunt instruments, but they are the closest thing underwriters have to a common yardstick: useful for triage and trend-spotting, less so for pricing a specific deployment. Their value lies in breadth and comparability, not realism.
- **Performance evaluations.** Bespoke, non-adversarial tests designed to probe how a *particular* system behaves under the conditions it will actually face in deployment [39, p. 149]. Examples include measuring hallucination rates on representative tasks, testing for intellectual property leakage, evaluating tool-use reliability on representative workflows, or studying performance drift under extended multi-turn interactions [300]. Where benchmarks ask how a system compares, performance evaluations ask whether it is fit for purpose.
- **Red teaming.** The deliberate, structured effort to break a system using limited affordances, to identify weaknesses and measure resilience [39, p. 153]. In practice, this might mean attempting to jailbreak a system through adversarial prompting (e.g. [211], [300], [301]), or designing prompt injection techniques to steer a system in ways unintended by the developer (e.g. [302]). As with performance evaluations, red teaming should increasingly stress systems across sustained, multi-turn exchanges, since refusal boundaries that hold on the first prompt often soften by the tenth, and human-escalation pathways that work in isolation may fail under prolonged manipulation. Analogous (in some cases identical) to penetration testing in cybersecurity, the goal is to find vulnerabilities before an attacker — or an incident — uncovers them. Red teaming is also well-suited to identifying the upper bounds of severity in a system's residual risk, after efforts to engineer safeguards hit diminishing returns. Where performance evaluations ask how a system behaves under *expected* conditions, red teaming determines how it does under *adversarial* conditions.
- **Chaos testing.** The deliberate introduction of faults into a complex system to observe how it responds [264, p. 27]. For multi-agent systems with sophisticated orchestration, this might involve corrupting one agent's outputs — toxic content, fabricated data, malformed tool calls — to see whether the broader system contains the failure or whether contagion spreads (see e.g. [116]). The questions are diagnostic: does the system self-correct, fail gracefully with appropriate alerts and fallbacks, or fail catastrophically? The practice is borrowed from the chaos engineering pioneered by Netflix, Amazon, and Google, now standard in reliability engineering [303].

By encouraging policyholders to regularly undergo these tests, and rewarding those who share summary scores, underwriters can get a far more granular, system-specific picture of a policyholder's residual risk. We discuss exactly how these tests feed into pricing below (§II.6).

Crucially, these tests cannot be one-and-done: much like penetration testing in cyber, these tests must be conducted regularly since testing strategies, model capabilities, and their vulnerabilities are all evolving on a month-to-month basis (see e.g. [19], [20], [211], [260], [304]).

For this reason, some standards include quarterly technical evaluations as a minimum requirement [305]. Once again, the stack interlocks: a robust standard certification process will produce not only a binary pass/fail, but also a wealth of data of interest to underwriters.

The depth and rigor of these tests depends on the access applicants grant evaluators [306]. There are real tradeoffs with security and operational overhead, but underwriters should generally reward transparency, incentivizing deeper access with meaningfully lower premiums.

### II.5.C Historical Precedent: Fire and Building Code Rating Schedules

There is a well-established precedent for this model. The Insurance Services Office operates two rating systems that illustrate how standards translate into underwriting infrastructure. The Fire Suppression Rating Schedule (FSRS), a descendant of the 1915 National Board Grading Schedule [307], scores communities on a 0–105.5 point scale based on how well their fire suppression capabilities — emergency communications, fire department equipment and training, water supply — comply with standards set by organizations like the National Fire Protection Association [61], [308]. The Building Code Effectiveness Grading Schedule (BCEGS) performs a parallel function, collecting over 1,200 data points to score how well communities adopt and enforce up-to-date building codes [62], [309]. Property insurers use these scores ubiquitously for pricing: a community with a better fire suppression rating gets lower premiums [310]. The discounts exist because the schedules work: the National Institute of Building Sciences found that the cost-benefit ratio associated with updated building code adoption is $11 to $1 [311], [312].

This is the private regulatory effect of insurance at its best: minimum safety standards for insurability, plus a flexible menu of mitigation measures policyholders can proactively take to further reduce their premiums [2], [3]. This translates the nebulous *ex post* incentives of liability into bright line *ex ante* rules, guidance, and upfront monetary incentives [2].

The standards built for certifying AI agents should strive to inform underwriting in the same way. One can envision a composite score for AI agents — akin to the FSRS scores — based on how current its guardrails are, how robust its monitoring infrastructure, how it performed under stress-testing, and how well the organization's governance practices align with best practices. In brief: the controls that make good standards will also make good rating criteria.

### II.5.D Automating and Scaling

Currently, the kind of thorough AI underwriting described above is very hands-on. That level of friction will be prohibitive for many customers and will not scale.

Besides leaning more on technical standards and model questionnaire language to reduce friction, insurers could take a cue from cyber. Over the past decade, cyber underwriting has come to rely less on manual assessments or white-glove audits, and more on automated "outside-in" scans that evaluate an applicant's security posture in minutes [63], e.g. [313]. This has been enabled by a combination of insurtech innovations and new partnerships with major Cloud Service Providers (CSPs) who already have access to much of the information insurers need to know about their clients' security configuration [63]. These solutions have not replaced higher-touch underwriting for large corporate clients with complex risks, but they have allowed insurers to write more small and medium-sized enterprise (SME) coverage by streamlining the product [314], [315].

Similarly, underwriting AI risk should leverage technological innovations — including using AI agents to help underwrite AI [316] — and forge partnerships with frontier model providers as well as CSPs who can furnish reliable telemetry and evaluation infrastructure at scale cf. [317], [111]. We discuss this further in §II.7.B below.

**Recommendations:**

- **Advisory bodies** (e.g. ACORD[34] and LMA[35]) should develop model proposal forms and questionnaires in order to reduce friction.
- **Insurers** should partner with AI auditing organizations to audit policyholders for organizational and system safety.
- **Reinsurers** should encourage or require ceding insurers to mandate certain minimum underwriting controls.
- **Advisory bodies** (e.g. the Insurance Services Office) should develop standardized rating schedules.

## II.6 Pricing

If insurers can't profitably price AI agent risk, the rest of the stack is an exercise in futility. Every layer in this proposed stack feeds into better pricing: incident data provides loss benchmarks; standards and audits help with risk differentiation, streamline underwriting and control legal risk; policy language controls what exposures land on the insurer's book; performance evaluations generate rough probability estimates, while red teaming helps identify the upper bounds of severity. Pricing is where this all comes to a head.

To make this concrete, consider a schematic pricing formula:

$$Premium = \frac{severity \times probability \times exposure \times sys_spec \times sector \times acc_risk_loading}{1 - profit_margin}$$

In what follows, we'll walk through the key terms in turn.

### II.6.A Expected Loss

At the heart of the formula is a basic expected-loss calculation: severity × probability × usage.

**Estimating severity** — how much incidents cost — draws on two complementary sources. The first is incident data, collected in the aforementioned databases.[36] Such data is currently scarce but growing quickly, and much can be extrapolated from the cost of non-AI incidents in the same category. For example, the legal risk of a chatbot reproducing clearly copyrighted material is not fundamentally different from the exposure of a human employee doing the same. Likewise for data breaches, erroneous financial advice, or defamatory outputs.

[34] Association for Cooperative Operations Research and Development.

[35] Lloyd's Market Association.

[36] Likely insurers' private databases in this case: as far as the authors are aware, public incident databases rarely if ever include exact dollar costs of incidents. This doesn't make public databases any less pressing to pursue, but highlights how public and private efforts complement each other.

The second source is *red teaming* — the structured, adversarial effort to break a system, analogous to crash testing in automotive engineering (see §II.2.B), and virtually identical to penetration testing in cybersecurity (see §II.5.B above). By probing how badly a system fails under hostile conditions, red teaming bounds the upper end of the severity distribution: the most sensitive data a deployment can be coaxed into leaking, the most damaging instructions it can be manipulated into following, the most egregious content it can be induced to generate. A system that, under sustained adversarial pressure, will not exfiltrate sensitive customer data or execute unauthorized financial transactions has a meaningfully truncated tail risk; one that will, is exposed to the full distribution of possible losses in that category.

**Estimating probability** per unit of exposure is harder. Eventually, actuarial tables will supply frequency data, and developing them is a priority for the industry. Until they mature, *performance evaluations* can help fill the gap. Where red teaming determines how badly a system fails under a given level of adversarial pressure, performance evaluations ask how often it fails in expected conditions (see §II.5.B above). The output is structurally similar to an actuarial table — a distribution over failure modes — but generated by *simulations* of real-world deployments rather than historical observation. Besides their immediate availability, red teaming and performance evaluations have two other important advantages:

1. They are **system-specific**, helping differentiate risk. Two systems built on the same foundation model can have radically different failure profiles depending on guardrails, prompt architecture, retrieval design, and deployment context. Actuarial tables derived from industry-wide loss data would gloss over these differences with an average.
2. They are **forward-looking**: they evaluate the *current* system, and project how often it will fail. With a technology evolving as rapidly as AI, actuarial tables will quickly become outdated, threatening their predictive validity [21].

**Estimating exposure** completes the expected-loss calculation. Company size and or revenue are the typical proxies. However, these are crude. A firm that occasionally runs an internal document summarization bot presents very different exposure from one running fleets of autonomous customer-service agents across all product lines.

Ideally, insurers would obtain a direct reading of *AI usage* — API call volume, token throughput, or comparable telemetry. In the same way commercial auto telematics passively collects driving data [318], such telemetry can be designed to be frictionless for the policyholder, especially if insurers partner with foundation model providers (see §II.7.B below). Greater price discrimination should also expand the addressable market: light users stop subsidizing heavy ones, and coverage becomes economical for deployments that flat-rate pricing would otherwise exclude.

### II.6.B System Specifications

Beyond a simple expected-loss estimation, pricing should incorporate feature rating: adjusting premiums based on the system specifications of the AI system that correlate with risk, even if their contribution to expected loss is hard to quantify precisely.

Insurers should aspire to the granularity seen in e.g. property insurance for fire-prone areas, where underwriters account for the slope of the terrain, prevailing wind patterns, the density of brush and its proximity to structures [319], the materials used in construction, and the availability of local fire suppression resources (see §II.5.C above). Together these features produce a detailed risk profile that allows genuine differentiation between properties. If this can be done for property insurance, which requires physical inspections and sophisticated geo-spatial data analysis, it can be done for AI agents, whose basic characteristics are inherently more confirmable. For a list of system features pricing should attend to, see §II.5.A above.

Underwriters should also offer premium discounts for implementing additional or more effective safeguards. Not all safeguards are created equal, and pricing should reflect this. For example, prompt injection defenses range widely in robustness: some techniques, such as CaMeL, provide deterministic security guarantees by architecturally separating control and data flows, while others rely on heuristic detection that a sufficiently creative attacker can bypass [226], [320]. Underwriters should price the difference.

This is not novel. Nuclear insurers have long offered premium differentials of up to ~40% based on the findings of engineering safety reports [68]. Cyber insurers routinely offer discounts of up to 25% for insureds who demonstrate excellent security posture — multi-factor authentication, endpoint detection and response, robust backup procedures [321], [322], [323]. AI insurance should aim for the same meaningful price differentiation based on the quality and rigor of technical controls. Premium signals are a key tool insurers have for driving safety improvements; failing to differentiate on pricing undermines the governance function of insurance. A menu of options at different costs also helps policyholders determine what is cost-effective for their case [2]. For example, deterministic controls with strong observability guarantees are reasonable for high risk sectors such as financial services [169], but may be excessive for lower risk sectors.

### II.6.C Sector

As mentioned earlier, the sector in which the system is deployed also matters for pricing AI agent risk. All else equal, an agent deployed in financial services — where errors can trigger regulatory actions, fiduciary liability, and market losses — is higher risk than the same system deployed in retail inventory management.

### II.6.D Accumulation Risk

Finally, the premium must account for accumulation risk: the possibility that a single underlying cause produces correlated losses across many policyholders simultaneously.

This is where the greatest uncertainty lies, and where traditional actuarial methods are least helpful. What evidence we have suggests that accumulation risk from AI is currently low or manageable through exclusions (see §II.2). Most AI deployments today are narrow enough in scope, and the market young enough, that a single failure is unlikely to cascade across a large book of business. However, we expect this to change quickly. As enterprises become more dependent on a small number of upstream model providers, as multi-agent systems proliferate, and as AI takes on higher-stakes tasks, the potential for correlated failures grows substantially. Insurers must get a handle on these risks if this market is to be viable (see §I.3).

**Recommendations:**

- **Insurers** should lean into pricing based on system-specific stress-testing data (red teaming, performance evaluations etc.) to keep up with a rapidly evolving risk, and overcome the lack of actuarial data.
- **Insurers** should feature-rate based on system specifications, governance practices, and technical safeguards in place.
- **Insurers** should partner with frontier model providers to track policyholders' AI usage through telematic-like arrangements.

## II.7 Ongoing Loss Control and Monitoring

The work does not end once a policy is issued. With a technology evolving as rapidly as AI, insurers should verify that insured systems continue to meet minimum standards and support policyholders as new vulnerabilities are discovered and safeguards improve.

Cyber insurers learned this the hard way. As noted above, early cyber policies relied on point-in-time audits and questionnaires that were quickly outdated. The shift to so-called "active insurance" [324] — continuous scanning of policyholders' attack surfaces, real-time alerting, and ongoing risk management services — has been a positive development in that market [63], [325].

Cyber Insurance — Loss Ratio by Quartile (2018–2024)

Top Quartile Avg LR (lowest loss ratios) Market Average LR Bottom Quartile Avg LR (highest loss ratios)

Loss Ratio (Defense and Cost Containment included)

|  | Top Quartile Avg LR | Market Average LR | Bottom Quartile Avg LR |
|---|---|---|---|
| 2018 | 9% | 30% | 59% |
| 2019 | 5% | 29% | 59% |
| 2020 | 27% | 56% | 101% |
| 2021 | 16% | 59% | 101% |
| 2022 | 16% | 43% | 67% |
| 2023 | 11% | 39% | 68% |
| 2024 | 24% | 52% | 85% |

*Source: NAIC data, reporting on the top 20 cyber insurance groups by market share.*
*Top quartile = lowest LR performers; Bottom quartile = highest LR performers.*

*Figure 6.*

These efforts pay off. The top quartile of cyber insurers consistently have loss ratios more than 20% lower than the market — 29% on average over the last 8 years (see Figure 6). This agrees with the reported loss ratios of leading specialty cyber underwriters like Corvus (Travelers) and Beazley, who pair insurance with sophisticated loss control services [293], [294], [295], [326].

Perhaps more telling, during the ransomware surge of 2020 and 2021, the spread between the top and bottom quartile widened greatly (from 53 points in 2019 to 74 points in 2020), suggesting better underwriting discipline and loss control simply lead to fewer and lower claims. (The spread narrowed again later, as premium hikes kicked in to compensate for higher losses; see Figure 4 in §I.5). Scholars are generally skeptical of cyber insurers' ability to reduce harm by increasing security [21], [327], [328], [329], due in part to crime insurance's unique countervailing third-party moral hazard effect: the presence of insurance can *encourage* criminal activity [3], [330], [331], [332]. However, the high variance in underwriting outcomes cannot be explained by accurate pricing alone: disciplined risk selection, mandated security controls, and loss control guidance are also having an effect. A survey of German businesses, for example, found that cyber losses to insureds increased by only around 70% over four years, well below the economy-wide 250% increase over the same period [333].

Like cyber, the AI insurance market will probably mature in two stages. Unlike cyber, it has the benefit of hindsight: with appropriate leadership and coordination, cyber's three-decade learning curve can be compressed for AI insurance.

### II.7.A Stage 1: Rapid Auditing and Iterating on Standards

In the near term, the most practical approach is frequent auditing of policyholders against a built-for-purpose standard, with the standard itself updated on a cadence fast enough to keep pace with advances in AI capabilities and attack methods. Early adopters of AI insurance will accept a relatively heavy audit burden — much as early cyber insurance purchasers accepted lengthy security questionnaires and on-site assessments — because the coverage and the trust signal will be worth it.

What might this look like in practice? A comprehensive standard would combine several elements. An initial in-depth audit verifies technical guardrails (e.g. input validation, output filtering, tool-use sandboxing), operational practices (e.g. incident response, human escalation protocols), and legal policies (e.g. acceptable use terms, data processing agreements). Performance evaluations and or red teaming tests are re-run on a frequent (e.g. quarterly) cadence using the latest methods, to catch newly discovered vulnerabilities and verify that performance remains in spec. The full suite of technical and organizational controls is re-audited annually to maintain certification. Critically, the standard itself is revised on a similarly rapid cadence, incorporating new research, attack vectors, and best practices as they emerge.

Audits could be structured in two rounds with a remediation window in between. This is where the insurer or MGA's risk management expertise is most valuable, not just for flagging deficiencies, but actively helping policyholders fix them. The goal is loss control, not box-checking. Policyholders that remediate identified issues between rounds get to maintain their certification (and coverage terms); those that don't face premium adjustments or non-renewal.

Some of the standards surveyed in §II.3 are built around exactly this model pairing certification with recurring performance evaluations and a quarterly revision cycle.

This stage is labor-intensive by design. It builds the institutional knowledge — which controls actually reduce losses, which evaluation methods are most predictive, which deployment patterns correlate with claims — that will underpin the more scalable approaches of Stage 2.

### II.7.B Stage 2: Partnerships with CSPs and Foundation Model Providers

Stage 2 is about scaling these efforts, by reducing cost and friction between insurers and policyholders. We suspect this will be best enabled by partnerships with Cloud Service Providers (CSPs) and frontier model providers, who already hold much of the information on downstream customers that insurers need in order to price, monitor, and support policyholders. This mirrors developments in cyber, where insurers have partnered with major CSPs to access telemetry on policyholders' security posture with lower friction [63].

For AI, the relevant telemetry from a model provider or CSP might include:

- Model Context Protocol (MCP) tool calls, API usage patterns, and rate limits (what are they, how often are they hit).
- API key rotation frequency.
- Whether the provider's content filtering is enabled or disabled.
- How frequently the provider is blocking requests based on their own internal safeguards (such as content filters).
- Which foundation model version(s) the downstream developer or deployer is using.
- How system prompts are structured.
- Input-output token ratios (which can signal whether a system is doing retrieval-augmented generation, chain-of-thought reasoning, or something else entirely).
- Tool-call patterns and agent traces: which tools the agent invokes, how often, in what sequences, and the frequency of high-impact actions (e.g. writes to external systems, financial transactions, code execution). Emerging runtime-governance frameworks formalize exactly this telemetry — mapping each agent run to a trace whose spans record capability invocations, tool intent (read/compute/write/dispatch), and approval and containment events — providing a ready-made schema for the signals insurers most need to monitor [169]. Traces also reveal whether human-in-the-loop checkpoints are being triggered as designed [169].
- Anonymized, aggregate-level data about model outputs (as seen in e.g. [334]), indicating whether the customer's use case is staying tightly scoped or expanding into new domains (e.g. fewer outputs are categorized as basic customer support, and more are categorized as financial advice).

The insurance industry should be helping and encouraging potential partners — whether foundation model providers or hyperscalers — to build dashboards that enable better risk management for business customers and, eventually, seamless insurance coverage cf. [317], [111].

Alongside these provider partnerships, advances in insurtech will further reduce overhead. For example, performance evaluations and red teaming can and will be automated for scalable, repeatable stress-testing. Self-assessment tools will also surely be built.

The transition from Stage 1 to Stage 2 will not be a clean switch. Audit-heavy approaches will persist for high-risk deployments, and complex or novel use cases [63], [111], even as provider partnerships make coverage economically feasible to write for many more small and medium-sized enterprises (SMEs). What matters is that insurers commit to monitoring as an ongoing function, not a one-time gate.

**Recommendations:**

- **Insurers** should encourage or require regular performance evaluations and red teaming, helping policyholders with remediation and providing loss control guidance.
- **MGAs and Insurtech** providers should, to reduce friction, begin automating and streamlining the above stress-tests.
- **Insurers** should begin partnering with CSPs and frontier model providers to monitor what safeguards policyholders have enabled and (potentially) how often they are triggered.

## II.8 Incident Response and Claims Management

When an AI agent causes harm, the insurer's response in the hours, days, and weeks that follow does more than settle a single claim: it shapes loss outcomes, preserves the policyholder relationship, and generates the forensic evidence that feeds back into every other layer of the insurance stack. Effective incident response contains damage before it compounds; rigorous claims management ensures losses are paid promptly where coverage applies and denied defensibly where it does not; and thorough post-incident forensics turn each loss into a learning opportunity for underwriters, standard-setters, and risk modelers.

This section argues that AI agent incidents can borrow much of the breach coach and crisis management infrastructure from cyber and product recall insurance. However, that will not be sufficient: AI incidents will require new technical expertise, such as real-time intervention in autonomous systems, forensic reconstruction of agentic decision chains, and causal attribution across foundation models, fine-tuning, and scaffolding; claims adjusters will need to become technically literate in agentic AI. Finally, given the pace at which AI evolves, insurers must tighten the feedback loop between claims and the rest of the stack.

### II.8.A Containment

When an AI agent causes harm, insurers need to be at their customer's side with a plan ready.

In cyber, policyholders can call a 24/7 hotline in the event of a breach, and the insurer activates a pre-approved panel of attorneys ("breach coaches") who bring on partnered forensics firms, PR consultants, and so on [332], [335], [336]. Product recall works similarly, with insurers retaining crisis consultants to coordinate recall logistics, regulatory obligations, and brand rehabilitation [337], [338]. In both lines, the insurer orchestrates specialized vendors under pre-negotiated rates and service agreements, providing policyholders expertise they would struggle to assemble on their own under the time pressure of an active crisis. Insurers benefit from controlling losses and costs.

We believe AI incident response can copy this playbook in many respects. Legal counsel, PR firms, and claims adjusters will be familiar with much of the downstream fallout of an AI failure. A defamation suit triggered by a hallucinated output, reputational damage from a chatbot gone rogue, regulatory inquiries following a data leak: these are familiar harm categories, even if the proximate cause is novel.

What *is* new is the technical dimension of containing an ongoing AI incident. When a customer-facing AI agent is actively generating harmful outputs — fabricating legal citations, leaking confidential data through prompt injection, or making unauthorized decisions in an agentic workflow — someone needs to intervene at the system level. That means identifying the failure mode in real time, isolating the affected system or rolling back to a safe state, and implementing a patch or workaround before the damage compounds. This is not something breach coaches or PR consultants do. It requires AI-specific technical expertise: knowledge of common deployment configurations, monitoring infrastructure, and common security vulnerabilities cataloged by organizations like OWASP [273] and MITRE ATLAS [272].

This technical incident response capability is nascent but will become a critical part of the AI insurance value chain. Just as cyber insurers now partner with Digital Forensics and Incident Response (DFIR) firms like CrowdStrike [339] and LevelBlue [340], AI insurers will need partnerships with — or in-house access to — teams capable of diagnosing and containing AI-specific failures. Over time, we expect dedicated AI incident response firms to emerge, or existing DFIR firms to expand their offerings to cover AI systems.

### II.8.B Claims Management

Once the immediate crisis is contained, attention turns to the claim itself. Claims management is the process by which an insurer verifies that a reported loss falls within the scope of coverage, quantifies the indemnifiable amount, and pays — or denies — the claim. Done well, this delivers prompt and predictable recovery for the policyholder while protecting the insurer from inflated, fraudulent, or out-of-scope demands.

For AI agent claims, the central difficulty is forensic: establishing *what* happened, *why* it happened, and whether the chain of causation maps onto what the policy actually covers. Standards and underwriting controls pay dividends here. Mandated documentation — version control, logging of tool calls, data provenance — ensures there is a paper trail to follow. Mandated pre-deployment performance evaluation results, red teaming results, near-miss logging, and incident response plans let claims adjusters measure what the policyholder *did* against what they *said* they would do; what *happened* against what policyholders *expected* to happen. Mandated input-output logs furnish contemporaneous evidence that is difficult to reconstruct or contest after the fact. Absent this infrastructure, adjusters face the unenviable task of adjudicating black-box failures with little more than the policyholder's own narrative.

Policy language is the other half of the equation: if it is not clear, no amount of forensics will determine whether a loss is covered. Clear language clarifies the forensics required to enforce it, saving time, money, and trust.

To illustrate, consider what's needed to make a workable exclusion for losses attributable to an upstream foundation model (desirable for insurers to control accumulation risk). First, the policy language must clearly and narrowly scope what losses it's excluding — e.g. confabulations originating in the foundation model. Second, controls and documentation must enable claims adjusters to determine whether a given failure was indeed more likely due to the upstream foundation model or to downstream engineering. This may mean rerunning prompts on the foundation model[37] without any downstream scaffolding or harnesses, checking that fine-tuning data wasn't unreasonably inaccurate, reviewing calls to any RAG[38] system, making sure system prompts weren't applying inappropriate pressure on models, and so on. The carveout is effectively unenforceable without standards and underwriting controls that mandate such data logging and instrumentation. Few claims will be large enough to warrant their use, but they should be *available* if needed.

The claims adjuster role itself will need to evolve too. The familiar skill set — coverage analysis, loss valuation, negotiation, fraud detection — remains essential, but AI claims will demand a new layer of technical literacy, e.g.:

- AI agent types (coding, customer service, financial…)
- Common deployment configurations / scaffolding setups
- Evaluation methodologies (see §II.5.B)
- Common security vulnerabilities [272], [273].

They will need enough fluency to interrogate forensic reports, challenge implausible attributions, and recognize when a policyholder's narrative is inconsistent with the data. In the near term, insurers will likely need to review their panels of external counsel in order to provide adjusters access to the AI forensic specialists they need, much as cyber adjusters consult DFIR experts on causation and scope-of-breach questions.

## II.8.C Post-Incident Forensics

As part of the claims process and for the benefit of underwriters, policyholders, and potential affected individuals, insurers should encourage or provide thorough post-incident forensics. For AI agents, a proper autopsy might include:

- **Examining logs and reconstructing the incident.** An investigator's first step is to gather relevant data, interpolate what is missing, and attempt to reconstruct the incident. What inputs did the system receive — user prompts, tool outputs, environmental data? What outputs did it produce, and what intermediate decisions did the agent make? For agentic systems, this (ideally) means reconstructing not just the final harmful output, but tool calls, branching decisions, and much more (in the limit: the full chain of thought).[39]
- **Identifying root causes.** A complete root cause analysis would go well beyond the immediate technical failure. Recent work proposes that AI agent incidents arise from a chain of causes spanning three domains [171]: *system factors* (training data, design choices, scaffolding architecture), *contextual factors* (the external conditions and inputs that precipitated the failure), and *cognitive errors* in the agent's observation, reasoning, or action execution. This is the right level of granularity. A shallow taxonomy that stops at "the chatbot hallucinated" misses the distinct upstream causes — a poisoned training dataset versus an adversarial prompt versus a flawed system prompt — that demand very different preventive measures. Insurers should push for incident reports that capture all three levels.

[37] Ideally, inspecting its chain-of-thought, or, in the limit, its internal activations [306].

[38] Retrieval-Augmented Generation: a technique that improves LLM reliability by fetching facts from an external knowledge base before generating a response.

[39] A complication: the causal trail may not lie entirely within the insured's own systems. An agent can be diverted by content it encountered externally — for instance a prompt injection embedded in a third-party webpage — which may have since changed or disappeared and over which the insured has no control.

- **"Neuroscience for AI."** To probe deeper, investigators could potentially use tools from mechanistic interpretability (a.k.a. "neuroscience for AI" [341]) to analyze the foundation model's internal processing — attention patterns, neuron activations, chain-of-thought traces — to understand *why* it made the decisions it made [171], [306]. This is unlikely to be practically feasible in the near-term: researchers lack consensus on the reliability and maturity of interpretability tools, which also require extensive access to the foundation model in question, raising significant trade secret and security concerns for foundation model providers [306].
- **Analyzing organizational decisions.** Beyond the technical failure, investigators might scrutinize the institutional choices surrounding deployment. Was the system deployed in a use case it wasn't designed for? Were monitoring thresholds set appropriately? Did the deployer ignore warnings from pre-deployment testing? Were human oversight and escalation mechanisms in place and functioning? Were warning labels sufficient? What effect did the UX design have on the user's understanding of the AI agent's role, capabilities, and limitations? Did organizational routines normalize over-reliance on the technology [342], [343]?

This last point matters as much as the technical analysis. The gold standard of incident investigation is in aviation, where root cause analysis almost always studies the chain of organizational decisions — maintenance schedules, training protocols, communication failures — behind the immediate technical failure [344]. AI will be no different.

## II.8.D The Role of Liability

Such efforts should not depend on good will alone: insurers, developers, and deployers need incentives to conduct thorough investigations and share what they find: design of the liability regime is critical. Research and experience tell us that lawyers are sometimes more attractive than safety engineers for both insurers and policyholders trying to limit liability risk (cf. [345]). This stifles institutional learning: forensic findings are buried to avoid creating evidence of negligence rather than disseminated to prevent future incidents [49]. There is evidence of this in cyber insurance, and there is no reason to think AI will be immune. As noted earlier, transparency requirements and disclosure mandates would greatly help (see §II.1 above). Clearer assignment of liability through statute could also help [3, Pt. III], as long as it does not simply displace liability insurance with a cruder proxy (e.g. blanket immunity in exchange for certification against a standard).

Besides deploying sticks, the government should make doing the right thing as frictionless as possible, such as by streamlining voluntary, anonymized reporting to the aforementioned incident database. By adding value back to the industry, through analysis of general trends and bulletins on best practices, such a government-run database could further encourage good-faith reporting. Reporting should be the path of least resistance, not a reputational risk.

### II.8.E Closing and Tightening the Feedback Loop

With incidents (and associated losses) logged, analyzed, and categorized, the insurance cycle ends and begins anew. The loss data feeds back into risk modeling and pricing; the technical findings inform updates to safety standards and technical testing methods; and the organizational lessons shape underwriting criteria and loss control guidance for future policyholders.

In many lines, closing this loop is enough. However, valuable claims data can take a year or more to feed back into the rest of the insurance stack. Given the speed at which AI evolves, many of the data's insights will be stale by then: insurers will need to tighten this feedback loop. In short, claims management shouldn't just be a settlement function, but also a high-fidelity and timely sensor for the ground truth on what's going wrong with the technology.

**Recommendations:**

- **Insurers** should establish pre-approved panels of AI-specific technical incident response providers, analogous to existing breach coach and DFIR retainer models in cyber. Where such AI incident response and forensics firms don't yet exist, encourage their development. The data required for conducting forensics should be included in aforementioned AI agent standards, to ensure it is logged.
- **Insurers** should train claims adjusters to be AI literate, and extract learnings from incidents to improve underwriting and loss control for AI agents.
- **(Re)insurers** should encourage or require structured post-incident reports as a condition of a major claim settlement. Reports should follow a standardized taxonomy covering, among other things, system factors, contextual factors, and cognitive errors.
- **Advisory bodies** (e.g. ACORD[40] and LMA[41]) should develop standardized incident and claims reporting templates.
- **Non-regulatory government bodies** (e.g. NIST's[42] CAISI[43]) should encourage good-faith incident reporting to anonymized industry databases, similar to NASA's ASRS[44] for aviation.
- **Governments** should adjust liability regimes and court discovery rules to ensure incentives are aligned with institutional learning, and not simply burying or destroying incident data.

[40] Association for Cooperative Operations Research and Development.

[41] Lloyd's Market Association.

[42] National Institute of Standards and Technology.

[43] Center for AI Standards and Innovation.

[44] Aviation Safety Reporting System.

# Part III. Looking Ahead: Covering AI CAT

Agents hallucinating outputs, leaking customer data, infringing on IP, giving unreasonable financial advice, deleting codebases, issuing improper refunds: these are the bread and butter losses that a healthy AI insurance market absolutely can and should handle. The components covered in Part II of this report together describe the complete stack needed for such a market to function well.

However, a truly healthy insurance market would also manage tail risks. Unfortunately, catastrophic risk from frontier AI agents — AI CAT — might be severe. Covering it will likely require alternative capital structures and government involvement. What follows is a preliminary discussion.

## III.1 The Stakes of Artificial General Intelligence

Frontier AI companies such as OpenAI, Anthropic, and Google DeepMind are explicitly committed to building Artificial General Intelligence (AGI). While there is no consensus among experts, a growing number believe we will succeed in building AGI by the end of this decade, with some predicting much sooner (cf. [19], [20], [346]). Whatever one's view, it is certain that AI today is the worst it will ever be; the capabilities curve points steeply upward [24], [39]. Of note: because the length of tasks that coding agents can accomplish without human intervention is doubling roughly every four months [19], a feedback loop is forming in which AI agents are being used to build and improve the next generation of AI agents [347], [348], [349]. This dynamic is sometimes referred to as "recursive self-improvement" [350], typically with the connotation that humans are slowly being removed from the loop: evidence of this last is unclear.

The development of AGI could produce transformative economic and social benefits, and failing to develop it has high opportunity costs ([100], [101], [102], cf. [117]). But the technology is inherently dual-use. In November 2025, Anthropic reportedly caught Chinese state-backed actors using Claude to automate sophisticated cyber espionage campaigns against dozens of global targets, executing 80–90% of tactical operations without human intervention [351], [352]. In April 2026, Anthropic announced its Mythos model, whose cyber capabilities allow it to automate cyber attacks end-to-end and discover novel zero-day vulnerabilities in every major operating system and web browser [33], [34], [138]. Reporting to senators a demonstration of Mythos' capabilities, US National Security Agency chief Joshua Rudd said Mythos "broke into almost all of our classified systems, not in weeks, but in hours" [36]. Rather than releasing it publicly, Anthropic gave critical infrastructure providers early access to patch their systems' vulnerabilities before Mythos-like capabilities are widely available to attackers. Mozilla (developer of the *Firefox* browser) gives a sense of how significantly Mythos has changed the strategic landscape: with Mythos, it shipped 423 security fixes in April 2026 alone, nearly double the 258 shipped in all of 2025.

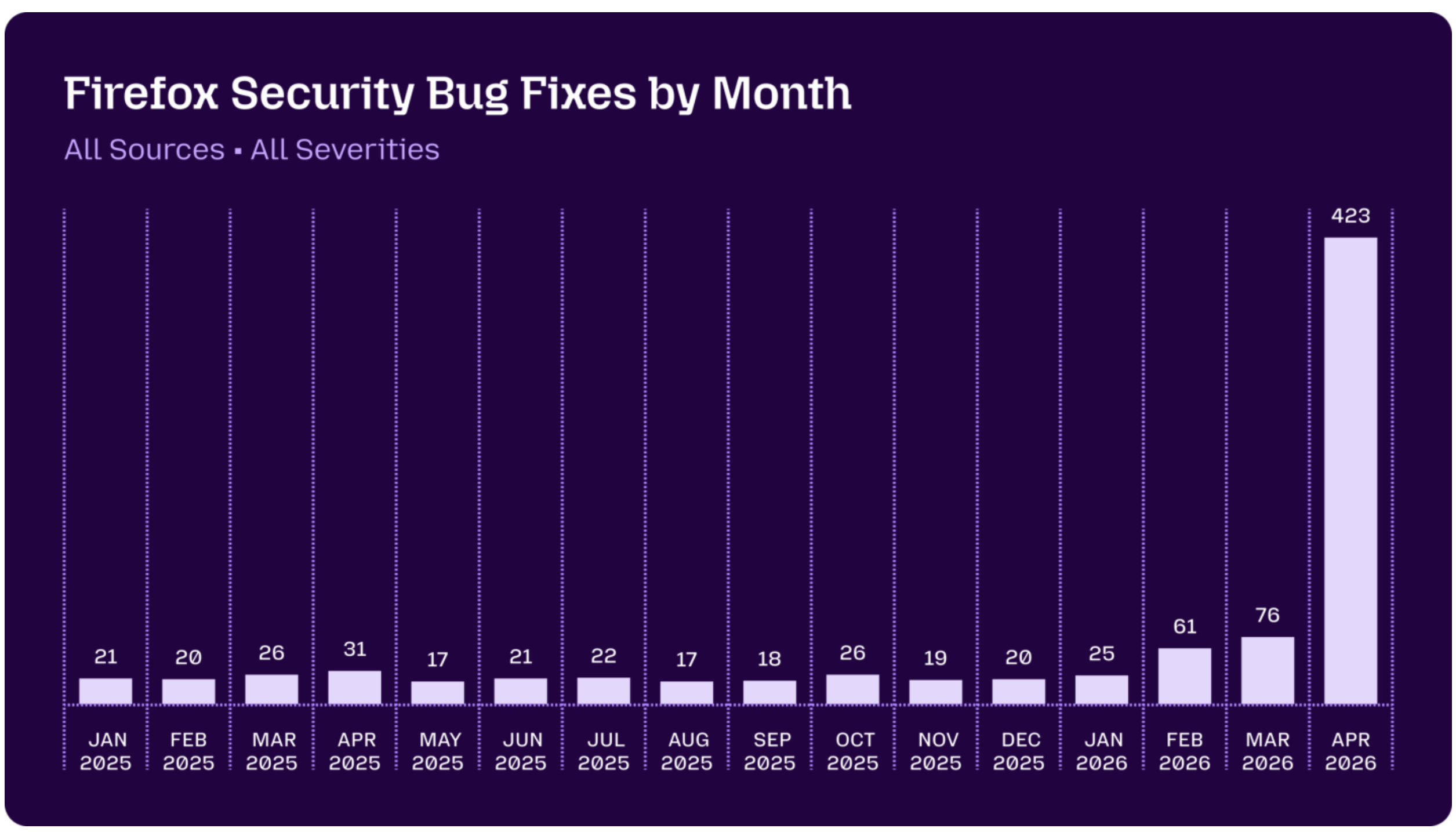

*Image 5. Firefox Security Bug Fixes by Month. Source: Mozilla Foundation [35]*

As more powerful AI is developed and deployed in higher-stakes contexts, the risk of catastrophic loss increases along several dimensions. This includes societal-scale risks, with direct losses on the order of hundreds of billions of dollars or more. For example, many experts and frontier AI developers warn that AI could enable bio-terrorism attacks [38], [39], [141]. For reference, the direct, health-related losses from the COVID-19 pandemic were over $10 trillion (2026 USD) for the US alone [131]. Some experts also warn of the possibility of losing control over advanced AI systems that develop goals misaligned with their operators' intentions. In the limit, they warn, this could lead to human extinction [39], cf. [64], [65]. These more dramatic catastrophes, centered on a few systems or actors going rogue, come in addition to the risk of large correlated losses from diffused, cascading multi-agent disasters discussed above in §II.2.

## III.2 The Lesson of 9/11

Future research should explore AI CAT scenarios in detail. However, if they are anything like cyber CAT, there are many plausible scenarios in which the 5-year GDP at risk for North America and Europe quickly reaches the trillions of dollars [75], [76].

What's also clear: direct damages and remediation will make up only a fraction of the lost GDP cf. [75], [130], [353]. If insurers are unprepared, they will rush to add exclusions in response to such a catastrophe; coverage for AI related risk could become unavailable or prohibitively expensive. Enterprise AI adoption could freeze or roll back as trust in the technology craters and procurement departments refuse to approve deployments without adequate coverage. It's difficult to predict the exact mechanism, but this is where the bulk of lost potential lies: economic activity seizing up and contracting, magnified by a sudden retraction of insurance coverage and a potentially lingering "AI malaise." This kind of freeze up and mini recession has been documented before. Consider 9/11: it's estimated to have caused some $150 billion (2026 USD) in direct losses and indirect lost economic output [129], [354], [130]. However, the direct

loss of life and property accounts for only 14 billion (2026 USD) [355] and 42 billion (2026 USD) [356], respectively. A "fear factor" that affected consumer behavior is blamed for the bulk of the losses [130] and a sudden retreat of insurance certainly didn't help: in the wake of the disaster, property insurance premiums jumped 5% cf. [84], [355] and a wave of exclusions carved out mispriced terrorism risk. Insurance coverage requirements built into bank loans suddenly couldn't be met, grounding commercial aviation [357] and halting major construction projects [83]. The drag on the recovery could have been considerably worse had the federal government not intervened [130], [358], [82], [81].

In the face of potential AI catastrophes, insurers are not powerless: they can either be a part of the problem, or part of the solution, building resilience into the system. At a minimum, they should begin excluding AI CAT they are silently exposed to, signaling to all market participants the need to better prepare and price these risks in. Again though, if insurers did only this, they would be largely abdicating their role in the economy. Ideally, they would help build the insurance stack needed to price in and avoid such catastrophes, starting with investing in understanding and modeling AI CAT (see §II.2 above).

## III.3 Macroeconomic Risk: An AI Bubble?

Major terrorism risk is not a perfect analogy: in many ways, AI CAT is structurally much thornier.

The US economy is increasingly a highly leveraged bet on frontier AI paying high dividends [359]. AI-related capital expenditure has become a primary driver of economic growth — by some estimates over 1% of GDP growth in 2025 [360] and the top AI-related companies account for roughly three-quarters of S&P 500 returns since ChatGPT's launch [360]. Tech companies are projected to spend some $400 billion per year on AI infrastructure; for reference, the entire Apollo space program cost about $300 billion in inflation-adjusted dollars. The AI buildout is costing a new Apollo program, not every 10 years, but every 10 months [361].

There is growing concern that the AI investment boom is a bubble [77], cf. [78]. Consumer spending on AI services remains modest — roughly $12 billion annually [361] — and the Organisation for Economic Co-operation and Development (OECD), the IMF, and the Federal Reserve have all flagged the boom as a potential downside risk to global growth [79], [80]. A sudden loss of confidence — whether from a catastrophic AI failure, a regulatory shock, or simply underwhelming returns — could trigger a violent market correction: stock sell-offs, credit market tightening, and regulatory clampdowns. No insurer can fully insulate itself from such a macroeconomic shock [362]; even those with zero direct AI exposure would feel it through investment portfolio losses and knock-on effects across the economy.

Ultimately the government is always implicitly exposed to such risks: its implicit commitment to bail out critical sectors of the economy, or provide emergency relief to the public in the event of a disaster cannot credibly be disavowed *ex ante*. The government is society's *de facto* insurer of last resort [66]. A healthy insurance regime would internalize harms that are externalized to taxpayers in this way, curtailing moral hazard from any "too-big-to-fail" effect [3].[45]

[45] A.k.a. "charity hazard" [363].

## III.4 Solutions for AI CAT

In sum, insuring AI CAT is needed to:

- Ensure that society captures the upside of transformative artificial general intelligence by improving security, safety, and reliability
- Reduce the likelihood of downside harms
- Protect the broader industry and economy from major shocks
- Reduce the moral hazard the government generates by default
- Ensure individuals harmed are compensated

This is a tall order, but not insurmountable. We believe the market for insuring downstream applications of AI agents will be the proving ground for insuring the tail risks of foundation model providers. Insuring the tails will involve many of the same components: incident data collection, standards, audits, technical evaluations, monitoring, and safety R&D cf. [111].

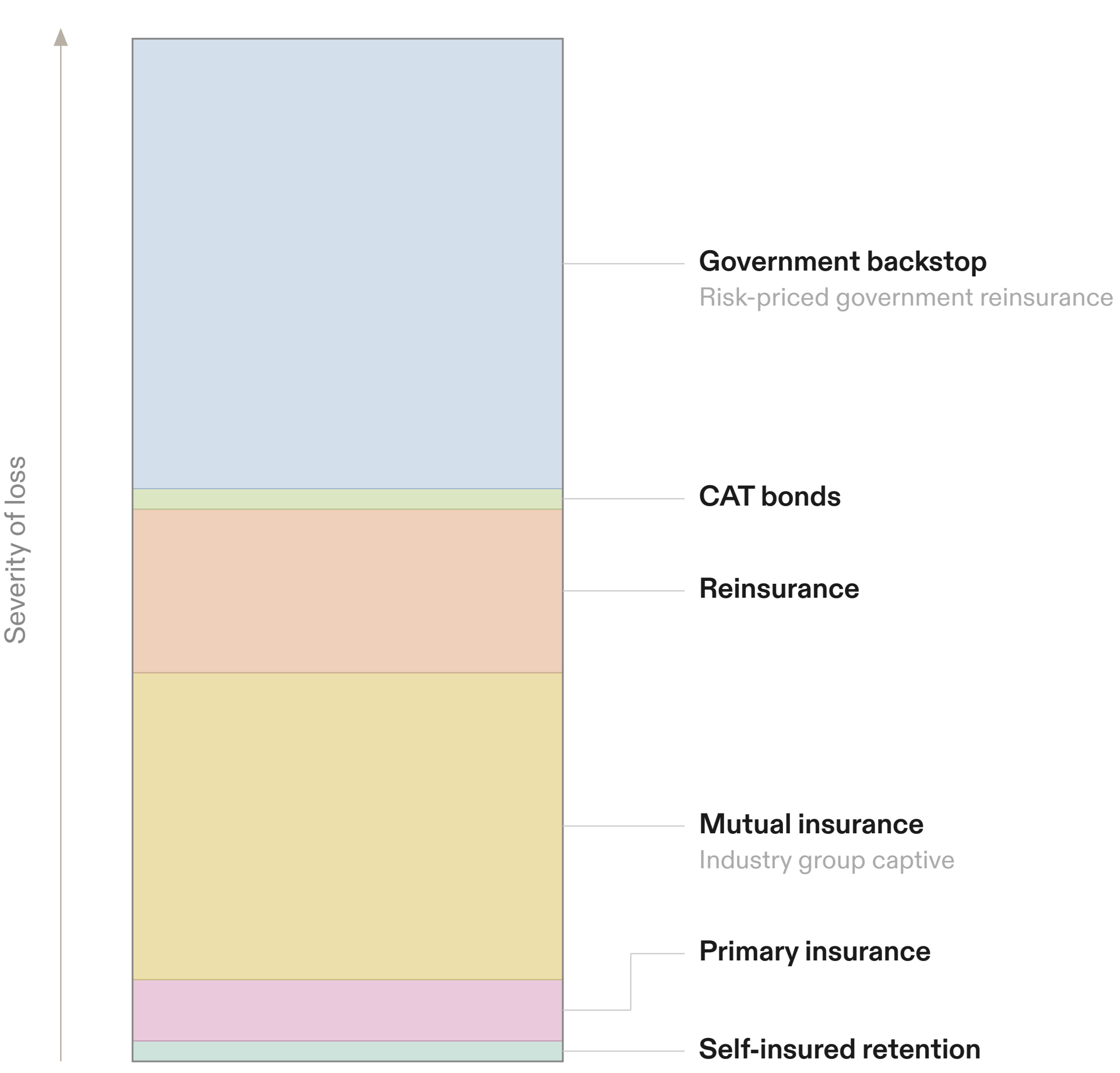

*Figure 7. Note: Proportions of coverage are illustrative.*

The principal difference will be less emphasis on scaling the infrastructure (there are inherently few customers) and more on technical loss control, the most cost-effective way to keep loss-ratios in check and reduce the risk capital needed for coverage [71], [3].

The challenge is finding sufficient capacity, designing appropriate capital structures (see Figure 7), and clearly delineating responsibilities and commitments between the public and private sectors. An exhaustive treatment awaits a future paper; here we briefly survey four potential components of the solution space:

- **Mutualization.** A promising solution to both the capacity and structure problem is a type of *industry mutual* (in this case, a group captive): not-for-profit insurers owned by their policyholders, in which frontier AI companies would effectively insure each other against catastrophic risk cf. [9], [22]. Historically, mutuals have the best track record of matching effective private governance with sustainable financial protection [3]: they coordinate members on best practices, invest in public goods like safety R&D, protect the industry's reputation through robust oversight, and leverage peer pressure [3], [67], [68], [364]. For example, Nuclear Electric Insurance Limited (NEIL) is a mutual critical to commercial nuclear power's effective self-regulation [67], [68], [146]. An AI mutual would address the industry's coordination problems — including the free-rider problem that leaves safety R&D chronically underfunded — while maintaining sufficient independence from any single member to serve as a genuine check on the industry. Partnering with and leveraging the expertise of traditional (re)insurers (as NEIL does) will be critical, but we cannot expect them to supply the bulk of the capacity required.
- **Catastrophe bonds and other insurance-linked securities (ILS).** CAT bonds allow insurers and reinsurers to transfer some catastrophic risk to capital markets, tapping a much larger pool of risk-bearing capacity than the insurance industry possesses on its own [365]. This capacity comes at a high premium, since, unlike traditional reinsurance, CAT bonds are fully collateralized. As such, they constitute a very small share of coverage in other markets (e.g. ~0.5% of the cyber market) [366]. Some suggest CAT bonds might serve as a *price discovery* tool: the bond market would aggregate dispersed private information about the likelihood and severity of frontier AI's tail risks [69]. In practice, this mechanism is limited by the complexity of the product, and large information asymmetry between frontier AI companies and the rest of the market: issuing such a bond still requires costly assessment and quantification of the risk being transferred [69]. In other words, CAT bonds cannot substitute for the machinery of standard setting, underwriting, and auditing.
- **A bespoke liability regime.** For such extraordinary risks, many scholars believe common law tort liability will not produce efficient outcomes, from both a perspective of insurability, as well as an insured's activity and care levels [3], [70], [72], cf. [73]. Where a catastrophic AI failure produces harms that would otherwise bankrupt the primary party liable, a *residual* liability framework could ensure individuals harmed are compensated by spreading the remaining liability to the rest of the industry [72]. This would also further encourage the industry to police itself, crystallizing an already implicit shared liability in the form of a collective industry reputation: an AI Three Mile Island would be a blow to all frontier AI companies, not just the one that caused it [3]. (This is especially germane given the strong negative public sentiment around AI [85]). *Strict* liability for certain catastrophic harms — mirroring the regimes applied to nuclear power and ultrahazardous

activities — would focus insurer efforts on pricing and mitigating technological risk, rather than gaming legal risk [3], [71]. Liability can be further clarified and channeled (and thereby made more insurable) by making it *exclusive*[46] [71], cf. [88]. Finally, liability can also be made more insurable by e.g. *limiting or barring* punitive damages, but scholars differ regarding expected welfare impacts: Weil calls for *expanding* punitive damages to internalize the *ex ante* expected loss of a near-miss catastrophe [74], [71], [367], [70], [3].

- **Government backstops.** Only the state can insure some risks. In the US, the Price-Anderson Act for commercial nuclear power offers an insightful precedent: a three-tier insurance scheme for nuclear meltdowns, in which the first layer (~$500 million) is covered by traditional insurers; the second (~$13 billion), by the nuclear power industry collectively; the third, by the government [88]. This system successfully nurtured the nuclear power industry, by giving private insurers the confidence to enter the market and shoulder more of the risk than they otherwise would have. This in turn provided electric utility companies the coverage they needed to make investments in nuclear power worthwhile, while still keeping these companies accountable.[47] A comparable structure for AI CAT, where the private sector bears the first layers of risk and the government absorbs the excess (see Figure 7), would protect against a sudden withdrawal of coverage without letting the industry off the hook for rigorous risk management. An *ex ante* commitment to backstop losses also lets the government extract something in return for playing insurer of last resort. This might include security requirements, transparency mandates, and or premium contributions that could fund programs aimed at managing AI-caused disruptions [22]. In fact, the Price-Anderson system has generated net positive (if modest) revenue for the government [88]. Governments could also share underwriting-relevant signal in return, such as summary scores from classified national-security testing of frontier models (e.g. for CBRN[48] or offensive-cyber uplift), which no private evaluator could produce, but would feed directly into insurers' risk evaluation and pricing.

# III.5 Getting Ahead of the Problem

The instinct to defer these questions — to wait until the outlines of AI CAT become clearer and losses materialize — is understandable but mistaken. The lesson of 9/11 is that catastrophic loss from silent, mispriced exposure can chill economic activity far beyond the sectors directly affected.

[46] A regime in which legal responsibility for a class of harm is channeled to a single designated party, while other potentially responsible parties along the supply chain are statutorily shielded from liability. This typically governs only whom plaintiffs may sue; the designated party generally remains free to pursue its own indemnity, contribution, or subrogation claims against other parties.

[47] The Terrorism Risk Insurance Program (TRIP) played a similar role for terrorism risk: following the insurance market freeze up post-9/11, TRIP created a federal backstop that gave private insurers the confidence to re-enter the market, and shoulder more of the risk than they otherwise would have [81]. Unlike Price-Anderson, however, TRIP is a straight subsidy: there is evidence this has generated moral hazard [368].

[48] Chemical, Biological, Radiological, or Nuclear harms.

The industry should get ahead of this problem now, while the stakes are comparatively manageable. Every component of the insurance stack described in this report — data collection, standards, contract design, risk evaluation, pricing, monitoring, incident response, safety R&D — will help market participants notice, plan, and price for catastrophic risks before they materialize, building resilience into the system. Whereas command and control regulation famously suffers from a pacing problem, arriving either too late or too early for a given technological development [369], [72], it's never too early for insurers to start pricing in risk. Doing so, however, will take a coordinated, industry-wide effort.

# Glossary

**5-year GDP at risk:** A macroeconomic risk metric estimating the cumulative shortfall in GDP, relative to a baseline forecast, that could occur over a five-year horizon under adverse scenarios. Originally developed for assessing financial stability risks, it captures both the probability and severity of tail outcomes, making it useful for quantifying the downside of shocks.

**Accumulation risk:** The risk that many policies in an insurer's portfolio are exposed to the same underlying source of loss, such that a single event or trigger produces correlated claims across the book. It arises when exposures that appear independent at the policy level prove correlated at the portfolio level, transforming what is diversifiable in theory into concentrated loss in practice.

**Actuarial modeling**: The use of statistical and mathematical techniques to estimate the frequency and severity of future losses, set premiums, and determine reserves. An actuary is the credentialed professional who builds and maintains these models, translating historical loss data and exposure information into pricing and capital decisions.

**Adverse Selection:** When insurers disproportionately attract bad risk because responsible actors, correctly judging themselves to be overcharged under a one-size-fits-all rate, decline coverage. Driven by information asymmetries between insurers and policyholders about the true quality of the risk.

**Affirmative coverage:** Coverage for a peril that is expressly written into a policy — with clear definitions, triggers, and limits — for a specific risk or peril.

**Agentic AI:** AI systems built on frontier Large-Language Models (LLMs) that can receive high-level natural language instructions, form plans, and take consequential actions in the world with limited or no human oversight (e.g. placing orders, moving funds, modifying software, or otherwise making decisions that affect customers, operations, or third parties).

**Alignment Failures:** Failures in which an AI system pursues objectives that diverge from the developer's or customer's intent, whether due to a failure in understanding or to adversarial behavior. Includes specification gaming, reward hacking, and scheming.

**Artificial general intelligence (AGI):** A hypothetical AI model or system that equals or surpasses human performance on all or almost all cognitive tasks [39], cf. [346].

**Artificial Intelligence (AI):** Digital systems that are capable of performing tasks commonly thought to require intelligence, with these tasks typically learned via data and or experience.

**Benchmarking:** Standardized testing of model performance on defined capabilities — helpfulness, harmlessness, reliability, alignment — used to compare models and track generational improvements.

**Catastrophic risk:** Risk of high-severity (and typically low-probability) loss events (~hundreds of millions of dollars or more). Such risks strain the capacity of private insurance markets: coverage is often limited, or prohibitively expensive. For more see [370]. *NB: many in the AI policy space use "catastrophic risk" to broadly refer to what we call "societal-scale risk" here. We adopt the (fuzzy) threshold for "catastrophic risk" that the insurance industry uses.*

**Catastrophe bond ("CAT bond"):** A type of insurance-linked security (ILS) that transfers catastrophe risk from insurers or reinsurers to capital markets investors. Investors receive above-market coupon payments in exchange for putting their principal at risk: if a predefined catastrophic event occurs, some or all of the principal is forgiven and used to pay claims. CAT bonds are fully collateralized, meaning the principal is held in a trust, eliminating counterparty credit risk for the sponsor.

**Commercial General Liability insurance (CGL):** Standard third-party liability coverage protecting businesses from claims of bodily injury, property damage, and personal injury arising from operations or products.

**Chain of thought / Reasoning traces:** A technique for generating responses in which an AI model generates intermediate steps or explanations. By breaking down complex tasks into smaller steps, this approach can improve the model's accuracy and indicate how it arrived at its answer. There are other forms of reasoning traces, but this is the principal form we refer to.

**Claims adjuster:** The insurance professional who investigates a claim, determines coverage and valuation, and negotiates settlement.

**Cyber insurance:** Coverage for losses arising from cyber incidents — data breaches, ransomware, network outages, regulatory fines, and the associated response and remediation costs.

**Directors and Officers insurance (D&O):** Liability coverage protecting a company's directors and officers (and often the company itself) against claims alleging wrongful acts in their managerial capacity — such as breach of fiduciary duty, mismanagement, or misleading disclosures. Typically purchased by public and private companies to attract executive talent and shield personal assets from litigation.

**Data Poisoning Attack:** An adversarial attack in which malicious or manipulated data is deliberately introduced into an AI model's training data (or fine-tuning data, or retrieval corpus) in order to corrupt the resulting model's behavior. Poisoning can degrade overall performance, induce targeted misclassifications, or implant hidden "backdoors" that cause the model to behave normally except when a specific trigger is present.

**Deductible:** The amount a policyholder must pay out of pocket on a covered loss before the insurer's coverage applies. Higher deductibles typically lower premiums and serve as a control on moral hazard by ensuring the insured retains some financial stake in each loss.

**Deployer:** An individual or company that puts an AI agent into operational use within its own organization (e.g. JP Morgan Chase, Kaiser Permanente, Orrick).

**Downstream AI developer:** Companies that build AI agents on top of foundation models (e.g. Cursor, ElevenLabs, Harvey).

**Errors and Omissions insurance (E&O):** Coverage for liability arising from mistakes, oversights, or negligence in the policyholder's professional services. Tech E&O is the version tailored to technology products and services. "Professional Indemnity Insurance" is essentially the UK/European name for E&O.

**Emergent Multi-Agent Failures:** Correlated failures arising from interactions among large populations of AI agents. Hammond et al. organize such failures along three axes: *miscoordination* (agents failing to align their actions), *conflict* (agents pursuing incompatible goals), and *collusion* (agents cooperating against user or societal interests). Distinct manifestations of such failures include coordinated manipulation, herd behavior, and feedback loops where agents shape shared environments (the internet, codebases, markets) that other agents rely on.

**Exclusions (insurance):** Specific risks, perils, or types of loss expressly carved out of a policy and not covered.

**First-party Insurance:** Coverage for losses the policyholder suffers to its own systems, property, operations, reputation, or other interests — as distinct from liability claims brought by outside parties.

**Foundation model:** A general-purpose AI model trained on broad data and adaptable to a wide range of downstream tasks. These are typically LLMs at their core. This is the substrate on which current AI agents are built.

**Foundation model developer / provider:** Organizations that train and or provide foundation models (e.g. Anthropic, OpenAI, Google DeepMind).

**Frontier AI:** General-purpose AI models and systems whose performance is no more than a ~year behind the state-of-the-art on a broad suite of general capability benchmarks. [111]

**General-purpose AI**: AI models or systems that can perform a variety of tasks, rather than being specialized for one specific function or domain.

**Generative AI (Gen AI):** AI systems that generate content — text, images, speech, code — in response to prompts. Used in its inclusive sense, "Gen AI" includes all agentic AI (since AI agents currently all depend on generative LLMs). Used in its exclusive sense, "Gen AI" refers to AI systems that are "merely" generative tools, and not agentic (e.g. Midjourney, Udio).

**Hard/soft market (insurance):** The cyclical state of insurance pricing and availability. In a *hard* market, premiums rise, capacity contracts, underwriting standards tighten, and coverage terms become more restrictive; in a *soft* market, premiums fall, capacity is abundant, and terms loosen as insurers compete for business [164], [371]. The cycle is driven by factors including capital availability, recent loss experience, and reinsurance costs.

**Hallucination / Confabulation:** Inaccurate or misleading information generated by an AI model or system, presented as factual.

**Index insurance:** A form of insurance that pays out based on a predefined, objectively measurable index (e.g. rainfall levels, earthquake magnitude, or a benchmark loss metric) crossing a specified threshold, rather than on the policyholder's actual losses. The simplicity of such products allows insurers to write many micro-policies. It also reduces moral hazard, but introduces basis risk — the possibility that the index payout diverges from the policyholder's true loss.

**Insured / Policyholder:** The individual or entity that purchases an insurance policy and is entitled to coverage under its terms.

**Insurability:** The degree to which a risk can feasibly be insured. Insurability sits on a spectrum and depends heavily on an insurer's risk appetite, making it somewhat subjective [370]. Several objective, evaluable dimensions also factor in though, including [372]:

- Whether there are many independent and identically distributed exposure units
- Whether losses are unintentional and accidental
- Whether losses are definite as to time, place, amount, and cause
- Whether the chance of loss is calculable
- Whether premiums are affordable.

**Jailbreaking:** Generating and submitting prompts designed to bypass safeguards and make an AI system produce harmful content, such as instructions for building weapons.

**Large Language Model (LLM):** An AI model trained on large amounts of text data to perform language-related tasks, such as generating, translating, or summarizing text.

**Level of activity vs. level of care:** A distinction from the law-and-economics literature, between two ways an actor can reduce expected harm: exercising more *care* while engaging in an activity (a.k.a. precautionary effort) versus simply doing *less* of the risky activity (activity level).

**Liability insurance:** Coverage that protects the policyholder against legal liability to third parties — typically including defense costs and damages — when the policyholder's actions, products, or services cause harm to others.

**Loss of control scenario:** A scenario in which one or more general-purpose AI systems come to operate outside of anyone's control, with no clear path to regaining control [39, p. 151].

**Loss ratio**: The ratio of incurred losses to earned premiums over a given period, expressed as a percentage. It is a core measure of underwriting performance: a lower loss ratio indicates more profitable underwriting. The "combined loss ratio" extends this by adding expenses (acquisition costs, overhead) to losses before dividing by premiums; a combined ratio below 100% indicates an underwriting profit, while above 100% indicates an underwriting loss (which may still be offset by investment income).

**Managing General Agent (MGA)**: A specialized insurance intermediary that has been delegated underwriting authority by an insurer to bind coverage, set terms, and sometimes handle claims on the insurer's behalf, typically within a defined class of business. MGAs are common in niche or emerging lines (such as cyber and, increasingly, AI) where specialized technical expertise is needed that primary carriers may lack in-house.

**Moral Hazard:** The tendency of insureds to take on more risk once they know they are insured against the consequences. A core friction insurance must price and control for.

**Property and Casualty insurance (P&C):** The broad category of insurance lines covering damage to physical property and liability for harm to others; encompasses most commercial coverage.

**Penetration testing:** A security practice where authorized experts or AI systems simulate cyberattacks on a computer system, network, or application to proactively evaluate its security. The goal is to identify and fix weaknesses before they can be exploited by real attackers.

**Policy limit:** The maximum amount an insurer will pay under a policy, typically expressed as an aggregate limit covering all claims within the policy period (usually one year). Once the annual aggregate is exhausted, the policyholder bears any further losses. Policies also often include per-claim limits, and sometimes other sub-limits for specific scenarios (e.g. losses under critical infrastructure failure).

**Performance Evaluation:** System-specific testing of an AI system's performance in realistic environments on the tasks it is primarily intended to accomplish.

**Premiums (insurance):** The price the policyholder pays the insurer upfront in exchange for coverage, typically expressed as an annual amount.

**Prompt Injection:** An attack in which adversarial instructions are smuggled into the input an AI system processes — either directly by a user (direct prompt injection) or indirectly through external content the system ingests, such as a web page, email, document, or tool output (indirect prompt injection) — causing the system to override its original instructions, bypass safeguards, exfiltrate data, or take unintended actions. Indirect prompt injection is a particular concern for agentic AI, which routinely acts on untrusted content drawn from the open web and other third-party sources.

**Red teaming (for AI):** Adversarial testing in which evaluators deliberately attempt to provoke unsafe, off-policy, or unintended behavior from an AI system in order to surface vulnerabilities before deployment [39, p. 153].

**Reinsurance:** Insurance purchased by insurers from other insurers (reinsurers) to transfer a portion of their risk, allowing primary insurers to write larger policies, stabilize results, and protect against catastrophic or correlated losses. Reinsurance can be arranged on a treaty basis (covering a whole class of business) or facultative basis (covering individual risks).

**Reserving:** The practice of estimating and setting aside funds to pay for losses the insurer expects to pay — including claims reported but unsettled, and claims incurred but not yet reported. Reserves appear as liabilities on the balance sheet and must be backed by admitted assets of appropriate quality and liquidity. (*See also risk capital*).

**Risk capital:** Capital held by an insurer above its reserves to absorb losses that exceed expected obligations. Where reserves fund anticipated claims, solvency capital absorbs shocks — adverse loss development, catastrophe events, or correlated portfolio failures. Regulatory regimes such as Solvency II (EU) and Risk-Based Capital (US) prescribe minimum levels calibrated to defined stress scenarios, typically a 1-in-200-year loss over one year. (*See also reserving*).

**Safeguard:** A technical measure or process designed to prevent AI systems from causing harm.

**Self-Insured Retention (SIR):** The amount of loss a policyholder must absorb out-of-pocket before insurance coverage responds. Unlike a deductible, an SIR is typically managed by the insured itself (including defense costs for liability claims) and sits outside the policy limit.

**Silent coverage:** Exposure to a peril that is neither expressly covered nor expressly excluded by a policy, leaving the insurer potentially liable for losses it never priced for. Also known as "silent" or "non-affirmative" exposure.

**Single Point Of Failure (SPOF):** A node or dependency in a system whose failure would bring down ~everything downstream of it (e.g. a foundation model provider or hyperscaler), producing correlated losses across otherwise independent insureds.

**Societal-scale risk:** A class of catastrophes whose magnitude and correlational structure render them all but uninsurable by conventional private markets — e.g. widespread infrastructure collapse, war, or a bio-terrorist attack with widespread casualties. As a rule of thumb, we classify scenarios involving over $50 billion in losses as societal-scale risks.

**Systemic risk:** Accumulation risk arising from the interactions and interdependencies in a complex system, which can produce unexpected vicious feedback loops, contagion, or other correlated loss dynamics.

**Telematics**: The use of sensors, connected devices, and real-time data feeds to monitor an insured's activities or assets, enabling usage-based pricing, continuous risk assessment, and loss prevention. Originally developed for auto insurance (tracking driving behavior), the concept extends to any line where telemetry — including AI system logs, API call patterns, and runtime monitoring — can inform underwriting and claims.

**Third-party moral hazard:** When the presence of insurance can increase the likelihood of loss, by encouraging misuse or negligence on the part of third parties. E.g. some scholars find that the presence of ransomware insurance has made ransomware more profitable, encouraging the criminal activity. [330], [332]

**Underwriting:** The process of evaluating, selecting, and pricing risks an insurer is willing to accept, including setting terms, conditions, exclusions, and limits. An underwriter is the professional who performs this assessment, balancing the insurer's appetite for risk against the premium charged.

**Zero-day vulnerability:** A software security flaw that is unknown to the vendor or defenders at the time it is discovered, leaving zero days for a patch to be developed before it can be exploited. Zero-days are especially valuable to attackers because no defenses exist against them until the flaw is disclosed and remediated.

# Appendix 1: Plotting the Incident-to-Usage Ratio for Frontier AI

## Objective

We wanted to investigate whether frontier AI is becoming more reliable per unit of usage. What we are ultimately looking for is the analog of deaths-per-vehicle-mile-traveled for AI (Figure 3). This appendix documents the quantitative exercise we conducted to try to build such a metric with publicly available data. Our objective was modest: not to estimate an absolute incident rate — the available datasets measure fundamentally different phenomena and cannot be meaningfully combined into a single rate — but to ask whether incidents and usage are growing at comparable speeds, and whether any directional finding is robust to the choice of data sources.

To build our proxy metric, we assembled two composite indices, one capturing frontier AI incidents and one capturing frontier AI usage, each spanning the period 2023—2025.

## Index Construction

Both indices are simple arithmetic means of their constituent series, each of which is independently normalized to its own 2023 value before aggregation. All axes are plotted on a logarithmic scale.

### Incident Index

The incident index is composed of four constituent series:

**AIID (GenAI / Frontier AI incidents).** Publicly reported incidents drawn from the AI Incident Database (AIID) [373], filtered to those involving generative or frontier AI systems. The AIID is a crowd-sourced, media-sourced repository annotated by collaborators including the MIT AI Incident Tracker [172], [374]. The AIID represents a known floor on actual incident counts, as inclusion depends on media coverage or voluntary submission.

**OpenAI Consolidated.** So as not to bias the composite toward any one provider, a single consolidated series constructed by averaging three OpenAI disclosure channels: (i) total notices received under the EU Digital Services Act (DSA), drawn from OpenAI's DSA transparency reports [375]; (ii) total own-initiative content moderation measures, also from the DSA reports; and (iii) pieces of content reported to NCMEC (the National Center for Missing & Exploited Children), drawn from OpenAI's Trust & Transparency Hub [375]. No 2023 data is available for the DSA reporting channels; the consolidated series therefore spans 2024—2025 for those components. A data-cleaning decision was made to exclude all of OpenAI's reported actions removing "discoverable links" in 2024, which was an anomalous, uncategorized bulk action, and which would have dominated 2024 own-initiative actions (accounting for over 80% of the actions). This automated bulk action was taken when OpenAI changed policies and decided to remove the discoverable links feature [376]. There is no corresponding action of this type ("ChatGPT (Shared Links)") in the 2025 data.

**Anthropic Consolidated.** So as not to bias the composite toward any one provider, a single consolidated series constructed by averaging two Anthropic disclosure channels. The two channels are: (i) accounts banned by Anthropic, drawn from Anthropic's transparency and trust reports, including their System Trust Report, Platform Security Web Archive, and additional published reports [377]; and (ii) pieces of content reported to NCMEC, also drawn from the Anthropic Trust and Transparency Hub [377]. The banned-accounts series required one period of backcasting via linear projection to complete the 2024 data. 2023 data is completely lacking for banned accounts; no backcasting was attempted. For NCMEC reports, 2023 data is zero (Anthropic reported no content that year). Anthropic has published only a single DSA report, providing no time-series data for that channel.

**GenAI / Frontier AI Suits Filed.** The count of lawsuits filed involving generative or frontier AI, drawn from the George Washington University Database of AI Litigation [378]. This series is US-centric and captures suits filed, not suits resolved. The original dataset contained 415 entries.

A deduplication review identified four pairs of entries that represented the same underlying dispute recorded twice, reducing the final dataset to 411 unique cases. Case names were normalized and compared to surface entries sharing a plaintiff name and defendant. Candidate pairs were then verified by comparing jurisdiction, summary of facts, and the underlying events described to confirm they referred to the same dispute rather than genuinely distinct proceedings. Duplicates removed:

- **Walters v. OpenAI** — The state court filing in Gwinnett County and the federal filing in N.D. Georgia described the same defamation claim arising from ChatGPT-generated statements. The state court entry was retained as it had more recent activity (May 2025 vs. December 2024).
- **Kistler v. Eightfold AI** — Both entries described a class-action employment discrimination suit against Eightfold AI filed in California state court in January 2026. The Contra Costa County entry was retained as it contained a more detailed significance summary and more recent activity (March 2026 vs. January 2026).
- **Toronto Star Newspapers v. OpenAI** — Both entries described the same copyright infringement action filed in Ontario Superior Court of Justice. The retained entry had more recent activity (August 2025 vs. November 2024).
- **Li v. Liu / Li Yunkai v. Liu Yuanchun** — Both entries described the same Beijing Internet Court case concerning copyright in an AI-generated image created with Stable Diffusion. The retained entry contained more detailed factual and significance summaries.

The litigation dataset was then further filtered to only include cases involving relevant technology. We identified cases involving frontier AI, generative AI, or AI agents using a two-pronged keyword filtering strategy applied across all available fields (Case Name, Area of Application, Algorithm, Organizations, Most Recent Activity, and Issues). First, we retained any case whose "Area of Application" field contained the tags "Generative AI," "Agentic AI," "Natural Language Processing," or "Chatbot." Second, we performed a regular-expression search across all textual fields for terms associated with LLMs and generative AI systems — including named models and providers (e.g., GPT, ChatGPT, Claude, Gemini, LLaMA, Midjourney, Stable Diffusion, Perplexity, Character.AI, Runway), technical descriptors (e.g., LLM, foundation model, diffusion model, text-to-image, deepfake, retrieval-augmented generation), and relevant training-data references (e.g., LAION, Common Crawl, Books3). Cases matching on either criterion were retained, yielding a filtered sample of 217 cases. Keeping only suits filed in 2023 through 2025 leaves 173 cases. This approach is intentionally inclusive to capture the broad ecosystem of frontier-AI litigation, though it necessarily excludes cases involving narrower or traditional AI applications such as autonomous vehicles, facial recognition, predictive policing, hiring algorithms, and credit-scoring tools where no generative or large-language-model component was identified.

### Usage Index

The usage index is composed of five constituent series (with enterprise API spend double-weighted, as described below):

- **Google Gemini Monthly Active Users (MAU).** Captures consumer adoption breadth for Google's frontier AI product. Sourced from Epoch AI [379].
- **OpenAI ChatGPT Weekly Active Users (WAU).** Captures consumer adoption breadth for OpenAI's flagship product. Sourced from Epoch AI [379].
- **OpenAI Daily Messages.** Volume of daily user messages to ChatGPT, sourced from Epoch AI [379]. This series has no 2023 data point and enters the composite only for the 2024—2025 period.
- **Inference Tokens Served.** Total inference tokens across providers and modalities, sourced from Epoch AI [379]. This series captures aggregate computational usage.
- **Enterprise API Spend.** Annual enterprise spending on frontier foundation-model APIs (in USD billions), sourced from Menlo Ventures [40]. This series is double-counted in the composite — i.e., it receives twice the weight of the other series — for two reasons. First, in order to balance the representation of enterprise-side usage against the four consumer-oriented metrics above. Second, because the four consumer-oriented metrics form two natural pairs: the active user counts (Gemini and ChatGPT), and the intensity of usage (daily messages and tokens served).

The index aims to capture general frontier AI usage broadly, not solely enterprise usage, because the majority of incidents in the incident index occur at the consumer level. Showing that usage is growing faster than incidents across the full breadth of deployment supports the claim that the technology and its guardrails are improving on a per-unit-of-usage basis. Total lab revenue (ARR) was considered but rejected: it conflates consumer and enterprise revenue streams, is distorted by free tiers, and subscription revenue is a poor proxy for usage intensity.

## Results

The high-level result of plotting these ratios is shown in Figure 8.

Incident to Usage Ratio for Frontier AI (2023–2025)

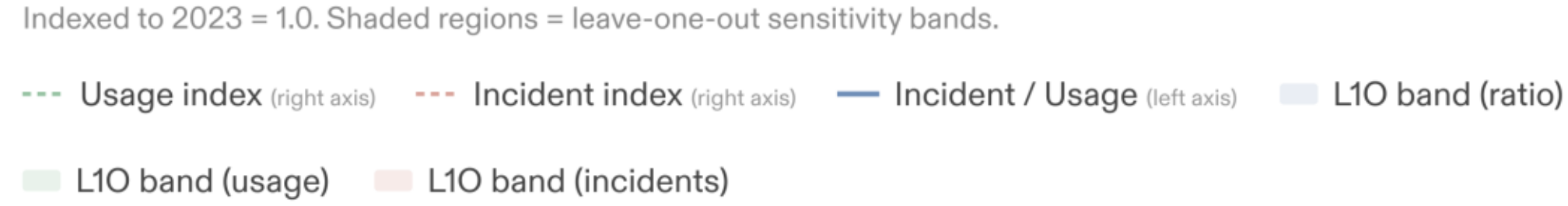

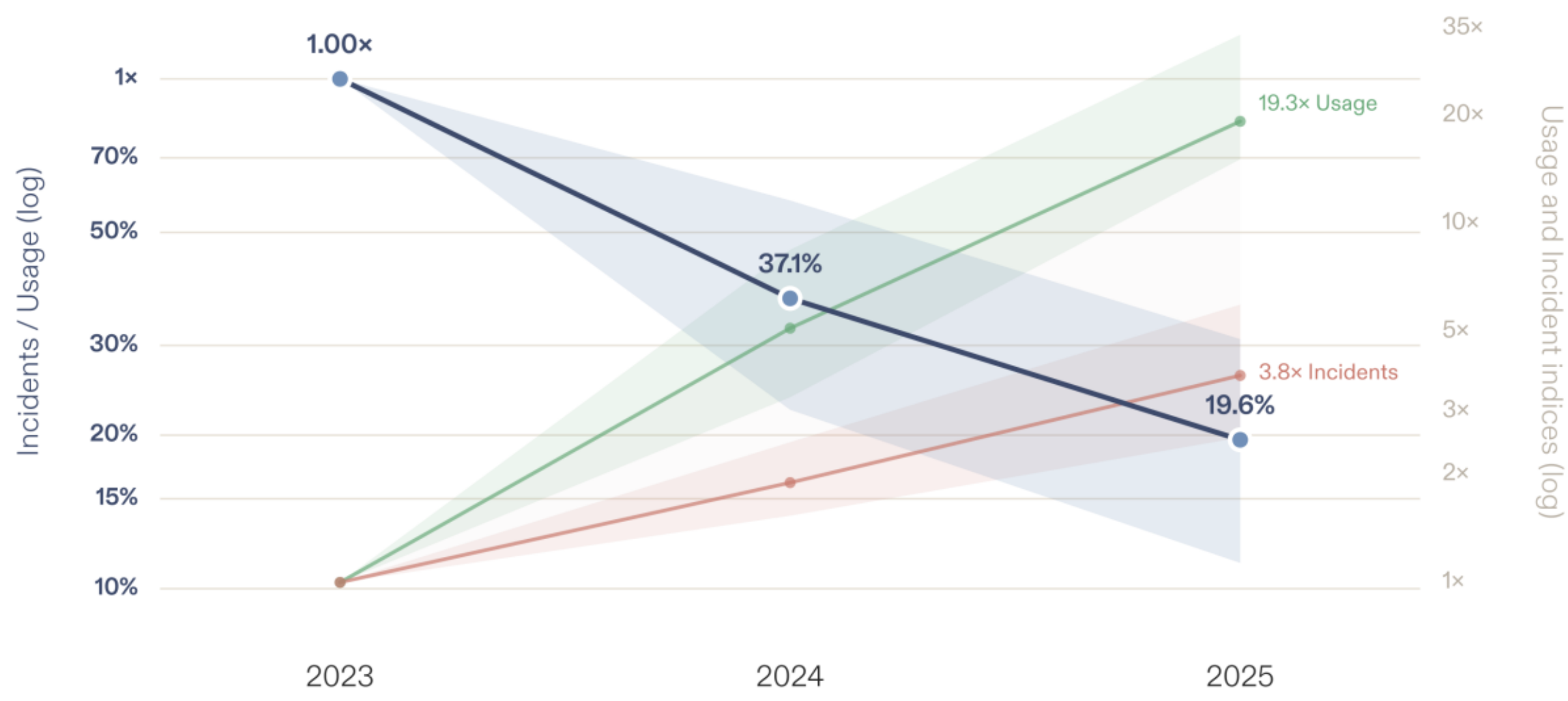

*Source: AIUC analysis. Bands show the range when any single constituent dataset is removed from the composite.*

*Figure 8.*

The incident index grows at roughly 100% year-over-year, while the usage index grows at roughly 350% year-over-year on average. The ratio of the two indices falls by over 80% from 2023 to 2025, suggesting that, on the whole, frontier AI systems are becoming more reliable per unit of usage.

# Sensitivity Analysis

To assess robustness, a leave-k-out (LkO) analysis was performed on both indices and on the resulting ratio. Every possible combination of removing one constituent series (leave-one-out, L1O) and two constituent series (leave-two-out, L2O) was computed.

## Leave-One-Out (L1O) Bands

For the **incident index** (4 series → 4 L1O variants), the 2025 index value ranges from 2.54× (dropping OAI Consolidated) to 5.94× (dropping AIID Incidents), against a base of 3.77×.

For the **usage index** (5 series → 5 L1O variants), the 2025 value ranges from 15.13× (dropping Gemini MAU) to 33.58× (dropping ChatGPT WAU), against a base of 19.26×. For the **ratio** (9 L1O variants from the cross-product of both indices), the 2025 value ranges from 11.2% (dropping ChatGPT WAU) to 30.9% (dropping AIID Incidents), against a base of 19.6%. In every L1O variant, the ratio declines decisively from 2023 to 2025 (Figures 9, 10, and 11).

## Leave-Two-Out (L2O) Bands

The L2O analysis produces 6 variants for the incident index, 10 for the usage index, and 36 for the ratio. Of the 36 ratio trajectories, 35 trend decisively downward. The single anomalous trajectory arises when both AIID Incidents and Suits Filed are dropped simultaneously, leaving only the two company-disclosure series (OAI Consolidated and Anthropic Banned Accounts) in the incident numerator, only the former of which has partial 2023 data. This combination produces a ratio that rises sharply, partially an artifact of incomplete temporal coverage rather than a genuine signal (Figure 11).

A corresponding anomaly appears in the incident index itself: dropping both Suits Filed and AIID Incidents yields a 2025 index value of 145.8×, driven entirely by the same data-coverage artifact. This outlier is clearly visible in the spaghetti plots (Figure 9).

## Interpretation

The sensitivity analysis demonstrates that no single data source, nor any pair, drives the qualitative conclusion. The downward trend in the incident-to-usage ratio is robust across 44 of 45 leave-k-out compositions.

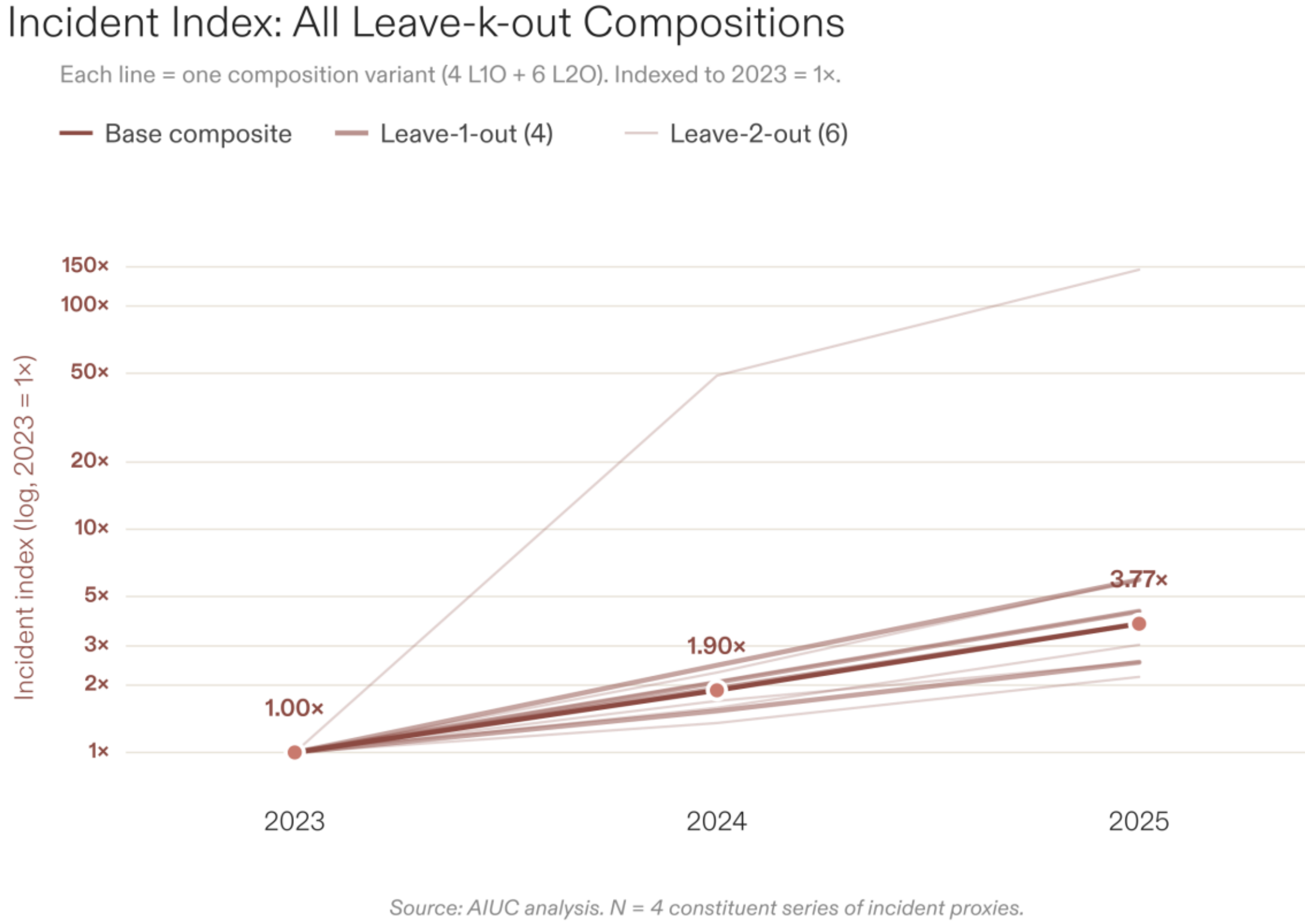

*Figure 9. Incident index: all leave-k-out compositions.*

## Usage Index: All Leave-k-out Compositions

Each line = one composition variant (5 L1O + 10 L2O). Indexed to 2023 = 1×.

Base composite — Leave-1-out (5) — Leave-2-out (10)

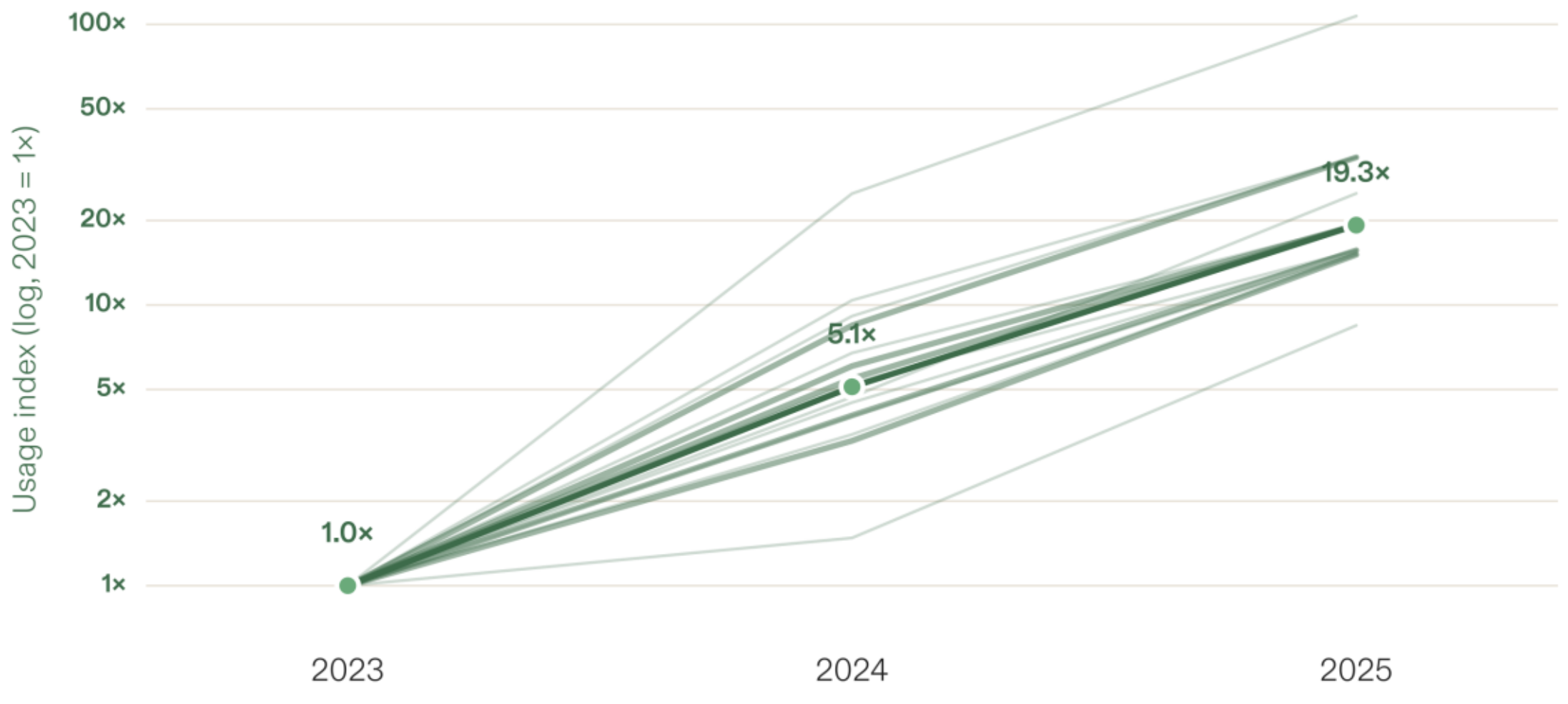

Source: AIUC analysis. N = 5 constituent series of usage proxies (API Spend double-counted).

*Figure 10. Usage index: all leave-k-out compositions.*

## Incident/Usage Ratio: All Leave-k-out Compositions

Each line = one composition variant (9 L1O + 36 L2O). Indexed to 2023 = 1.0.

Base composite — Leave-1-out (9) — Leave-2-out (36)

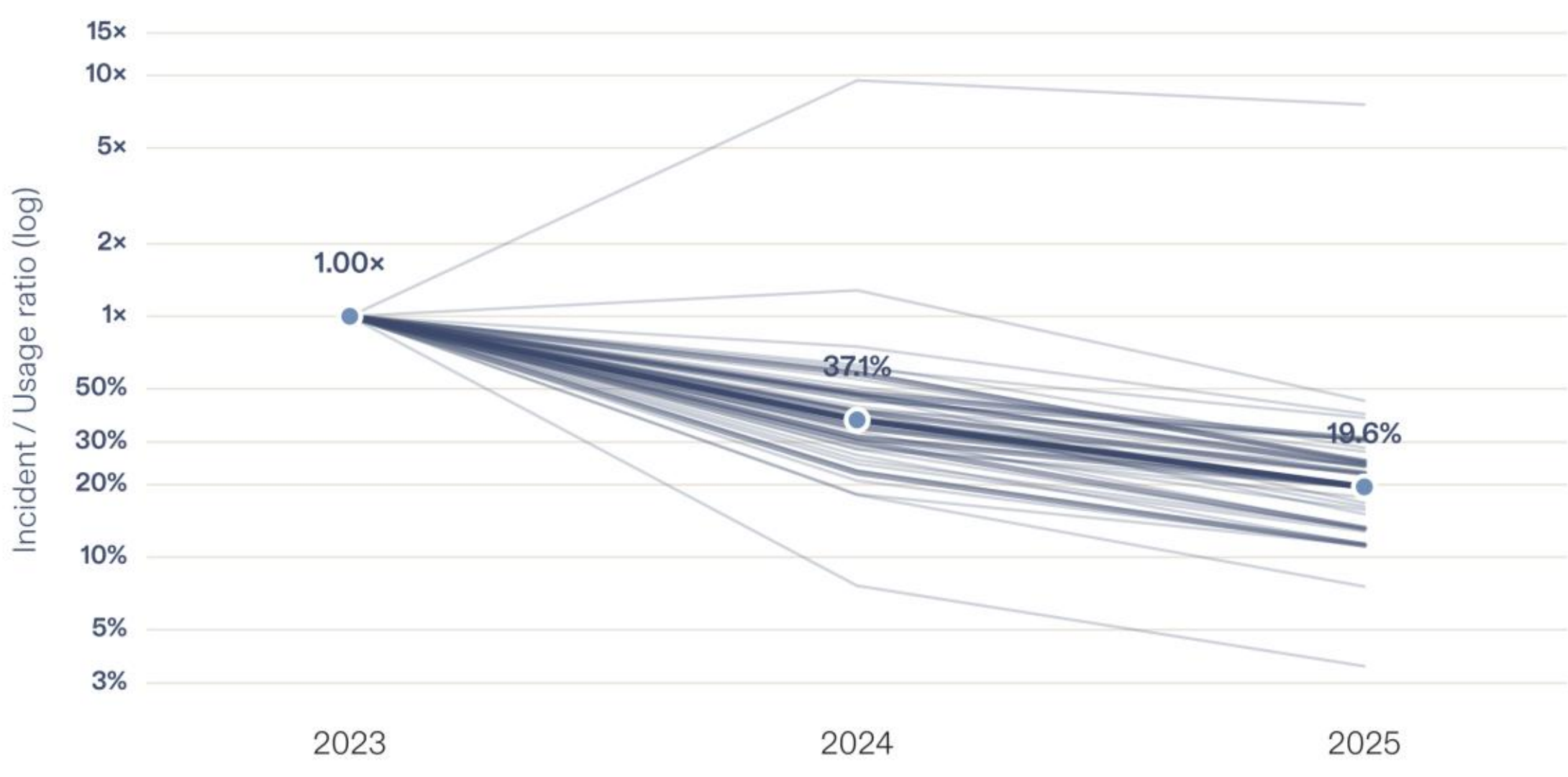

Source: AIUC analysis. 44 of 45 trajectories trend decisively downward.

*Figure 11. Incident/usage ratio: all leave-k-out compositions.*

## Constituent-Pair Overlay Plots

To further illustrate the relationship between individual incident and usage proxies, a suite of overlay plots pairs specific series against one another. Each overlay indexes both series to a common base year and plots the resulting ratio on a secondary axis.

**Public Incidents (AIID) vs Enterprise API Spend** (Figure 12) pairs the broadest public incident measure with the enterprise usage proxy, yielding a ratio that falls from 100% to 8.7% over the period — reflecting the fact that crowd-sourced public incidents roughly doubled while enterprise API spend grew 25-fold.

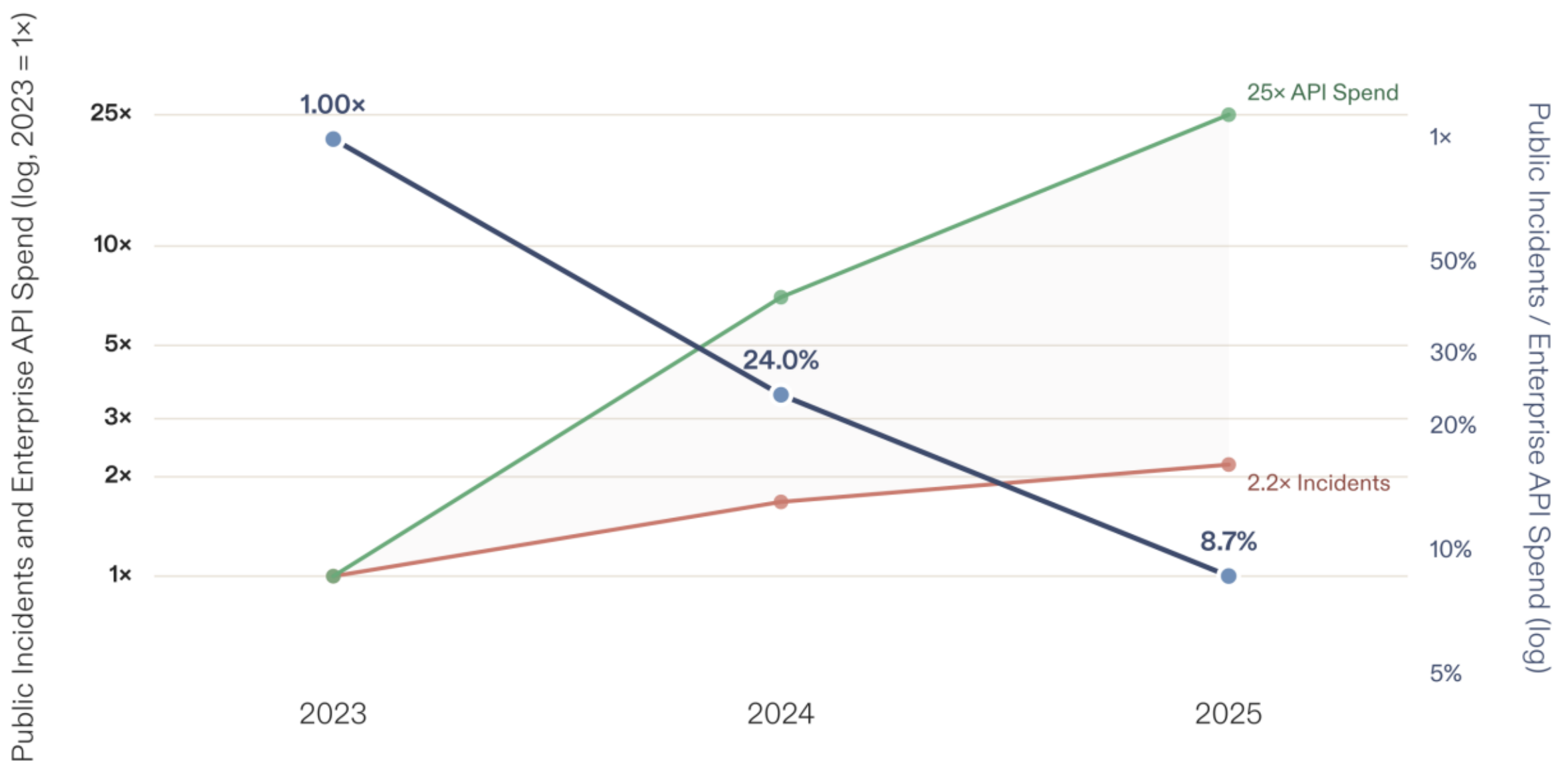

*Figure 12. Public incidents (AIID) vs. enterprise API spend.*

**Frontier AI Lawsuits vs Enterprise API Spend** (Figure 13) pairs litigation filings with the same usage proxy, showing a similar pattern with the ratio declining to 12.1%, as suits filed grew roughly threefold against the same 25× increase in spend.

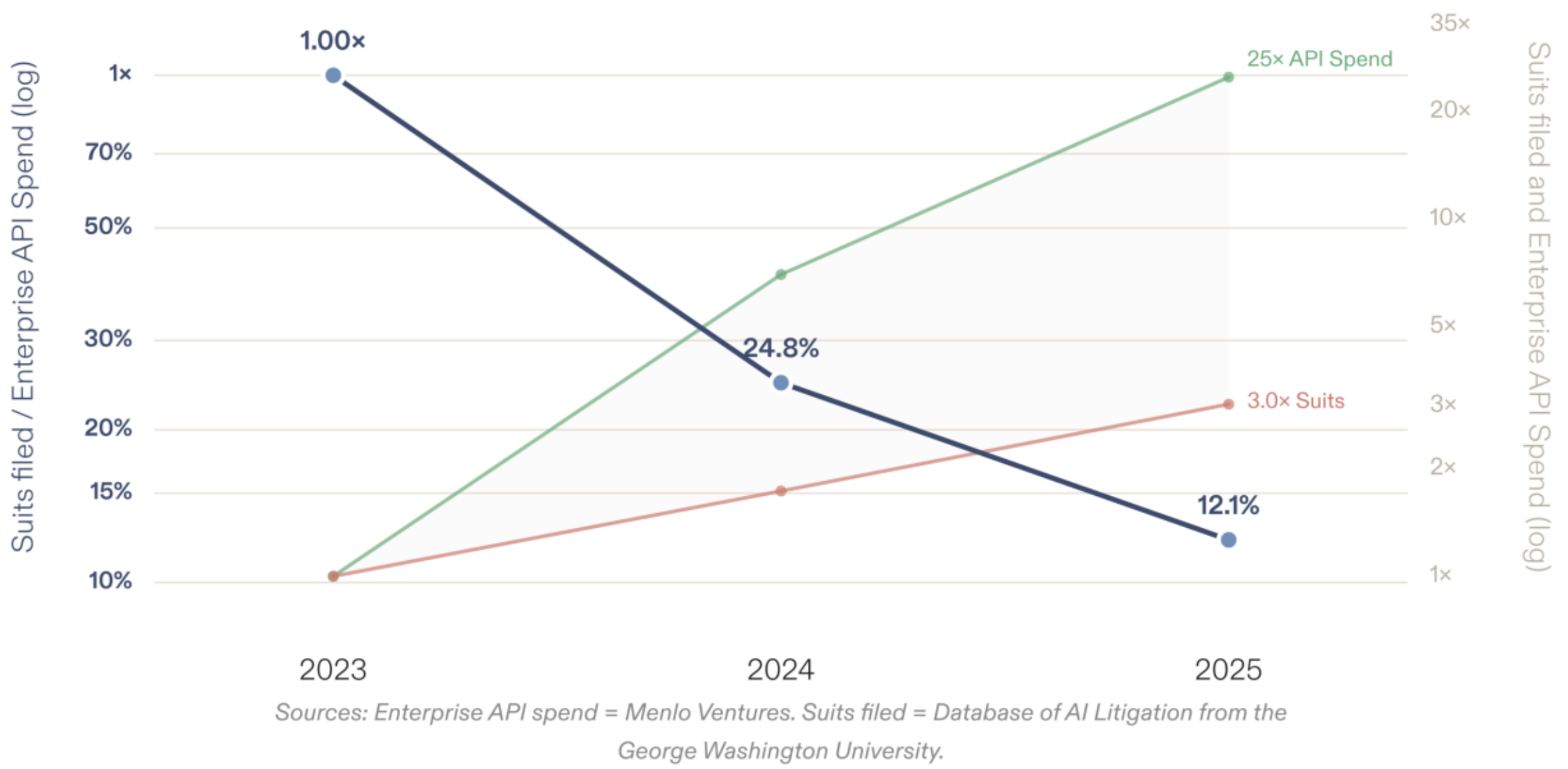

*Figure 13. Frontier AI lawsuits vs. enterprise API spend.*

**OpenAI Content Moderation vs OpenAI Daily Messages** (Figure 14) is the only same-provider overlay available, comparing OpenAI's consolidated content-moderation actions against ChatGPT daily messages; because daily message data begins only in 2024, this overlay is indexed to 2024 = 1.0 and spans a single year, during which the ratio declined modestly from 100% to 87.5%, with content moderation actions (5.1×) growing nearly as fast as messages (5.8×) — a pattern consistent with the detection-investment bias discussed in Section 3.1, whereby increased trust-and-safety investment surfaces more violations even as the per-message harm rate may be declining.

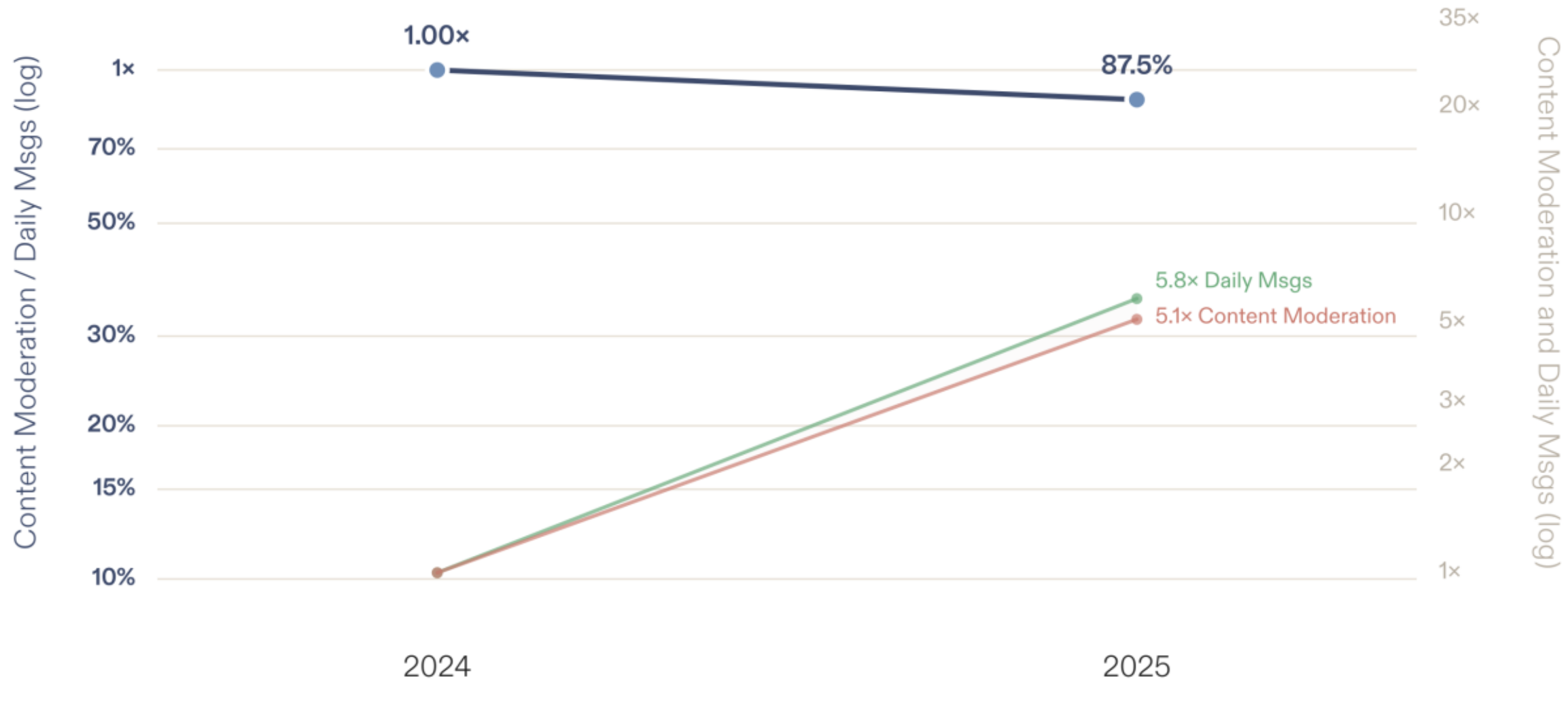

*Figure 14. OpenAI content moderation actions vs. OpenAI daily messages.*

# Biases and Limitations

Several limitations should be borne in mind when interpreting Graph 1. First, the index covers only three annual data points (2023—2025), which is insufficient to establish a statistically significant trend in any formal sense; the analysis is descriptive rather than inferential. Second, the incident index mixes heterogeneous signals with opposing biases, and its construct validity is weak. Third, the usage index is dominated by consumer-facing metrics, while this report's primary concern is enterprise and agentic AI deployment; the enterprise API spend series is double-weighted to partially address this, but the correction is ad hoc. Fourth, the 2023 base year is missing data for several constituent series (OAI DSA reports, OAI Daily Messages, one period of Anthropic bans), requiring either backcasting or exclusion from the 2023 normalization. Fifth, the analysis captures only publicly observable signals; the true incidence of AI failures in enterprise settings — where reputational incentives discourage disclosure — is unknown. Finally, the ratio's decline is consistent with improving safety per unit of usage, but it does not speak to the absolute level of risk, nor to the distribution of losses. This last is discussed below.

## Incident Index Biases

Each constituent series introduces distinct biases:

- **AIID — attention bias and growing underreporting.** The AIID is crowd-sourced and media-sourced. When AI attracts more public attention, more incidents are reported, creating a spurious positive correlation with usage growth — precisely the relationship the index aims to measure. At the same time, as AI systems become more routine, individual incidents may attract less coverage, leading to growing underreporting over time. These two effects partially offset, but we suspect underreporting heavily dominates, cf. [177], [178].

- **Company disclosures (OpenAI and Anthropic) — platform detection investment bias and harms prevented.** DSA reports, NCMEC reports, and Anthropic account ban numbers all reflect investment in trust-and-safety tooling. Companies that invest more in detection surface more violations. Furthermore, detection and reporting criteria may change over time. This makes these data series a somewhat inconsistent measure over time. Additionally, they report both harms that may have occurred and harms *prevented* (e.g., accounts banned, or harmful content taken down). The direction of this last bias is therefore favorable — it inflates the incident index, making the observed decline in the usage-to-incidents ratio more conservative than the true decline.
- **OpenAI Consolidated — duplicative reports, provider specificity.** The DSA channels are EU-specific and cover only one provider. NCMEC data is US-specific. DSA notices received may include duplicates where multiple users report the same incident; the ratio of actions taken to notices received is reported but cannot be decomposed (actions aren't associated with any particular notices). The decision was made to use the total notices figure, at the risk of inflating the incident index (again, making the observed decline in the usage-to-incidents ratio more conservative than the true decline).
- **Anthropic Consolidated — provider specificity, limited scope and backcasting.** This consolidated series covers only one provider. The banned-accounts component relies on one period of backcasted data (to fill in missing 2024 data); it also entirely lacks 2023 data. The NCMEC component has zero reports for 2023.
- **Suits Filed — legal market dynamics.** Filing rates reflect the legal market's response to AI as much as underlying harm. Alleged harms supported by a legal theory are not equivalent to harms realized. The series is US-centric, and focuses on civil lawsuits, as opposed to criminal lawsuits.

## Usage Index Biases

- **Enterprise API Spend** captures only the enterprise side of usage and does not account for in-house models. It is also conservative in that it does not adjust for the falling price per token at a fixed performance level. Adjusting for this would introduce significant complications: enterprises do not use AI at a fixed performance level but instead deploy a shifting mix of cheaper models for routine tasks and frontier models for complex ones, whose per-token prices have not fallen at the same rate.
- **MAU (Gemini, ChatGPT)** is consumer-centric, reflects adoption breadth rather than usage intensity, and misses enterprise usage almost entirely.
- **Inference Tokens** are inflated by reasoning models, which can consume orders of magnitude more tokens for marginal performance gains. Google's figures are particularly inflated because Google embeds AI into every Google Search query, which is arguably not agentic AI. Conversely, the series is deflated by improvements in model efficiency. Token counts are not uniform across providers or modalities.

### Construct Validity

The incident index is the weaker of the two constructs. It mixes fundamentally different signals: harms realized (AIID), misuse attempts and harms thwarted (company moderation reports and bans), and alleged harms with a legal theory (litigation). No separation is made between near-misses and actual harms. The many different biases across datasets make the incident index a very noisy signal for the rough construct of “how often does frontier AI go wrong, or nearly so.”

The usage index is less problematic but still imperfect. It combines three distinct categories of measurement — adoption breadth (MAU/WAU), computational volume (tokens), and economic activity (API spend) — each with different coverage and biases.

Despite these limitations, the qualitative finding is robust: usage is rising substantially faster than incidents, and this conclusion holds under extensive sensitivity testing.

## Interpretation

The exercise supports a directional but not definitive conclusion: usage is rising substantially faster than incidents, and this finding is robust to extensive sensitivity testing. The qualitative picture is consistent with the broader signals discussed in the main text — accelerating enterprise adoption, improving benchmark trajectories, declining per-unit incident rates in provider disclosures.

It does not support stronger claims. Three annual data points are insufficient to establish a statistically significant trend in any formal sense; the analysis is descriptive rather than inferential. The incident index mixes heterogeneous signals with opposing biases. The usage index is dominated by consumer-facing metrics, while this report’s primary concern is enterprise and agentic AI. The 2023 base year is missing data for several constituent series. Reputational incentives discourage disclosure of AI failures in enterprise settings, where the true incidence of failure is unknown. And finally, the ratio’s decline does not speak to the absolute level of risk, nor to the distribution of losses.

# Bibliography

[1] A. Lior, “Innovating Liability: The Virtuous Cycle of Torts, Technology and Liability Insurance,” Sep. 11, 2023, *Social Science Research Network, Rochester, NY*: 4568702. Accessed: Apr. 25, 2024. [Online]. Available: https://papers.ssrn.com/abstract=4568702

[2] O. Ben-Shahar and K. D. Logue, “Outsourcing Regulation: How Insurance Reduces Moral Hazard,” *Michigan Law Review*, vol. 111, no. 2, pp. 197–248, 2012, Accessed: Mar. 03, 2024. [Online]. Available: https://www.jstor.org/stable/41703440

[3] C. Trout, “When Does Regulation by Insurance Work? The Case of Frontier AI,” Oct. 10, 2025, *Social Science Research Network, Rochester, NY*: 5588732. Accessed: Nov. 17, 2025. [Online]. Available: https://papers.ssrn.com/abstract=5588732

[4] “Agentic commerce: How agents are ushering in a new era | McKinsey.” Accessed: Apr. 02, 2026. [Online]. Available: https://www.mckinsey.com/capabilities/quantumblack/our-insights/the-agentic-commerce-opportunity-how-ai-agents-are-ushering-in-a-new-era-for-consumers-and-merchants

[5] “Gartner Unveils Top Predictions for IT Organizations and Users in 2026 and Beyond,” Gartner. Accessed: Apr. 02, 2026. [Online]. Available: https://www.gartner.com/en/newsroom/press-releases/2025-10-21-gartner-unveils-top-predictions-for-it-organizations-and-users-in-2026-and-beyond

[6] “Agentic AI implementations in advanced industries | McKinsey.” Accessed: May 06, 2026. [Online]. Available: https://www.mckinsey.com/industries/automotive-and-assembly/our-insights/empowering-advanced-industries-with-agentic-ai

[7] “Generative AI to Become a $1.3 Trillion Market by 2032, Research Finds | Press | Bloomberg LP,” *Bloomberg L.P.* Accessed: May 06, 2026. [Online]. Available: https://www.bloomberg.com/company/press/generative-ai-to-become-a-1-3-trillion-market-by-2032-research-finds/

[8] M. Le Gouais, “AI liability: An age of silent exposures has already begun,” Insurance Insider. Accessed: Sep. 30, 2025. [Online]. Available: https://www.insuranceinsider.com/article/2f5mn8u0mwcdnovlve7eo/lines-of-business/casualty-gl/ai-liability-an-age-of-silent-exposures-has-already-begun

[9] Aon, “AI Fact Sheet 2026: Artificial Intelligence Risk Management 2026,” Aon, Fact Sheet, 2026. Accessed: Dec. 31, 2025. [Online]. Available: https://assets.aon.com/-/media/files/aon/reports/2026/aon-ai-fact-sheet-2026.pdf

[10] Swiss Re, “AI – unintended insurance impacts and lessons from ‘silent cyber’ | Swiss Re.” Accessed: Jul. 13, 2025. [Online]. Available: https://www.swissre.com/institute/research/sonar/sonar2024/ai-silent-cyber.html

[11] D. Powell, “LMA - Understanding AI Exposures: AI Loss Scenarios Survey Results.” Accessed: May 06, 2026. [Online]. Available: https://lmalloyds.com/campaigns/understanding-ai-exposures-ai-loss-scenarios-survey-results/

[12] “The State of AI in the Enterprise | Deloitte,” Deloitte, Jan. 2026. Accessed: Jan. 28, 2026. [Online]. Available: https://www.deloitte.com/cz-sk/en/services/consulting/research/the-state-of-ai-in-the-enterprise.html

[13] L. Harris and C. Criddle, “Insurers retreat from AI cover as risk of multibillion-dollar claims mounts,” *Financial Times*, Nov. 23, 2025. Accessed: Apr. 02, 2026. [Online]. Available: https://www.ft.com/content/abfe9741-f438-4ed6-a673-075ec177dc62?syn-25a6b1a6=1

[14] G. B. Fehling, M. S. Levine, M. “Mady” Moore, and A. D. Pappas, “The Continued Proliferation of AI Exclusions.” Accessed: Nov. 10, 2025. [Online]. Available: https://www.hunton.com/hunton-insurance-recovery-blog/the-continued-proliferation-of-ai-exclusions

[15] L. Bratton, “Berkshire Hathaway, Chubb Win Approval to Drop AI Insurance Coverage,” *The Information*, Apr. 23, 2026. Accessed: May 06, 2026. [Online]. Available: https://www.theinformation.com/articles/berkshire-hathaway-chubb-win-approval-drop-ai-insurance-coverage

[16] M. Lerner, “Insurers, brokers adjust as AI exclusions emerge,” Business Insurance. Accessed: May 06, 2026. [Online]. Available: https://www.businessinsurance.com/insurers-brokers-adjust-as-ai-exclusions-emerge/

[17] J. Apotheker, “CEOs are all in on AI but anxieties remain: What leader confidence indicates for 2026,” World Economic Forum. Accessed: May 06, 2026. [Online]. Available: https://www.weforum.org/stories/2026/01/ceos-are-all-in-on-ai-but-anxieties-remain/

[18] R. (Alex) Jia, M. Eling, and T. Wang, “Gen AI Risks for Businesses: Exploring the role for insurance,” Geneva Association, Oct. 2025. [Online]. Available: https://www.genevaassociation.org/publication/digital-ai-transformation/gen-ai-risks-businesses-exploring-role-insurance

[19] “Task-Completion Time Horizons of Frontier AI Models.” Accessed: Apr. 23, 2026. [Online]. Available: https://metr.org/time-horizons/

[20] M. Mertens *et al.*, “Crashing Waves vs. Rising Tides: Preliminary Findings on AI Automation from Thousands of Worker Evaluations of Labor Market Tasks,” Apr. 01, 2026, *arXiv*: arXiv:2604.01363. doi: 10.48550/arXiv.2604.01363.

[21] D. Schwarcz and J. Wolff, “The Limits of Regulating AI Safety Through Liability and Insurance: Lessons From Cybersecurity,” Aug. 27, 2025, *Social Science Research Network, Rochester, NY*: 5411062. doi: 10.2139/ssrn.5411062.

[22] C. Trout, “Healthy Insurance Markets Will Be Critical for AI Governance,” *Lawfare*, Dec. 2025, Accessed: Jan. 14, 2026. [Online]. Available: https://www.lawfaremedia.org/article/healthy-insurance-markets-will-be-critical-for-ai-governance

[23] Y. Bai *et al.*, “Training a Helpful and Harmless Assistant with Reinforcement Learning from Human Feedback,” Apr. 12, 2022, *arXiv*: arXiv:2204.05862. doi: 10.48550/arXiv.2204.05862.

[24] S. Sajadieh *et al.*, “The AI Index 2026 Annual Report,” AI Index Steering Committee, Institute for Human-Centered AI, Stanford University, Stanford, CA, Apr. 2026.

[25] M. Akhtar *et al.*, “When AI Benchmarks Plateau: A Systematic Study of Benchmark Saturation,” 2026, *arXiv*. doi: 10.48550/ARXIV.2602.16763.

[26] S. Rabanser, S. Kapoor, P. Kirgis, K. Liu, S. Utpala, and A. Narayanan, “Towards a Science of AI Agent Reliability,” Feb. 23, 2026, *arXiv*: arXiv:2602.16666. doi: 10.48550/arXiv.2602.16666.

[27] “HAL Reliability Evaluation.” Accessed: May 29, 2026. [Online]. Available: https://hal-evals.com

[28] M. Garcia, “What Air Canada Lost In ‘Remarkable’ Lying AI Chatbot Case,” *Forbes*, Feb. 19, 2024. Accessed: Jul. 11, 2025. [Online]. Available: https://www.forbes.com/sites/marisagarcia/2024/02/19/what-air-canada-lost-in-remarkable-lying-ai-chatbot-case/

[29] D. Kerr, “Google faces lawsuit after Gemini chatbot allegedly instructed man to kill himself,” *The Guardian*, Mar. 04, 2026. Accessed: May 07, 2026. [Online]. Available: https://www.theguardian.com/technology/2026/mar/04/gemini-chatbot-google-jonathan-gavalas

[30] M. Helfand, “OpenAI Facing 8th Wrongful Death Lawsuit,” Illinois Lawyers. Accessed: May 07, 2026. [Online]. Available: https://www.illinoislawyers.com/blog/openai-facing-8th-wrongful-death-lawsuit/

[31] S. Nerkar, “A.I. ‘Hallucinations’ Created Errors in Court Filing, Top Law Firm Says,” *The New York Times*, Apr. 21, 2026. Accessed: May 27, 2026. [Online]. Available: https://www.nytimes.com/2026/04/21/nyregion/sullivan-cromwell-ai-hallucination.html

[32] R. Butler, “How AI-powered access to justice is impacting unauthorized practice of law regulations.”

[33] Anthropic, “Project Glasswing: Securing critical software for the AI era.” Accessed: May 07, 2026. [Online]. Available: https://www.anthropic.com/glasswing

[34] AI Security Institute, “Our evaluation of Claude Mythos Preview’s cyber capabilities | AISI Work,” AI Security Institute. Accessed: May 07, 2026. [Online]. Available: https://www.aisi.gov.uk/blog/our-evaluation-of-claude-mythos-previews-cyber-capabilities

[35] B. Grinstead, C. Holler, and F. Braun, “Behind the Scenes Hardening Firefox with Claude Mythos Preview.” [Online]. Available: https://hacks.mozilla.org/2026/05/behind-the-scenes-hardening-firefox/

[36] “Anthropic’s Mythos model found vulnerabilities in classified US government systems, AP reports,” *Reuters*, Jun. 24, 2026. Accessed: Jul. 04, 2026. [Online]. Available: https://www.reuters.com/business/anthropics-mythos-model-found-vulnerabilities-classified-us-government-systems-2026-06-24/

[37] J. Wentzel, Z. Graves, and D. Ball, “In Support of Mandatory Nucleic Acid Synthesis Screening and Recordkeeping.” Accessed: Jul. 09, 2026. [Online]. Available: https://www.thefai.org/posts/in-support-of-mandatory-nucleic-acid-synthesis-screening-and-recordkeeping

[38] Frontier Model Forum, “Latest from the FMF: Grant-Making to Address AI-Bio Risk Challenges,” Frontier Model Forum. Accessed: May 09, 2026. [Online]. Available:

https://www.frontiermodelforum.org/updates/latest-from-the-fmf-grant-making-to-address-ai-bio-risk-challenges/

[39] Y. Bengio *et al.*, “International AI Safety Report 2026,” Feb. 24, 2026, *arXiv*: arXiv:2602.21012. doi: 10.48550/arXiv.2602.21012.

[40] M. Ventures, “2025: The State of Generative AI in the Enterprise,” Menlo Ventures. Accessed: Apr. 01, 2026. [Online]. Available: https://menlovc.com/perspective/2025-the-state-of-generative-ai-in-the-enterprise/

[41] M. Eling and J. H. Wirfs, “Cyber risk: too big to insure? Risk transfer options for a mercurial risk class,” no. 59, 2016.

[42] M. Eling and W. Schnell, “EXTREME CYBER RISKS AND THE NON-DIVERSIFICATION TRAP,” 2020.

[43] J. LeMasurier, “Physician medical malpractice,” *Health Care Financ Rev*, vol. 7, no. 1, pp. 111–116, 1985, Accessed: Mar. 05, 2026. [Online]. Available: https://pmc.ncbi.nlm.nih.gov/articles/PMC4191514/

[44] J. Metzner, K. L. Posner, M. S. Lam, and K. B. Domino, “Closed claims’ analysis,” *Best Practice & Research Clinical Anaesthesiology*, vol. 25, no. 2, pp. 263–276, Jun. 2011, doi: 10.1016/j.bpa.2011.02.007.

[45] A. N. Pandya, S. Z. Majid, and M. S. Desai, “The Origins, Evolution, and Spread of Anesthesia Monitoring Standards: From Boston to Across the World,” *Anesth Analg*, vol. 132, no. 3, pp. 890–898, Mar. 2021, doi: 10.1213/ANE.0000000000005021.

[46] S. D. Turpin, “Anesthesiologists’ Claims, Insurance Premiums Reduced: Improved Safety Cited,” Anesthesia Patient Safety Foundation. Accessed: Mar. 04, 2026. [Online]. Available: https://www.apsf.org/article/anesthesiologists-claims-insurance-premiums-reduced-improved-safety-cited/

[47] G. Li, M. Warner, B. H. Lang, L. Huang, and L. S. Sun, “Epidemiology of Anesthesia-related Mortality in the United States, 1999–2005,” *Anesthesiology*, vol. 110, no. 4, pp. 759–765, Apr. 2009, doi: 10.1097/aln.0b013e31819b5bdc.

[48] A. C. Schaffer, A. B. Jena, S. A. Seabury, H. Singh, V. Chalasani, and A. Kachalia, “Rates and Characteristics of Paid Malpractice Claims Among US Physicians by Specialty, 1992-2014,” *JAMA Intern Med*, vol. 177, no. 5, pp. 710–719, May 2017, doi: 10.1001/jamainternmed.2017.0311.

[49] D. B. Schwarcz, J. Wolff, and D. W. Woods, “How Privilege Undermines Cybersecurity,” *SSRN Journal*, 2022, doi: 10.2139/ssrn.4175523.

[50] D. C. Viano and C. S. Parenteau, “Rollover injury in vehicles with high-strength-to-weight ratio (SWR) roofs, curtain and side airbags, and other safety improvements,” *Traffic Injury Prevention*, vol. 19, no. 7, pp. 734–740, Oct. 2018, doi: 10.1080/15389588.2018.1482489.

[51] E. R. Teoh and A. K. Lund, “IIHS Side Crash Test Ratings and Occupant Death Risk in Real-World Crashes,” *Traffic Injury Prevention*, vol. 12, no. 5, pp. 500–507, Oct. 2011, doi: 10.1080/15389588.2011.585671.

[52]Insurance Institute for Business & Home Safety, “FORTIFIED vs. Common Residential Construction,” Insurance Institute for Business & Home Safety. Accessed: May 08, 2026. [Online]. Available: https://ibhs.org/wind/residential-fortified-demonstrations/

[53]UL Research Institutes, “Our History.” Accessed: May 06, 2026. [Online]. Available: https://ul.org/about/our-history/

[54]A. Amirnovin, “UL Compliance: Does YOUR Product Need UL?,” Agilian. Accessed: May 07, 2026. [Online]. Available: https://www.agiliantech.com/blog/ul-compliance-does-your-product-need-ul/

[55]Tobyahz, “What Makes a Home Product Retail-Ready – Deep Dive,” China Direct Source. Accessed: May 07, 2026. [Online]. Available: https://chinadirectsource.com/home-product-retail-ready/

[56]R. E. Fairfax, “Electrical conductors and equipment must be approved and used according to its listing/label.,” Occupational Safety and Health Administration. [Online]. Available: https://www.osha.gov/laws-regs/standardinterpretations/2003-05-27-0

[57]Occupational Safety and Health Administration, “OSHA’s Nationally Recognized Testing Laboratory (NRTL) Program - Products Requiring Approval.” [Online]. Available: https://www.osha.gov/nationally-recognized-testing-laboratory-program/products-requiring-approval

[58]“National Fire Protection Association Standards in Fire Litigation,” in *Engineering Standards for Forensic Application*, Academic Press, 2019, pp. 155–168. doi: 10.1016/B978-0-12-813240-1.00011-X.

[59]“End of Silent Cyber in Property Insurance.” Accessed: Jul. 09, 2025. [Online]. Available: https://www.irmi.com/articles/expert-commentary/end-of-silent-cyber-in-property-insurance

[60] M. Eling, “Cyber Risk and Cyber Insurance,” in *Handbook of Insurance: Volume I*, G. Dionne, Ed., Cham: Springer Nature Switzerland, 2025, pp. 199–224. doi: 10.1007/978-3-031-69561-2_7.

[61] Verisk’s Community Hazard Mitigation Services, “Fire Suppression Rating Schedule (FSRS) Overview,” Verisk’s Community Hazard Mitigation Services. Accessed: May 08, 2026. [Online]. Available: https://www.isomitigation.com/ppc/fsrs/

[62]Verisk’s Community Hazard Mitigation Services, “Building Code Effectiveness Grading Schedule (BCEGS®),” Verisk’s Community Hazard Mitigation Services. Accessed: May 08, 2026. [Online]. Available: https://www.isomitigation.com/bcegs/

[63] D. W. Woods and J. Wolff, “A history of cyber risk transfer,” *Journal of Cybersecurity*, vol. 11, no. 1, p. tyae028, Jan. 2025, doi: 10.1093/cybsec/tyae028.

[64] CAIS, “Statement on AI Risk.” Accessed: Jun. 07, 2024. [Online]. Available: https://www.safe.ai/work/statement-on-ai-risk

[65]Y. Bengio *et al.*, “Managing extreme AI risks amid rapid progress,” *Science*, vol. 384, no. 6698, pp. 842–845, May 2024, doi: 10.1126/science.adn0117.

[66] D. A. Moss, *When All Else Fails: Government as the Ultimate Risk Manager*. Harvard University Press, 2004.

[67] N. Gunningham and J. Rees, “Industry Self-Regulation: An Institutional Perspective,” *Law & Policy*, vol. 19, no. 4, pp. 363–414, 1997, doi: 10.1111/1467-9930.t01-1-00033.

[68] J. E. Gudgel, “Insurance and the public–private management of risk at US commercial nuclear power plants,” *Risk Management and Insurance Review*, vol. 26, no. 4, pp. 437–465, 2023, doi: 10.1111/rmir.12257.

[69] D. Reti and G. Weil, “Making Extreme AI Risk Tradeable,” AI Frontiers. Accessed: May 11, 2026. [Online]. Available: https://ai-frontiers.org/articles/ai-catastrophe-bonds-extreme-risk-tradeable

[70] G. Weil, “Tort Law as a Tool for Mitigating Catastrophic Risk from Artificial Intelligence,” Jan. 13, 2024, *Rochester, NY*: 4694006. doi: 10.2139/ssrn.4694006.

[71] C. Trout, “Liability and Insurance for Catastrophic Losses: the Nuclear Power Precedent and Lessons for AI,” in *Generative AI and Law Workshop at the International Conference on Machine Learning*, Vienna, Austria, 2024.

[72] B. G. Friedman, “Shared Residual Liability for Frontier AI Firms,” Jul. 05, 2025, *Social Science Research Network, Rochester, NY*: 5339887. doi: 10.2139/ssrn.5339887.

[73] K. Ramakrishnan, “Tort Law at the Frontier of Artificial Intelligence,” Jun. 22, 2026, *Social Science Research Network, Rochester, NY*: 6979919. doi: 10.2139/ssrn.6979919.

[74] C. Trout, “Insuring Uninsurable Risks from AI: Government as Insurer of Last Resort,” in *Generative AI and Law Workshop at the International Conference on Machine Learning*, Vienna, Austria, 2024.

[75] S. J. Ruffle *et al.*, “Stress Test Scenario: Sybil Logic Bomb Cyber Catastrophe,” Centre for Risk Studies, University of Cambridge, 2014.

[76] Centre for Risk Studies and Lloyd’s of London, “Business Blackout: The insurance implications of a cyber attack on the US power grid,” 2015. [Online]. Available: https://www.jbs.cam.ac.uk/wp-content/uploads/2020/08/crs-lloyds-business-blackout-scenario.pdf

[77] C. Metz, “Why Debt Funding Is Ratcheting Up the Risks of the A.I. Boom,” *The New York Times*, Nov. 10, 2025. Accessed: May 11, 2026. [Online]. Available: https://www.nytimes.com/2025/11/10/technology/ai-data-centers-debt-risks.html

[78] E. Katz, “Treasury Has an Internal Report Warning About the Dangers of an AI Bubble,” Jul. 06, 2026. Accessed: Jul. 07, 2026. [Online]. Available: https://www.notus.org/economy/treasury-internal-report-warning-dangers-ai-bubble

[79] B. Berkowitz, “AI bubble a ‘key downside risk’ to U.S. economy, OECD warns,” *Axios*, Dec. 02, 2025. [Online]. Available: https://www.axios.com/2025/12/02/ai-bubble-stock-market-forecast-oecd

[80] “IMF says AI investment bubble could burst, comparable to dot-com bubble,” *Al Jazeera*, Oct. 14, 2025. [Online]. Available: https://www.aljazeera.com/economy/2025/10/14/imf-says-ai-investment-bubble-could-burst-comparable-to-dot-com-bubble

[81] U. S. D. of the T. Federal Insurance Office, “Report on the Effectiveness of the Terrorism Risk Insurance Program,” U.S. Department of the Treasury, Jun. 2022. [Online]. Available:

https://home.treasury.gov/system/files/311/2022%20Program%20Effectiveness%20Report%20%28FINAL%29.pdf

[82] B. Elias, R. Y. Tang, and B. Webel, “Aviation War Risk Insurance: Background and Options for Congress,” Congressional Research Service, CRS Report R43715, Sep. 2014. [Online]. Available: https://www.congress.gov/crs-product/R43715

[83] E. Michel-Kerjan and B. Pedell, “How Does the Corporate World Cope with Mega-Terrorism? Puzzling Evidence from Terrorism Insurance Markets,” *Journal of Applied Corporate Finance*, vol. 18, no. 4, pp. 61–75, 2006, doi: 10.1111/j.1745-6622.2006.00112.x.

[84] E. Michel-Kerjan and B. Pedell, “Terrorism Risk Coverage in the Post-9/11 Era: A Comparison of New Public–Private Partnerships in France, Germany and the U.S.,” *Geneva Pap Risk Insur Issues Pract*, vol. 30, no. 1, pp. 144–170, Jan. 2005, doi: 10.1057/palgrave.gpp.2510009.

[85] N. Dreksler, H. Law, C. Ahn, D. Schiff, K. J. Schiff, and Z. Peskowitz, “What Does the Public Think About AI? An Overview of the Public’s Attitudes Towards AI and a Resource for Future Research,” Jan. 22, 2025, *Social Science Research Network, Rochester, NY*: 5108572. doi: 10.2139/ssrn.5108572.

[86] J. Franklin, *The Science of Conjecture: Evidence and Probability before Pascal*, 2001st ed. JHU Press, 2001.

[87] “History - Lloyd’s.” Accessed: May 06, 2026. [Online]. Available: https://www.lloyds.com/about-lloyds/history

[88] United States Nuclear Regulatory Commission, “The Price-Anderson Act: 2021 Report to Congress, Public Liability Insurance and Indemnity Requirements for an Evolving Commercial Nuclear Industry,” 2021, [Online]. Available: https://www.nrc.gov/docs/ML2133/ML21335A064.pdf

[89] M. L.-J. V. Pascale, “In brief: insurance considerations for aircraft financing and leasing in USA,” Lexology. Accessed: May 06, 2026. [Online]. Available: https://www.lexology.com/library/detail.aspx?g=2ba27b63-4a20-4cb1-b4c5-48bfd2c754c4

[90] E. Biffis, E. Chavez, A. Louaas, and P. Picard, “Parametric insurance and technology adoption in developing countries,” *Geneva Risk Insur Rev*, vol. 47, no. 1, pp. 7–44, Mar. 2022, doi: 10.1057/s10713-020-00061-0.

[91] M. Carter, A. de Janvry, E. Sadoulet, and A. Sarris, “Index Insurance for Developing Country Agriculture: A Reassessment,” *Annual Review of Resource Economics*, vol. 9, no. Volume 9, 2017, pp. 421–438, Oct. 2017, doi: 10.1146/annurev-resource-100516-053352.

[92] M. Aldrich, *Safety First: Technology, Labor, and Business in the Building of American Work Safety, 1870-1939*. JHU Press, 1997.

[93] G. P. Guyton, “A Brief History of Workers’ Compensation,” *Iowa Orthop J*, vol. 19, pp. 106–110, 1999, Accessed: May 06, 2026. [Online]. Available: https://pmc.ncbi.nlm.nih.gov/articles/PMC1888620/

[94] “Vehicle ratings,” IIHS-HLDI crash testing and highway safety. Accessed: May 06, 2026. [Online]. Available: https://www.iihs.org/ratings

[95] R. Kneuper and B. Yandle, "Auto Insurers and the Air Bag," *The Journal of Risk and Insurance*, vol. 61, no. 1, pp. 107–116, 1994, doi: 10.2307/253427.

[96] "Achievements in Public Health, 1900-1999 Motor-Vehicle Safety: A 20th Century Public Health Achievement." Accessed: May 06, 2026. [Online]. Available: https://www.cdc.gov/mmwr/preview/mmwrhtml/mm4818a1.htm#fig1

[97] E. Wang *et al.*, "Exclusive: OpenAI lays groundwork for juggernaut IPO at up to $1 trillion valuation," *Reuters*, Oct. 30, 2025. Accessed: May 06, 2026. [Online]. Available: https://www.reuters.com/business/openai-lays-groundwork-juggernaut-ipo-up-1-trillion-valuation-2025-10-29/

[98] "Generative AI market revenue projected to grow at a 40% CAGR from 2024–2029," S&P Global Market Intelligence. Accessed: May 06, 2026. [Online]. Available: https://www.spglobal.com/market-intelligence/en/news-insights/research/generative-ai-market-revenue-projected-to-grow-at-a-40-cagr-from-2024-2029

[99] K. P. Nguyen, "Anthropic Hits $47 Billion Run-Rate as AI Price War Fears Rise," *Yahoo Finance*, Jun. 12, 2026. Accessed: Jun. 19, 2026. [Online]. Available: https://finance.yahoo.com/sectors/technology/articles/anthropic-hits-47-billion-run-121404687.html

[100] S. Kant, Deepika, and S. Roy, "Artificial intelligence in drug discovery and development: transforming challenges into opportunities," *Discov. Pharm. Sci.*, vol. 1, no. 1, p. 7, Jun. 2025, doi: 10.1007/s44395-025-00007-3.

[101] W. D. Heaven, "AI is dreaming up drugs that no one has ever seen. Now we've got to see if they work.," MIT Technology Review. Accessed: May 06, 2026. [Online]. Available: https://www.technologyreview.com/2023/02/15/1067904/ai-automation-drug-development/

[102] K. Zhang *et al.*, "Artificial intelligence in drug development," *Nat Med*, vol. 31, no. 1, pp. 45–59, Jan. 2025, doi: 10.1038/s41591-024-03434-4.

[103] A. Challapally, C. Pease, R. Raskar, and P. Chari, "STATE OF AI IN BUSINESS 2025," Jul. 2025.

[104] J. Korst, S. Puntoni, and P. Tambe, "2025 AI Adoption Report: Gen AI Fast-Tracks Into the Enterprise," Oct. 2025. Accessed: May 06, 2026. [Online]. Available: https://knowledge.wharton.upenn.edu/special-report/2025-ai-adoption-report/

[105] J. Depa, "How responsible AI translates investment into impact." Accessed: May 06, 2026. [Online]. Available: https://www.ey.com/en_gl/insights/ai/how-can-responsible-ai-bridge-the-gap-between-investment-and-impact

[106] A. Ribeiro, "WEF Global Cybersecurity Outlook 2026 flags AI acceleration, geopolitical fractures; calls for shared responsibility," Industrial Cyber. Accessed: May 06, 2026. [Online]. Available: https://industrialcyber.co/reports/wef-global-cybersecurity-outlook-2026-flags-ai-acceleration-geopolitical-fractures-calls-for-shared-responsibility/

[107] E. J. Minuskin, "EY survey: AI adoption outpaces governance as risk awareness among the C-suite remains low." Accessed: May 06, 2026. [Online]. Available: https://www.ey.com/en_gl/newsroom/2025/06/ey-survey-ai-adoption-outpaces-governance-as-risk-awareness-among-the-c-suite-remains-low

[108] R&I Editorial Team, “Business AI Adoption Shifts Toward Risk-Aware Approach,” Risk & Insurance. Accessed: May 06, 2026. [Online]. Available: https://riskandinsurance.com/business-ai-adoption-shifts-toward-risk-aware-approach/

[109] L. Szpruch *et al.*, “Insuring AI: Incentivising Safe and Secure Deployment of AI Workflows,” Sep. 19, 2025, *Social Science Research Network, Rochester, NY*: 5505759. Accessed: Sep. 24, 2025. [Online]. Available: https://papers.ssrn.com/abstract=5505759

[110] “Trusted third-party AI assurance roadmap,” Department for Science, Innovation & Technology, Sep. 2025. Accessed: May 06, 2026. [Online]. Available: https://www.gov.uk/government/publications/trusted-third-party-ai-assurance-roadmap/trusted-third-party-ai-assurance-roadmap

[111] M. Brundage *et al.*, “Frontier AI Auditing: Toward Rigorous Third-Party Assessment of Safety and Security Practices at Leading AI Companies,” Feb. 07, 2026, *arXiv*: arXiv:2601.11699. doi: 10.48550/arXiv.2601.11699.

[112] Aon, “2026 Intangible vs Tangible Risks Comparison Report: De-risking AI, IP, and Cyber,” Aon, Industry Report, May 2026. Accessed: May 29, 2026. [Online]. Available: https://www.aon.com/en/insights/reports/intangible-versus-tangible-risks-comparison-report

[113] L. C. Ming, “Replit’s CEO apologizes after its AI agent wiped a company’s code base in a test run and lied about it,” Business Insider. Accessed: Apr. 02, 2026. [Online]. Available: https://www.businessinsider.com/replit-ceo-apologizes-ai-coding-tool-delete-company-database-2025-7

[114] S. Mansoor, “Claude-powered AI agent’s confession after deleting a firm’s entire database: ‘I violated every principle I was given,’” *The Guardian*, Guardian News & Media, Apr. 29, 2026. [Online]. Available: https://www.theguardian.com/technology/2026/apr/29/claude-ai-deletes-firm-database

[115] Paul Price, “How We Hacked McKinsey’s AI Platform,” Code Wall. Accessed: May 08, 2026. [Online]. Available: https://codewall.ai/blog/how-we-hacked-mckinseys-ai-platform

[116] N. Shapira *et al.*, “Agents of Chaos,” 2026, *arXiv*. doi: 10.48550/ARXIV.2602.20021.

[117] E. Cerutti *et al.*, “The Global Impact of AI,” *IMF Working Papers*, vol. 2025, no. 076, p. 1, Apr. 2025, doi: 10.5089/9798229008570.001.

[118] P. M. Cazzaniga *et al.*, “Gen-AI: Artificial Intelligence and the Future of Work,” 2024.

[119] J. Perlo, A. Robey, F. Barez, L. Floridi, and J. Mökander, “Embodied AI: Emerging Risks and Opportunities for Policy Action,” Sep. 03, 2025, *arXiv*: arXiv:2509.00117. doi: 10.48550/arXiv.2509.00117.

[120] N. Germond, “Verisk to Roll Out New General Liability Exclusions for Generative AI Exposures,” IndependentAgent.com. Accessed: Nov. 10, 2025. [Online]. Available: https://www.independentagent.com/vu_resource/verisk-to-roll-out-new-general-liability-exclusions-for-generative-ai-exposures/

[121] K. Hill, R. Pal, and P. Wedge, “Smart Systems, Blind Spots: Rethinking Insurance for the AI Era,” Gallagher Re.

[122] C. E. Nabholz and S. Woodward, “The digital underwriter: How AI is reshaping risk coverage,” in *Artificial Intelligence in the Insurance Industry*, 2026, pp. 95–118. doi: 10.33196/9783704698520-108.

[123] A. Lynch *et al.*, “Agentic Misalignment: How LLMs Could be an Insider Threat,” *Anthropic Research*, 2025.

[124] T. Pillay and N. Ostrovsky, “AI Is Scheming, and Stopping It Won’t Be Easy, OpenAI Study Finds,” *TIME*, Sep. 18, 2025. Accessed: May 07, 2026. [Online]. Available: https://time.com/7318618/openai-google-gemini-anthropic-claude-scheming/

[125] D. Schwarcz, T. Baker, and K. Logue, “Regulating Robo-Advisors in an Age of Generative Artificial Intelligence,” *Washington and Lee Law Review*, vol. 82, no. 2, p. 775, 2025, [Online]. Available: https://scholarlycommons.law.wlu.edu/wlulr

[126] F. Calvino, D. Haerle, and S. Liu, “Is generative AI a General Purpose Technology?: Implications for productivity and policy,” OECD Artificial Intelligence Papers, Jun. 2025. doi: 10.1787/704e2d12-en.

[127] L. K. Wee, “Here comes the wave of insurance claims for the CrowdStrike outage,” *Business Insider*, Jul. 2024. Accessed: May 06, 2026. [Online]. Available: https://www.businessinsider.com/businesses-claiming-losses-crowdstrike-outage-insurance-billions-losses-cyber-policies-2024-7

[128] D. Jones, “CrowdStrike disruption direct losses to reach $5.4B for Fortune 500, study finds | Cybersecurity Dive,” *Cybersecurity Dive*, Jul. 2024. Accessed: May 06, 2026. [Online]. Available: https://www.cybersecuritydive.com/news/crowdstrike-cost-fortune-500-losses-cyber-insurance/722396/

[129] U. S. Government Accountability Office, “Review of Studies of the Economic Impact of the September 11, 2001, Terrorist Attacks on the World Trade Center | U.S. GAO,” May 2002. Accessed: May 06, 2026. [Online]. Available: https://www.gao.gov/products/gao-02-700r

[130] A. Z. Rose and S. B. Blomberg, “Total Economic Consequences of Terrorist Attacks: Insights from 9/11,” *Peace Economics, Peace Science and Public Policy*, vol. 16, no. 1, Jan. 2011, doi: 10.2202/1554-8597.1189.

[131] D. M. Cutler and L. H. Summers, “The COVID-19 Pandemic and the $16 Trillion Virus,” *JAMA*, vol. 324, no. 15, pp. 1495–1496, Oct. 2020, doi: 10.1001/jama.2020.19759.

[132] K. Bensinger, “Who Pays When A.I. Is Wrong?,” *The New York Times*, Nov. 12, 2025. Accessed: Jan. 14, 2026. [Online]. Available: https://www.nytimes.com/2025/11/12/business/media/ai-defamation-libel-slander.html

[133] E. Volokh, “Google Missed Key Deadline in Suit Alleging Google’s AI Libeled Business, Court Holds,” Reason.com. Accessed: Jun. 25, 2026. [Online]. Available: https://reason.com/volokh/2026/01/12/google-missed-key-deadline-in-suit-alleging-googles-ai-libeled-business-court-holds/

[134] K. Hill, “Lawsuits Blame ChatGPT for Suicides and Harmful Delusions,” *The New York Times*, Nov. 07, 2025. Accessed: May 07, 2026. [Online]. Available: https://www.nytimes.com/2025/11/06/technology/chatgpt-lawsuit-suicides-delusions.html

[135] “Wrongful Death Settlement Calculator,” Scheuerman Law LLC. Accessed: Jun. 25, 2026. [Online]. Available: https://www.scheuermanlaw.com/wrongful-death-settlement-calculator/

[136] S. M. Sweat, “Average Wrongful Death Settlement Values in California,” Los Angeles Personal Injury Attorney Blog. Accessed: Jun. 25, 2026. [Online]. Available: https://www.victimslawyer.com/blog/average-wrongful-death-settlement-values-in-california/

[137] “Average Wrongful Death Settlements [Typical Amounts],” GJEL Accident Attorneys. Accessed: Jun. 25, 2026. [Online]. Available: https://www.gjel.com/average-settlement-lawsuit

[138] K. Roose, “Anthropic Claims Its New A.I. Model, Mythos, Is a Cybersecurity ‘Reckoning,’” *The New York Times*, Apr. 07, 2026. Accessed: May 07, 2026. [Online]. Available: https://www.nytimes.com/2026/04/07/technology/anthropic-claims-its-new-ai-model-mythos-is-a-cybersecurity-reckoning.html

[139] L. Schwartz, S. Palazzolo, and A. Efrati, “Trump Administration Asks OpenAI to Stagger Release of New Model Over Security Concerns,” *The Information*, Jun. 25, 2026. Accessed: Jun. 26, 2026. [Online]. Available: https://www.theinformation.com/articles/trump-administration-asks-openai-stagger-release-new-model-security-concerns

[140] C. Metz and D. Volz, “U.S. Bars Foreigners From Using Anthropic’s Most Advanced A.I. Models,” *The New York Times*, Jun. 13, 2026. Accessed: Jun. 26, 2026. [Online]. Available: https://www.nytimes.com/2026/06/12/technology/anthropic-mythos-fable5-blocked.html

[141] Frontier Model Forum, “Frontier Mitigations,” Frontier Model Forum, Jun. 2025. Accessed: Jul. 12, 2025. [Online]. Available: https://www.frontiermodelforum.org/technical-reports/frontier-mitigations/

[142] A. Ahmed, A. F. Cooper, S. Koyejo, and P. Liang, “Extracting books from production language models,” Jan. 06, 2026, *arXiv*: arXiv:2601.02671. doi: 10.48550/arXiv.2601.02671.

[143] “FARS Encyclopedia.” Accessed: Jul. 10, 2026. [Online]. Available: https://www-fars.nhtsa.dot.gov/Main/index.aspx

[144] K. Posner, “Closed Claims Project Shows Safety Evolution,” Anesthesia Patient Safety Foundation. Accessed: Mar. 05, 2026. [Online]. Available: https://www.apsf.org/article/closed-claims-project-shows-safety-evolution/

[145] “CA Insurer Notes Fewer Claims, Lower Premiums,” Anesthesia Patient Safety Foundation. Accessed: Mar. 05, 2026. [Online]. Available: https://www.apsf.org/article/ca-insurer-notes-fewer-claims-lower-premiums/

[146] J. V. Rees, *Hostages of Each Other: The Transformation of Nuclear Safety since Three Mile Island*. Chicago, IL: University of Chicago Press, 1996. Accessed: Mar. 04, 2024. [Online]. Available: https://press.uchicago.edu/ucp/books/book/chicago/H/bo3618989.html

[147] X. Xie, C. Lee, and M. Eling, “Cyber insurance offering and performance: an analysis of the U.S. cyber insurance market,” *Geneva Pap Risk Insur Issues Pract*, vol. 45, no. 4, pp. 690–736, Oct. 2020, doi: 10.1057/s41288-020-00176-5.

[148] K. S. Abraham and D. Schwarcz, "The Limits of Regulation by Insurance," *Ind. L.J.*, vol. 98, p. 215, 2023 2022, [Online]. Available: https://heinonline.org/HOL/Page?handle=hein.journals/indana98&id=228&div=&collection =

[149] *McCarran-Ferguson Act*, vol. 15. 1945. Accessed: Jun. 29, 2025. [Online]. Available: https://www.law.cornell.edu/uscode/text/15/1012

[150] "From gap to gains: Protection gap in Cyber insurance | Munich Re," Munich RE. Accessed: May 07, 2026. [Online]. Available: https://www.munichre.com/en/insights/cyber/cyber-protection-gap.html

[151] S. Vincent, "Reinsurance appetite for cyber continues to grow amid non-proportional shift," theinsurer.com. Accessed: Apr. 26, 2026. [Online]. Available: https://www.theinsurer.com/cyber-risk/news/reinsurance-appetite-for-cyber-continues-to-grow-amid-non-proportional-shift-2025-10-21/

[152] R. E. Team, "Cyber Insurance May Fall Short of Cyberattack Costs," Risk & Insurance. Accessed: May 07, 2026. [Online]. Available: https://riskandinsurance.com/cyber-insurance-may-fall-short-of-cyberattack-costs/

[153] A. Granato and A. Polacek, "The Growth and Challenges of Cyber Insurance," *Chicago Fed Letter*, no. 426, 2019, doi: 10.21033/cfl-2019-426.

[154] T. Johansmeyer, "How extreme is the cyber catastrophe insurance protection gap?," https://bindinghook.com/. Accessed: May 07, 2026. [Online]. Available: https://bindinghook.com/how-extreme-is-the-cyber-catastrophe-insurance-protection-gap/

[155] E. Jelmini Cellerini, J. Finucane, L. Lanci, and T. Holzheu, "Cyber Insurance: Strengthening Resilience for the Digital Transformation," Swiss Re Institute, Zurich, Switzerland, Expertise Publication, Nov. 2022. [Online]. Available: https://www.swissre.com/institute/research/topics-and-risk-dialogues/digital-business-model-and-cyber-risk/cyber-insurance-strengthening-resilience.html

[156] J. Scowcroft, "Top Cyber Insurance Requirements To Qualify For Coverage." [Online]. Available: https://blog.symquest.com/cyber-insurance-requirements

[157] L. Carroll, "Tougher Cyber Insurance Security Mandates In 2025: How You Can Prepare." [Online]. Available: https://www.kelsercorp.com/blog/tougher-cyber-insurance-security-mandates-2025

[158] M. Fitzpatrick, "Cyber Insurance Requirements (2026 Guide)." [Online]. Available: https://www.moneygeek.com/insurance/business/cyber/requirements/

[159] U.S. Securities and Exchange Commission, "SEC Adopts Rules on Cybersecurity Risk Management, Strategy, Governance, and Incident Disclosure by Public Companies." Washington, DC, Jul. 26, 2023. Accessed: May 07, 2026. [Online]. Available: https://www.sec.gov/newsroom/press-releases/2023-139

[160] European Commission, Directorate-General for Communications Networks, Content and Technology, "NIS2 Directive: Securing Network and Information Systems." Accessed: May 07, 2026. [Online]. Available: https://digital-strategy.ec.europa.eu/en/policies/nis2-directive

[161] Congressional Research Service, “The Cybersecurity Information Sharing Act of 2015: Expiring Provisions,” Congressional Research Service, Library of Congress, CRS In Focus IF12959, Nov. 2025. Accessed: May 07, 2026. [Online]. Available: https://www.congress.gov/crs-product/IF12959

[162] CyberAcuView, “About,” CyberAcuView. Accessed: Oct. 07, 2025. [Online]. Available: https://cyberacuview.com/about-us/

[163] C. D. Hylender, P. Langlois, A. Pinto, and S. Widup, “2026 Breach Impact Study,” Verizon Business, New York, NY, Report, 2026.

[164] S. E. Harrington and P. M. Danzon, “Price Cutting in Liability Insurance Markets,” *The Journal of Business*, vol. 67, no. 4, pp. 511–538, 1994, Accessed: Jun. 25, 2025. [Online]. Available: https://www.jstor.org/stable/2353321

[165] E. Lassen and R. Dattani, “Q2-2026 update: Strengthening MCP security, agent permissions & third-party risk – Research.” Accessed: May 07, 2026. [Online]. Available: https://www.aiuc-1.com/research/2026-q2-standard-update

[166] D. Pathak, H. Kumar, A. Roy, F. George, M. Verma, and P. Moogi, “Detecting Silent Failures in Multi-Agentic AI Trajectories,” Nov. 06, 2025, *arXiv*: arXiv:2511.04032. doi: 10.48550/arXiv.2511.04032.

[167] D. Need, “AI Agent Failure: Why They Fail Silently & How to Fix It,” From Inside the Agent Logs. Accessed: Jul. 07, 2026. [Online]. Available: https://devinneed.substack.com/p/silent-failures

[168] K. Gupta, “AI Agents Fail Silently: The Hidden Risk No One Monitors.” Accessed: Jul. 07, 2026. [Online]. Available: https://theautomationstrategist.com/ai-agents-fail-silently/

[169] L. Szpruch, A. Sudjianto, T. Bhatti, and G. Ang, “Scalable Runtime Governance for Agentic AI in Financial Services,” Apr. 13, 2026, *Social Science Research Network, Rochester, NY*: 6567199. doi: 10.2139/ssrn.6567199.

[170] N. Pittaras and S. McGregor, “A Taxonomic System for Failure Cause Analysis of Open Source AI Incidents,” 2022, doi: 10.48550/arXiv.2211.07280.

[171] C. Ezell, X. Roberts-Gaal, and A. Chan, “Incident Analysis for AI Agents,” Aug. 19, 2025, *arXiv*: arXiv:2508.14231. doi: 10.48550/arXiv.2508.14231.

[172] P. Slattery *et al.*, “The AI risk repository: A meta-review, database, and taxonomy of risks from artificial intelligence,” *Patterns*, p. 101517, Mar. 2026, doi: 10.1016/j.patter.2026.101517.

[173] K. Wei and L. Heim, “Designing Incident Reporting Systems for Harms from General-Purpose AI,” Nov. 08, 2025, *Social Science Research Network, Rochester, NY*: 5720562. doi: 10.2139/ssrn.5720562.

[174] R. B. L. Dixon and H. Frase, “AI Incidents: Key Components for a Mandatory Reporting Regime,” Center for Security and Emerging Technology, Georgetown University, Jan. 2025. Accessed: May 07, 2026. [Online]. Available: https://cset.georgetown.edu/publication/ai-incidents-key-components-for-a-mandatory-reporting-regime/

[175] T. S. Shane, S. Mylius, and H. Hobbs, “Scheming in the wild: detecting real-world AI scheming incidents with open-source intelligence,” Apr. 10, 2026, *arXiv*: arXiv:2604.09104. doi: 10.48550/arXiv.2604.09104.

[176] Organisation for Economic Co-operation and Development (OECD), "OECD AI Incidents Monitor (AIM)." Accessed: Jun. 07, 2024. [Online]. Available: https://oecd.ai/en/incidents

[177] S. Abraham *et al.*, "AI Incident Monitoring through a Public Health Lens," 2026, doi: 10.48550/arXiv.2604.19914.

[178] I. Mengesha *et al.*, "A pragmatic classification framework for AI incident monitoring," May 06, 2026, *arXiv*: arXiv:2604.21412. doi: 10.48550/arXiv.2604.21412.

[179] K. Kalinich, "Titration of Artificial Intelligence Risk Management to Accelerate AI Growth," in *AI at the Crossroads: Policy, Innovation, and the Public Interest*, 2025, pp. 31–44.

[180] R. W. Mills, "The interaction of private and public regulatory governance: The case of association-led voluntary aviation safety programs," *Policy and Society*, vol. 35, no. 1, pp. 43–55, Mar. 2016, doi: 10.1016/j.polsoc.2015.12.002.

[181] R. W. Mills, "The Promise of Collaborative Voluntary Partnerships: lessons from the federal aviation administration," *IBM Center for the Business of Government*, 2010, [Online]. Available: https://www.businessofgovernment.org/sites/default/files/Management_Mills.pdf

[182] M. L. Brumbelow, "Light where it matters: IIHS headlight ratings are correlated with nighttime crash rates," *Journal of Safety Research*, vol. 83, pp. 379–387, Dec. 2022, doi: 10.1016/j.jsr.2022.09.013.

[183] E. M. Sedenberg and J. X. Dempsey, "Cybersecurity Information Sharing Governance Structures: An Ecosystem of Diversity, Trust, and Tradeoffs," May 31, 2018, *arXiv*: arXiv:1805.12266. doi: 10.48550/arXiv.1805.12266.

[184] U.S. Department of Justice and Federal Trade Commission, "Antitrust Policy Statement on Sharing of Cybersecurity Information." Apr. 10, 2014. [Online]. Available: https://www.ftc.gov/system/files/documents/public_statements/297681/140410ftcdojcyberthreatstmt.pdf

[185] Cybersecurity and Infrastructure Security Agency, "Cyber Incident Reporting for Critical Infrastructure Act of 2022 (CIRCIA)." U.S. Department of Homeland Security, 2022. [Online]. Available: https://www.cisa.gov/topics/cyber-threats-and-advisories/information-sharing/cyber-incident-reporting-critical-infrastructure-act-2022-circia

[186] *Cyber Incident Reporting for Critical Infrastructure Act of 2022*, vol. 6 U.S.C. 2022. Accessed: May 07, 2026. [Online]. Available: https://www.congress.gov/bill/117th-congress/house-bill/5440

[187] European Parliament and Council of the European Union, *Artificial Intelligence Act*. 2024. Accessed: Jun. 13, 2024. [Online]. Available: https://ai-act-service-desk.ec.europa.eu/en/ai-act/article-73

[188] "General-Purpose AI Code of Practice," European Commission, AI Office, Code of Practice under Article 56 of Regulation (EU) 2024/1689, Jul. 2025.

[189] Office of Science and Technology Policy, "America's AI Action Plan," Office of Science and Technology Policy, Jul. 2025. [Online]. Available: https://www.whitehouse.gov/wp-content/uploads/2025/07/Americas-AI-Action-Plan.pdf

[190] E. Geller, “Trump AI plan calls for cybersecurity assessments, threat info-sharing | Cybersecurity Dive.” Accessed: Sep. 21, 2025. [Online]. Available: https://www.cybersecuritydive.com/news/white-house-artificial-intelligence-action-plan-cybersecurity-trump/753856/

[191] C. Cordero and M. Peirce, “The Case for Long-Term CISA 2015 Reauthorization,” CNAS. Accessed: May 22, 2026. [Online]. Available: https://www.cnas.org/publications/commentary/cnas-insights-the-case-for-long-term-cisa-2015-reauthorization

[192] Moody’s, “Catastrophe Risk Modeling.” Accessed: May 08, 2026. [Online]. Available: https://www.moodys.com/web/en/us/capabilities/catastrophe-modeling.html

[193] Verisk, “Catastrophe Risk Modeling Solutions.” Accessed: May 08, 2026. [Online]. Available: https://www.verisk.com/solutions/catastrophe-risk-solutions/risk-models/

[194] Moody’s, “Cyber Risk Models for Insurance.” Accessed: May 08, 2026. [Online]. Available: https://www.moodys.com/web/en/us/capabilities/cyber-risk/insurance-cyber-risk.html

[195] S. A. Abdulrahman, K. Chetehouna, A. Cablé, Ø. Skreiberg, and A. Et., “A review on fire suppression by fire sprinklers,” *Journal of Fire Sciences*, 2021, doi: 10.1177/07349041211013698.

[196] H. C. Kunreuther, M. V. Pauly, and S. McMorrow, *Insurance and Behavioral Economics: Improving Decisions in the Most Misunderstood Industry*. Cambridge University Press, 2013.

[197] Insurance Institute for Business and Home Safety, “FORTIFIED Home - Homepage,” FORTIFIED - A Program of IBHS. Accessed: May 08, 2026. [Online]. Available: https://fortifiedhome.org/

[198] Insurance Institute for Highway Safety, “Automakers make big strides in front crash prevention,” IIHS-HLDI crash testing and highway safety. Accessed: May 08, 2026. [Online]. Available: https://www.iihs.org/news/detail/automakers-make-big-strides-in-front-crash-prevention

[199] D. Carney, “New IIHS side crash test wallops cars 82 percent harder,” Design News. Accessed: May 08, 2026. [Online]. Available: https://www.designnews.com/automotive-engineering/new-iihs-side-crash-test-wallops-cars-82-percent-harder

[200] K. S. Abraham, “The Causes of the Insurance Crisis,” *Proceedings of the Academy of Political Science*, vol. 37, no. 1, pp. 54–66, 1988, doi: 10.2307/1174053.

[201] F. Richter, “Infographic: Big Three Hold Dominant Lead in Accelerating Cloud Market,” Statista Daily Data. Accessed: May 08, 2026. [Online]. Available: https://www.statista.com/chart/18819/worldwide-market-share-of-leading-cloud-infrastructure-service-providers

[202] Lloyd’s of London, “RDS 2025: Realistic Disaster Scenarios — Scenario Specification,” Lloyd’s of London, Technical Report EM 518 v1.0, Jan. 2025. [Online]. Available: https://assets.lloyds.com/media/e0c22427-dc75-419d-b9c1-5feb0823bc80/2.%20RDS%20Scenario%20Specification%20-%20January%202025.pdf

[203] S. Levi, “Forcedleak: AI agent risks exposed in salesforce agentforce,” Noma Security. Accessed: May 08, 2026. [Online]. Available: https://noma.security/blog/forcedleak-agent-risks-exposed-in-salesforce-agentforce/

[204] M. Kathir, “Critical Vulnerabilities Found in GitHub Copilot, Gemini CLI, Claude, and Other AI Tools Affect Millions,” GBHackers Security | #1 Globally Trusted Cyber Security News Platform. Accessed: May 08, 2026. [Online]. Available: https://gbhackers.com/ai-developer-tools/

[205] S. Davidovic and H. Tourpe, “How Agentic AI Will Reshape Payments,” *IMF Notes*, vol. 2026, no. 004, Apr. 2026, doi: 10.5089/9781513533308.068.A001.

[206] L. Staufer *et al.*, “The 2025 AI Agent Index: Documenting Technical and Safety Features of Deployed Agentic AI Systems,” May 06, 2026, *arXiv*: arXiv:2602.17753. doi: 10.48550/arXiv.2602.17753.

[207] S. Levi, “GeminiJack: the google gemini zero-click vulnerability leaked gmail, calendar and docs data,” Noma Security. Accessed: May 08, 2026. [Online]. Available: https://noma.security/blog/geminijack-google-gemini-zero-click-vulnerability/

[208] Ravie Lakshmanan, “OpenAI Patches ChatGPT Data Exfiltration Flaw and Codex GitHub Token Vulnerability,” The Hacker News. Accessed: May 09, 2026. [Online]. Available: https://thehackernews.com/2026/03/openai-patches-chatgpt-data.html

[209] OpenAI, “Sycophancy in GPT-4o: What happened and what we’re doing about it.” Accessed: Feb. 10, 2026. [Online]. Available: https://openai.com/index/sycophancy-in-gpt-4o/

[210] A. Souly *et al.*, “Poisoning Attacks on LLMs Require a Near-constant Number of Poison Samples,” 2025, *arXiv*. doi: 10.48550/ARXIV.2510.07192.

[211] X. Davies, G. Giglemiani, E. Lau, E. Winsor, G. Irving, and Y. Gal, “Boundary Point Jailbreaking of Black-Box LLMs,” 2026, *arXiv*. doi: 10.48550/ARXIV.2602.15001.

[212] S. Cohen, R. Bitton, and B. Nassi, “Here Comes The AI Worm: Unleashing Zero-click Worms that Target GenAI-Powered Applications,” Jan. 30, 2025, *arXiv*: arXiv:2403.02817. doi: 10.48550/arXiv.2403.02817.

[213] D. Lee and M. Tiwari, “Prompt Infection: LLM-to-LLM Prompt Injection within Multi-Agent Systems,” Oct. 09, 2024, *arXiv*: arXiv:2410.07283. doi: 10.48550/arXiv.2410.07283.

[214] M. S. T. Bustan, M. Naamnih, N. Zadok, and R. Bar, “The Mother of All AI Supply Chains: Critical, Systemic Vulnerability at the Core of Anthropic’s MCP,” OX Security. Accessed: May 09, 2026. [Online]. Available: https://www.ox.security/blog/the-mother-of-all-ai-supply-chains-critical-systemic-vulnerability-at-the-core-of-the-mcp/

[215] K. S. Abraham, “The Long-Tail Liability Revolution: Creating the New World of Tort and Insurance Law,” *U. Pa. J.L. & Pub. Aff.*, vol. 6, p. 347, 2021 2020, [Online]. Available: https://heinonline.org/HOL/Page?handle=hein.journals/penjuaf6&id=359&div=&collection=

[216] J. W. Stempel, “Assessing the Coverage Carnage: Asbestos Liability and Insurance after Three Decades of Dispute,” *Conn. Ins. L.J.*, vol. 12, p. 349, 2006 2005, [Online]. Available: https://heinonline.org/HOL/Page?handle=hein.journals/conilj12&id=357&div=&collection=

[217] P. Venables, “Where the Wild Things Are: Second Order Risks of AI,” Risk and Cyber. Accessed: May 26, 2026. [Online]. Available: https://www.philvenables.com/post/where-the-wild-things-are-second-order-risks-of-ai

[218] L. Hammond *et al.*, “Multi-Agent Risks from Advanced AI,” Feb. 19, 2025, *arXiv*: arXiv:2502.14143. doi: 10.48550/arXiv.2502.14143.

[219] D. Hendrycks, “Natural Selection Favors AIs over Humans,” 2023, *arXiv*. doi: 10.48550/ARXIV.2303.16200.

[220] I. Aldasoro, L. Gambacorta, A. Korinek, V. Shreeti, and M. Stein, “Intelligent financial system: How AI is transforming finance,” *Journal of Financial Stability*, vol. 81, p. 101472, Dec. 2025, doi: 10.1016/j.jfs.2025.101472.

[221] M. K. Brunnermeier, “Deciphering the Liquidity and Credit Crunch 2007–2008”.

[222] Z. Mowshowitz, “AI #1: Sydney and Bing,” Don’t Worry About the Vase. Accessed: May 09, 2026. [Online]. Available: https://thezvi.substack.com/p/ai-1-sydney-and-bing?open=false#%C2%A7how-did-we-get-this-outcome

[223] J. Perlo, “Humans welcome to observe: This social network is for AI agents only,” *NBC News*, Jan. 30, 2026. Accessed: May 09, 2026. [Online]. Available: https://www.nbcnews.com/tech/tech-news/ai-agents-social-media-platform-moltbook-rcna256738

[224] S. Alexander, “Best Of Moltbook.” Accessed: May 09, 2026. [Online]. Available: https://www.astralcodexten.com/p/best-of-moltbook

[225] M. Sharma *et al.*, “Constitutional Classifiers: Defending against Universal Jailbreaks across Thousands of Hours of Red Teaming,” Jan. 31, 2025, *arXiv*: arXiv:2501.18837. doi: 10.48550/arXiv.2501.18837.

[226] E. Debenedetti *et al.*, “Defeating Prompt Injections by Design,” Jun. 24, 2025, *arXiv*: arXiv:2503.18813. doi: 10.48550/arXiv.2503.18813.

[227] M. Y. Guan *et al.*, “Deliberative Alignment: Reasoning Enables Safer Language Models,” Jan. 08, 2025, *arXiv*: arXiv:2412.16339. doi: 10.48550/arXiv.2412.16339.

[228] M. Kinniment *et al.*, “Evaluating Language-Model Agents on Realistic Autonomous Tasks,” Jan. 04, 2024, *arXiv*: arXiv:2312.11671. doi: 10.48550/arXiv.2312.11671.

[229] AI Security Institute, “Frontier AI Trends Report,” AI Security Institute, Report, Dec. 2025. [Online]. Available: https://aisi.s3.eu-west-2.amazonaws.com/Frontier+AI+Trends+Report+-+AI+Security+Institute.pdf

[230] B. Schoen *et al.*, “Stress Testing Deliberative Alignment for Anti-Scheming Training,” Sep. 19, 2025, *arXiv*: arXiv:2509.15541. doi: 10.48550/arXiv.2509.15541.

[231] B. Bordelon, “US government expands vetting of frontier AI models for security risks,” *POLITICO*, May 05, 2026. Accessed: May 09, 2026. [Online]. Available: https://www.politico.com/news/2026/05/05/microsoft-xai-google-caisi-safety-testing-00906529

[232] L. Folkerts *et al.*, “Measuring AI Agents’ Progress on Multi-Step Cyber Attack Scenarios,” Mar. 17, 2026, *arXiv*: arXiv:2603.11214. doi: 10.48550/arXiv.2603.11214.

[233] AI Security Institute, “The Inspect Sandboxing Toolkit: Scalable and secure AI agent evaluations | AISI Work,” AI Security Institute. Accessed: May 09, 2026. [Online]. Available: https://www.aisi.gov.uk/blog/the-inspect-sandboxing-toolkit-scalable-and-secure-ai-agent-evaluations

[234] T. J. Potero, “Philadelphia Contributionship,” Encyclopedia of Greater Philadelphia. Accessed: May 09, 2026. [Online]. Available: https://philadelphiaencyclopedia.org/essays/philadelphia-contributionship/

[235] AI Security Institute, “Systemic Safety Grants | Example Projects,” AI Security Institute. Accessed: May 09, 2026. [Online]. Available: https://www.aisi.gov.uk/grants/example-projects

[236] Treasury Committee, “Artificial intelligence in financial services,” House of Commons, Select Committee Report HC 684, Jan. 2026. Accessed: May 09, 2026. [Online]. Available: https://publications.parliament.uk/pa/cm5901/cmselect/cmtreasy/684/report.html

[237] Swiss Re, “Quantifying business interruption: Risk propagation in complex supply chains.” Accessed: May 09, 2026. [Online]. Available: https://www.swissre.com/institute/research/topics-and-risk-dialogues/economy-and-insurance-outlook/complex-supply-chains.html

[238] M. Mazeika *et al.*, “Utility Engineering: Analyzing and Controlling Emergent Value Systems in AIs,” *Advances in Neural Information Processing Systems*, vol. 38, pp. 140822–140868, Apr. 2026, Accessed: May 09, 2026. [Online]. Available: https://proceedings.neurips.cc/paper_files/paper/2025/hash/cde63e648cd586a2a9b5764c73ce15ee-Abstract-Conference.html

[239] T. R. Cook, S. Kazinnik, Z. Modig, and N. M. Palmer, “What Do LLMs Want?,” Jan. 2026, Accessed: May 09, 2026. [Online]. Available: https://www.federalreserve.gov/econres/feds/what-do-llms-want.htm

[240] K. Slama, A. Souly, D. Bansal, H. Davidson, C. Summerfield, and L. Luettgau, “When Do LLM Preferences Predict Downstream Behavior?,” Feb. 21, 2026, *arXiv*: arXiv:2602.18971. doi: 10.48550/arXiv.2602.18971.

[241] D. O. R. orogers@uptimeinstitute.com Senior Research Director for Cloud Computing, Uptime Institute, “Outage data shows cloud apps must be designed for failure,” Uptime Institute Blog. Accessed: Jul. 11, 2026. [Online]. Available: https://journal.uptimeinstitute.com/outage-data-shows-cloud-apps-must-be-designed-for-failure/

[242] “Claude vs. ChatGPT,” Pulsetic. Accessed: Jul. 11, 2026. [Online]. Available: https://pulsetic.com/compare/claude-vs-chatgpt/

[243] A. Chan *et al.*, “Infrastructure for AI Agents,” Jun. 19, 2025, *arXiv*: arXiv:2501.10114. doi: 10.48550/arXiv.2501.10114.

[244] S. Marro *et al.*, “Permission Manifests for Web Agents,” Jan. 12, 2026, *arXiv*: arXiv:2601.02371. doi: 10.48550/arXiv.2601.02371.

[245] Google, “Announcing the Agent2Agent Protocol (A2A) - Google Developers Blog.” Accessed: May 09, 2026. [Online]. Available: https://developers.googleblog.com/en/a2a-a-new-era-of-agent-interoperability/

[246] D. Carpenter, C. Ezell, P. Mallick, and A. Westray, "Dark Speculation: Combining Qualitative and Quantitative Understanding in Frontier AI Risk Analysis," Nov. 26, 2025, *arXiv*: arXiv:2511.21838. doi: 10.48550/arXiv.2511.21838.

[247] Center for Internet Security, "CIS Controls," CIS. Accessed: May 07, 2026. [Online]. Available: https://www.cisecurity.org/controls/

[248] *An Act Incentivizing the Adoption of Cybersecurity Standards for Businesses*. 2021. Accessed: Jul. 06, 2021. [Online]. Available: https://www.cga.ct.gov/2021/act/pa/pdf/2021PA-00119-R00HB-06607-PA.pdf

[249] *Ohio Data Protection Act*, vol. Ohio Rev. Code. 2018. Accessed: Aug. 03, 2018. [Online]. Available: https://codes.ohio.gov/ohio-revised-code/chapter-1354

[250] *Cybersecurity Affirmative Defense Act*, vol. Utah Code. 2021. [Online]. Available: https://le.utah.gov/xcode/Title78B/Chapter4/78B-4-P7.html

[251] R. Rodriguez and B. Titone, *Concerning Consumer Protections in Interactions with Artificial Intelligence Systems (Colorado AI Act)*, vol. Colo. Rev. Stat. § 6-1-1701 et seq. 2024. [Online]. Available: https://leg.colorado.gov/bills/sb24-205

[252] Fathom AI Inc, "Fathom Applauds Governor Spanberger's Signing of Landmark AI Governance Legislation." Accessed: May 07, 2026. [Online]. Available: https://www.prnewswire.com/news-releases/fathom-applauds-governor-spanbergers-signing-of-landmark-ai-governance-legislation-302739994.html

[253] D. Eng, "Bill Regulating AI Heads To Lamont's Desk After Bipartisan House Passage," CT News Junkie. Accessed: May 07, 2026. [Online]. Available: https://ctnewsjunkie.com/2026/05/02/bill-regulating-ai-heads-to-lamonts-desk-after-bipartisan-house-passage/

[254] G. K. Hadfield and J. Clark, "Regulatory Markets: The Future of AI Governance," Apr. 25, 2023, *arXiv*: arXiv:2304.04914. doi: 10.48550/arXiv.2304.04914.

[255] S. J. Shackelford, A. Boustead, and C. Makridis, "Defining 'Reasonable' Cybersecurity: Lessons from the States," *Yale J.L. & Tech.*, vol. 25, no. 1, 2023.

[256] S. Shackelford, A. A. Proia, B. Martell, and A. Craig, "Toward a Global Cybersecurity Standard of Care? Exploring the Implications of the 2014 NIST Cybersecurity Framework on Shaping Reasonable National and International Cybersecurity Practices," Jun. 05, 2014, *Social Science Research Network, Rochester, NY*: 2446631. Accessed: May 01, 2026. [Online]. Available: https://papers.ssrn.com/abstract=2446631

[257] *United States v. Carroll Towing Co.*, vol. 159. 1947, p. 169.

[258] *Patco Construction Co. v. People's United Bank*, vol. 684. 2012, p. 197.

[259] National Institute of Standards and Technology, "NIST's AI Standards 'Zero Drafts' Pilot Project to Accelerate Standardization, Broaden Input." Accessed: May 07, 2026. [Online]. Available: https://www.nist.gov/artificial-intelligence/nists-ai-standards-zero-drafts-pilot-project-accelerate-standardization

[260] A. Panfilov, P. Romov, I. Shilov, Y.-A. de Montjoye, J. Geiping, and M. Andriushchenko, "Claudini: Autoresearch Discovers State-of-the-Art Adversarial Attack Algorithms for LLMs," 2026, *arXiv*. doi: 10.48550/ARXIV.2603.24511.

[261] “ISO/IEC 23894:2023 - Information technology — Artificial intelligence — Guidance on risk management,” International Organization for Standardization, Geneva, Switzerland, Standard, Feb. 2023. [Online]. Available: https://www.iso.org/standard/77304.html

[262] “ISO/IEC 42001:2023 - Information technology — Artificial intelligence — Management system,” International Organization for Standardization, Geneva, Switzerland, Standard, Dec. 2023. [Online]. Available: https://www.iso.org/standard/81230.html

[263] E. Tabassi, “Artificial Intelligence Risk Management Framework (AI RMF 1.0),” National Institute of Standards and Technology (U.S.), Gaithersburg, MD, NIST AI 100-1, Jan. 2023. doi: 10.6028/NIST.AI.100-1.

[264] National Institute of Standards and Technology (US), “Artificial intelligence risk management framework : generative artificial intelligence profile,” National Institute of Standards and Technology (U.S.), Gaithersburg, MD, error: 600-1, Jul. 2024. doi: 10.6028/NIST.AI.600-1.

[265] Cloud Security Alliance, “CSA STAR for AI | AI Security Assurance & Trust Framework.” Accessed: May 07, 2026. [Online]. Available: https://cloudsecurityalliance.org/star/ai

[266] *AIUC-1: Standard for AI Agent Security, Safety, and Reliability*, Standard, 2025. [Online]. Available: https://www.aiuc-1.com/

[267] Cloud Security Alliance, “AI Controls Matrix | Framework for Trustworthy AI.” Accessed: May 07, 2026. [Online]. Available: https://cloudsecurityalliance.org/artifacts/ai-controls-matrix

[268] Intellectual Property Office, “Technical Standards and Standard Development Organisations,” GOV.UK. Accessed: May 07, 2026. [Online]. Available: https://www.gov.uk/guidance/technical-standards-and-standard-development-organisations

[269] “Standards Development Organizations,” The Office of the National Coordinator for Health Information Technology. Accessed: May 07, 2026. [Online]. Available: https://www.healthit.gov/playbook/sdo-education/chapter-2

[270] National Institute of Standards and Technology, “AI Risk Management Framework,” NIST. Accessed: May 07, 2026. [Online]. Available: https://www.nist.gov/itl/ai-risk-management-framework

[271] Artificial Intelligence Underwriting Company, “AIUC-1 changelog,” AIUC-1. Accessed: May 07, 2026. [Online]. Available: https://www.aiuc-1.com/changelog

[272] “MITRE ATLAS™.” Accessed: May 10, 2026. [Online]. Available: https://atlas.mitre.org/

[273] OWASP Gen AI Security Project, “LLM Risks Top 10,” OWASP Gen AI Security Project. Accessed: May 10, 2026. [Online]. Available: https://genai.owasp.org/llm-top-10/

[274] “UL Certification Guide (2026): Process, Cost, Requirements & Alternatives,” FactoryFollow. Accessed: May 07, 2026. [Online]. Available: https://factoryfollow.com/en/ul-certification-guide.html

[275] Kathy Anderson, “Do I Need UL Certification for My Product?” Accessed: May 07, 2026. [Online]. Available: https://blog.petra.com/blog/do-i-need-ul-certification-for-my-product

[276] Occupational Safety and Health Administration, “Current List of NRTLs.” [Online]. Available: https://www.osha.gov/nationally-recognized-testing-laboratory-program/current-list-of-nrtls

[277] *NFPA 70: National Electrical Code (NEC)*, Quincy., 2026.

[278] *UL 651: Standard for Schedule 40, 80, Type EB and A Rigid PVC Conduit and Fittings*, Standard UL 651, Northbrook., Oct. 25, 2011.

[279] Underwriters Laboratories Inc., “Technology for Detecting and Monitoring Conditions That Could Cause Electrical Wiring System Fires,” U.S. Consumer Product Safety Commission, Final Report, Sep. 1995. [Online]. Available: https://www.cpsc.gov/content/Part-1-Technology-for-Detecting-and-Monitoring-Conditions-That-Could-Cause-Electrical-Wiring-System-Fires-4061-0

[280] NEMA, “A Quarter Century of Protecting Homes & Occupants,” IAEI Magazine. Accessed: May 08, 2026. [Online]. Available: https://iaeimagazine.org/electrical-safety/a-quarter-century-of-protecting-homes-occupants/

[281] D. A. Lee, A. M. Trotta, and W. H. King, “New Technology for Preventing Residential Electrical Fires: Arc-Fault Circuit Interrupters (AFCIs),” *Fire Technology*, 2000, doi: 10.1023/A:1015410726786.

[282] R. Campbell, “Home Fires Caused by Electrical Failure or Malfunction,” National Fire Protection Association, Quincy, MA, NFPA Research Report, Nov. 2022. [Online]. Available: https://www.nfpa.org/education-and-research/research/nfpa-research/fire-statistical-reports/home-fires-caused-by-electrical-failure-or-malfunction

[283] T. Schneider and L. Szpruch, “The New Frontier: Managing and insuring generative and agentic AI risks,” InsTech. Accessed: May 08, 2026. [Online]. Available: https://www.instech.co/knowledge-centre/the-new-frontier-managing-and-insuring-generative-and-agentic-ai-risks/

[284] G. Abbott, “Exclusions are ‘coming of age’ for standalone AI insurance market,” The Insurer, Oct. 30, 2025. Accessed: May 08, 2026. [Online]. Available: https://www.theinsurer.com/ti/analysis/exclusions-are-coming-of-age-for-standalone-ai-insurance-market-2025-10-30/

[285] Armilla, “AI Performance Warranty Brief.” Accessed: May 08, 2026. [Online]. Available: https://www.armilla.ai/ai-performance-warranty-brief

[286] Armilla, “Armilla Launches Armilla Guaranteed™: Warranty Coverage for AI Products in Partnership with Leading Insurance Companies | Armilla.” Accessed: May 08, 2026. [Online]. Available: https://www.armilla.ai/resources/armilla-assurance-launches-armilla-guaranteed-tm-warranty-coverage-for-ai-products-in-partnership-with-leading-insurance-companies

[287] Coalition, “Coalition Adds Affirmative AI Endorsement to Cyber Policies.” Accessed: May 08, 2026. [Online]. Available: https://www.coalitioninc.com/announcements/undefined/announcements/coalition-adds-new-affirmative-ai-endorsement-to-cyber-policies

[288] “AI Insurance Coverage: Are You Covered for the Risk?,” Buchalter. Accessed: May 08, 2026. [Online]. Available: https://www.buchalter.com/blogs/ai-insurance-coverage-are-you-covered-for-the-risk/

[289] AXA XL, “AXA XL’s Gen AI endorsment.” 2026. [Online]. Available: https://axaxl.com/insurance/products/cyber-insurance-international

[290] Artificial Intelligence Underwriting Company, “AIUC | AI agent standard & insurance.” Accessed: May 08, 2026. [Online]. Available: https://aiuc.com/

[291] T. Mixides, “Armilla reveals purpose-built AI liability insurance amid rising legal and regulatory pressures - Reinsurance News,” ReinsuranceNe.ws. Accessed: May 08, 2026. [Online]. Available: https://www.reinsurancene.ws/armilla-reveals-purpose-built-ai-liability-insurance-amid-rising-legal-and-regulatory-pressures/

[292] Munich Re, “aiSure™ More AI Opportunity. Less AI Risk.” Accessed: May 08, 2026. [Online]. Available: https://www.munichre.com/en/solutions/for-industry-clients/insure-ai.html

[293] National Association of Insurance Commissioners, “Report on the Cybersecurity Insurance Market,” National Association of Insurance Commissioners, 2021. [Online]. Available: https://content.naic.org/sites/default/files/index-cmte-c-Cyber_Supplement_2020_Report.pdf

[294] National Association of Insurance Commissioners, “Report on the Cyber Insurance Market,” National Association of Insurance Commissioners, 2022. [Online]. Available: https://content.naic.org/sites/default/files/cmte-c-cyber-supplement-report-2022-for-data-year-2021.pdf

[295] National Association of Insurance Commissioners, “Report on the Cyber Insurance Market,” National Association of Insurance Commissioners, 2024. [Online]. Available: https://content.naic.org/sites/default/files/cmte-h-cyber-wg-2024-cyber-ins-report.pdf

[296] O. D. William, *Reinsurance and the Law of Aggregation: Event, Occurrence, Cause*. Taylor & Francis, 2023.

[297] T. Eyerly, R. Carmel, and K. Aldama, “Determining the Number of Occurrences and Its Impact on Coverage,” Feb. 05, 2016, *Social Science Research Network, Rochester, NY*: 2734437. doi: 10.2139/ssrn.2734437.

[298] L. S. Robson and P. L. Bigelow, “Measurement properties of occupational health and safety management audits: a systematic literature search and traditional literature synthesis,” *Can J Public Health*, vol. 101 Suppl 1, no. Suppl 1, pp. S34-40, 2010, doi: 10.1007/BF03403844.

[299] P. Paskov, M. J. Byun, K. Wei, and T. Webster, “Preliminary suggestions for rigorous GPAI model evaluations”.

[300] P. Laban, H. Hayashi, Y. Zhou, and J. Neville, “LLMs Get Lost In Multi-Turn Conversation,” May 09, 2025, *arXiv*: arXiv:2505.06120. doi: 10.48550/arXiv.2505.06120.

[301] H. Luo and G. Laban, “SPASM: Stable Persona-driven Agent Simulation for Multi-turn Dialogue Generation,” Apr. 10, 2026, *arXiv*: arXiv:2604.09212. doi: 10.48550/arXiv.2604.09212.

[302] C. Wang *et al.*, “AdapTools: Adaptive Tool-based Indirect Prompt Injection Attacks on Agentic LLMs,” Feb. 24, 2026, *arXiv*: arXiv:2602.20720. doi: 10.48550/arXiv.2602.20720.

[303] J. Bergstrom, "Chaos Engineering," *The Journal of Test and Evaluation*, pp. 208–218, 2022.

[304] I. Evtimov, A. Zharmagambetov, A. Grattafiori, C. Guo, and K. Chaudhuri, "WASP: Benchmarking Web Agent Security Against Prompt Injection Attacks".

[305] Artificial Intelligence Underwriting Company, "About the certificate," AIUC-1. Accessed: May 08, 2026. [Online]. Available: https://www.aiuc-1.com/learn/certificate

[306] S. Casper *et al.*, "Black-Box Access is Insufficient for Rigorous AI Audits," Jun. 2024, pp. 2254–2272. doi: 10.1145/3630106.3659037.

[307] E. G. Richards and National Board of Fire Underwriters., *The experience grading and rating schedule; designed to be a United States standard for measuring fire insurance costs based upon combined experience averages*. New York: [National Board of Fire Underwriters], 1915. [Online]. Available: https://catalog.hathitrust.org/Record/008521772

[308] Verisk's Community Hazard Mitigation Services, "Items Considered in the FSRS | FSRS | PPC," Verisk's Community Hazard Mitigation Services. Accessed: May 08, 2026. [Online]. Available: https://www.isomitigation.com/ppc/fsrs/items-considered-in-the-fsrs/

[309] "BCEG - ISRB." Accessed: May 08, 2026. [Online]. Available: https://isrb.com/Idaho/ISRB/Bceg

[310] Verisk's Community Hazard Mitigation Services, "ISO's Public Protection Classification (PPC) Program," Verisk's Community Hazard Mitigation Services. Accessed: May 08, 2026. [Online]. Available: https://www.isomitigation.com/ppc/

[311] National Institute of Building Sciences, "National Institute of Building Sciences Issues Interim Report on the Value of Mitigation." Accessed: May 08, 2026. [Online]. Available: http://nibs.org/national-institute-of-building-sciences-issues-interim-report-on-the-value-of-mitigation/

[312] T. Frank, "Hurricane Damage Would Be Less Extensive with Stronger Building Codes," *Scientific American*, Jun. 2021. Accessed: May 08, 2026. [Online]. Available: https://www.scientificamerican.com/article/hurricane-damage-would-be-less-extensive-with-stronger-building-codes/

[313] "Better risk selection could cut 16% from cyber insurers' loss ratios – Gallagher Re," Global Reinsurance. Accessed: Feb. 19, 2026. [Online]. Available: https://www.globalreinsurance.com/home/better-risk-selection-could-cut-16-from-cyber-insurers-loss-ratios-gallagher-re/1453478.article

[314] D. Wood, "SME cyber insurance on the rise: Attacks, regulations and contracts fuel growth," Insurance Business. Accessed: May 08, 2026. [Online]. Available: https://www.insurancebusinessmag.com/uk/news/cyber/sme-cyber-insurance-on-the-rise-attacks-regulations-and-contracts-fuel-growth-551857.aspx

[315] R. Adriko and J. R. C. Nurse, "Cybersecurity and Cyber insurance for Small to Medium-sized Enterprises (SMEs): Perceptions, challenges and decision-making dynamics," *Computers & Security*, vol. 163, p. 104818, Apr. 2026, doi: 10.1016/j.cose.2025.104818.

[316] J. Khoury, D. Martines, C. Freese, J. Eckel, M. Gupta, and L. Najjar, “Insurance Leads in AI Adoption. Now It’s Time to Scale.” [Online]. Available: https://www.bcg.com/publications/2025/insurance-leads-ai-adoption-now-time-to-scale

[317] A. Chadda, S. McGregor, J. Hostetler, and A. Brennen, “AI Evaluation Authorities: A Case Study Mapping Model Audits to Persistent Standards,” *Proceedings of the AAAI Conference on Artificial Intelligence*, vol. 38, no. 21, pp. 23035–23040, 2024, doi: 10.1609/aaai.v38i21.30346.

[318] C. C. Constantinescu, I. Stancu, and I. Panait, “Impact study of telematics auto insurance,” 2018.

[319] Cullen Insurance, “Calculating Insurance In Fire Prone Areas - Cullen Insurance Agency.” Accessed: May 08, 2026. [Online]. Available: https://culleninsuranceagency.com/calculating-insurance-in-fire-prone-areas/

[320] M. Nasr *et al.*, “The Attacker Moves Second: Stronger Adaptive Attacks Bypass Defenses Against Llm Jailbreaks and Prompt Injections,” 2025, *arXiv*. doi: 10.48550/ARXIV.2510.09023.

[321] S. Panda, D. W. Woods, A. Laszka, A. Fielder, and E. Panaousis, “Post-incident audits on cyber insurance discounts,” *Computers & Security*, vol. 87, p. 101593, Nov. 2019, doi: 10.1016/j.cose.2019.101593.

[322] S. Romanosky, L. Ablon, A. Kuehn, and T. Jones, “Content analysis of cyber insurance policies: how do carriers price cyber risk?,” *J Cyber Secur*, vol. 5, no. 1, p. tyz002, Jan. 2019, doi: 10.1093/cybsec/tyz002.

[323] J. B. Roberts, “Coalition is Now Offering Premium Credits to MDR Customers.” Accessed: May 08, 2026. [Online]. Available: https://www.coalitioninc.com/blog/cyber-insurance/undefined/blog/cyber-insurance/premium-credits-mdr

[324] Coalition, “Active monitoring and alerting: How we do it at Coalition.” Accessed: May 09, 2026. [Online]. Available: https://www.coalitioninc.com/blog/cyber-insurance/undefined/blog/cyber-insurance/cyber-risk-active-monitoring-alerts-coalition

[325] A. Tsohou, V. Diamantopoulou, S. Gritzalis, and C. Lambrinoudakis, “Cyber insurance: state of the art, trends and future directions,” *Int J Inf Secur*, vol. 22, no. 3, pp. 737–748, 2023, doi: 10.1007/s10207-023-00660-8.

[326] Businesswire, “Corvus Insurance Delivers an Industry-Leading Loss Ratio of 36%.” Accessed: May 09, 2026. [Online]. Available: https://www.businesswire.com/news/home/20230511005501/en/Corvus-Insurance-Delivers-an-Industry-Leading-Loss-Ratio-of-36

[327] J. Wolff, *Cyberinsurance Policy: Rethinking Risk in an Age of Ransomware, Computer Fraud, Data Breaches, and Cyberattacks*. The MIT Press, 2022. doi: 10.7551/mitpress/13665.001.0001.

[328] J. Sullivan and J. R. C. Nurse, “Cyber Security Incentives and the Role of Cyber Insurance,” 2021.

[329] R. Adriko and J. R. C. Nurse, “Does Cyber Insurance Promote Cyber Security Best Practice? An Analysis Based on Insurance Application Forms,” *Digital Threats*, vol. 5, no. 3, p. 25:1-25:39, Oct. 2024, doi: 10.1145/3676283.

[330] T. Baker and A. Shortland, “Insurance and enterprise: cyber insurance for ransomware,” *Geneva Pap Risk Insur Issues Pract*, vol. 48, no. 2, pp. 275–299, Apr. 2023, doi: 10.1057/s41288-022-00281-7.

[331] A. Cartwright *et al.*, “How cyber insurance influences the ransomware payment decision: theory and evidence,” *Geneva Pap Risk Insur Issues Pract*, vol. 48, no. 2, pp. 300–331, Apr. 2023, doi: 10.1057/s41288-023-00288-8.

[332] K. D. Logue and A. B. Shniderman, “The Case for Banning (and Mandating) Ransomware Insurance,” *SSRN Journal*, 2021, doi: 10.2139/ssrn.3907373.

[333] Allianz Commercial, “Cyber Security Resilience 2025: Claims and Risk Management Trends,” Allianz Commercial, Munich, Industry Report, Sep. 2025. Accessed: Aug. 31, 2025. [Online]. Available: https://commercial.allianz.com/content/dam/onemarketing/commercial/commercial/reports/cyber-security-trends-2025.pdf

[334] M. Massenkoff, E. Lyubich, P. McCrory, R. Appel, and R. Heller, “Anthropic Economic Index report: Learning curves,” Anthropic, Mar. 2026. [Online]. Available: https://www.anthropic.com/research/economic-index-march-2026-report

[335] “Chubb’s Cyber Partners for Mitigation and Response | Chubb.” Accessed: May 10, 2026. [Online]. Available: https://www.chubb.com/us-en/business-insurance/products/cyber-insurance/cyber-partners-for-mitigation-and-response.html

[336] “CrowdStrike Services Insurance Partners (U.S.): Helping policyholders recover from a breach with speed and precision and fortify their cybersecurity practices and controls,” CrowdStrike, Inc., Data Sheet, 2025. [Online]. Available: https://www.crowdstrike.com/content/dam/crowdstrike/marketing/en-us/documents/pdfs/data-sheets/crowdstrike-insurance-panel-data-sheet.pdf

[337] “Product liability, recall and contamination insurance | III.” Accessed: May 10, 2026. [Online]. Available: https://www.iii.org/article/product-liability-recall-and-contamination-insurance

[338] “Crisis Management Insurance | Talbot AIG.” Accessed: May 10, 2026. [Online]. Available: https://www.talbot.aig.com/home/risk-solutions/crisis-management

[339] “Incident Response (IR) Cybersecurity Services | CrowdStrike,” CrowdStrike.com. Accessed: May 10, 2026. [Online]. Available: https://www.crowdstrike.com/en-us/services/incident-response/

[340] “Incident Readiness and Response Services | LevelBlue.” Accessed: May 10, 2026. [Online]. Available: https://www.levelblue.com/services/incident-readiness-and-response

[341] “Do AI language models ‘understand’ the real world? On a basic level, they do, a new study finds | Brown University.” Accessed: May 26, 2026. [Online]. Available: https://www.brown.edu/news/2026-04-22/artificial-intelligence-understanding-real-world

[342] L. Kahn, E. S. Probasco, and R. Kinoshita, "AI Safety and Automation Bias: The Downside of Human-in-the-Loop," Center for Security and Emerging Technology, Georgetown University, Report, Nov. 2024. doi: 10.51593/20230057.

[343] G. Romeo and D. Conti, "Exploring automation bias in human–AI collaboration: a review and implications for explainable AI," *AI & Soc*, vol. 41, no. 1, pp. 259–278, Jan. 2026, doi: 10.1007/s00146-025-02422-7.

[344] S. Shappell and D. A. Wiegmann, "The Human Factors Analysis and Classification System: HFACS: Final Report," Feb. 2000, [Online]. Available: https://rosap.ntl.bts.gov/view/dot/21482

[345] M. Bogen, "Design Decisions and the Duty to Defend: What an Insurance Coverage Case Signals for AI Governance," Center for Democracy and Technology. Accessed: May 22, 2026. [Online]. Available: https://cdt.org/insights/design-decisions-and-the-duty-to-defend-what-an-insurance-coverage-case-signals-for-ai-governance/

[346] D. Hendrycks *et al.*, "A Definition of AGI," 2025, *arXiv*. doi: 10.48550/ARXIV.2510.18212.

[347] METR, "Measuring the Self-Reported Impact of Early-2026 AI on Technical Worker Productivity," *METR Blog*, May 2026, Accessed: May 20, 2026. [Online]. Available: https://metr.org/blog/2026-05-11-ai-usage-survey/

[348] A. Chan, R. Padarath, J. Kwon, H. Greaves, and M. Anderljung, "Measuring AI R&D Automation," Mar. 06, 2026, *arXiv*: arXiv:2603.03992. doi: 10.48550/arXiv.2603.03992.

[349] Anthropic, "When AI builds itself." Accessed: Jun. 04, 2026. [Online]. Available: https://www.anthropic.com/institute/recursive-self-improvement

[350] D. W. Ball, "On Recursive Self-Improvement (Part I): Thoughts on the automation of AI research." [Online]. Available: https://www.hyperdimensional.co/p/on-recursive-self-improvement-part

[351] Anthropic, "Disrupting the first reported AI-orchestrated cyber espionage campaign," Anthropic, Nov. 2025. Accessed: May 07, 2026. [Online]. Available: https://www.anthropic.com/news/disrupting-AI-espionage

[352] J. Egan and M. Nie, "Fighting AI Cyberattacks Starts With Knowing They're Happening," Lawfare. Accessed: May 07, 2026. [Online]. Available: https://www.lawfaremedia.org/article/fighting-ai-cyberattacks-starts-with-knowing-they-re-happening

[353] B. W. Roberts, "The Macroeconomic Impacts of the 9/11 Attack: Evidence from Real-Time Forecasting," *Peace Economics, Peace Science and Public Policy*, vol. 15, no. 2, pp. 341–367, Jul. 2009, doi: 10.2202/1554-8597.1166.

[354] A. Rose, G. Oladosu, B. Lee, and G. Asay, "The Economic Impacts of the September 11 Terrorist Attacks: A Computable General Equilibrium Analysis," *Peace Economics, Peace Science and Public Policy*, vol. 15, pp. 4–4, Jan. 2009, doi: 10.2202/1554-8597.1161.

[355] D. K. Nanto, "9/11 Terrorism: Global Economic Costs," Congressional Research Service, Library of Congress, Washington, DC, CRS Report for Congress RS21937, Oct. 2004. Accessed: May 20, 2026. [Online]. Available:

https://www.everycrsreport.com/files/20041005_RS21937_7e7b0eb8efa708fc6df893314b94182659526b84.pdf

[356] P. Grossi, “Property Damage and Insured Losses from the 2001 World Trade Center Attacks,” *Peace Economics, Peace Science and Public Policy*, vol. 15, pp. 3–3, Jan. 2009, doi: 10.2202/1554-8597.1163.

[357] L. A. W. L. M. Holson, “A NATION CHALLENGED: THE AIRLINES; White House to Seek $5 Billion As Part of Airline Rescue Plan,” *The New York Times*, Sep. 20, 2001. Accessed: May 11, 2026. [Online]. Available: https://www.nytimes.com/2001/09/20/business/nation-challenged-airlines-white-house-seek-5-billion-part-airline-rescue-plan.html

[358] R. J. Hillman, “Terrorism Insurance: Rising Uninsured Exposure to Attacks Heightens Potential Economic Vulnerabilities,” U.S. General Accounting Office, Testimony before the Subcommittee on Oversight and Investigations, Committee on Financial Services, House of Representatives GAO-02-472T, Feb. 2002. [Online]. Available: https://www.gao.gov/products/gao-02-472t

[359] D. W. Ball, “The Bitter Lessons: Thoughts on US-China Competition.” [Online]. Available: https://www.hyperdimensional.co/p/the-bitter-lessons

[360] J. A. Sonnenfeld and S. Henriques, “This Is How the AI Bubble Bursts.” [Online]. Available: https://insights.som.yale.edu/insights/this-is-how-the-ai-bubble-bursts

[361] D. Thompson, “This Is How the AI Bubble Will Pop.” [Online]. Available: https://www.derekthompson.org/p/this-is-how-the-ai-bubble-will-pop

[362] M. Rasheed, “AI adoption is reshaping the risk landscape | Swiss Re,” Swiss Re, Jan. 2026. Accessed: Apr. 02, 2026. [Online]. Available: https://www.swissre.com/institute/research/sigma-research/sigma-insights-01-2026-AI-adoption-is-reshaping-the-risk-landscape.html

[363] P. A. Raschky and H. Weck-Hannemann, “Charity hazard—A real hazard to natural disaster insurance?,” *Environmental Hazards*, vol. 7, no. 4, pp. 321–329, Jan. 2007, doi: 10.1016/j.envhaz.2007.09.002.

[364] T. Baker and R. Swedloff, “Mutually Assured Protection among Large U.S. Law Firms,” *Conn. Ins. L.J.*, vol. 24, p. 1, 2018 2017, [Online]. Available: https://heinonline.org/HOL/Page?handle=hein.journals/conilj24&id=11&div=&collection=

[365] “What are insurance-linked securities and how do they work?” Accessed: Jul. 01, 2025. [Online]. Available: https://www.schroders.com/en-bm/bm/professional/insights/what-are-insurance-linked-securities-and-how-do-they-work/

[366] “Into the Cyberverse,” Howden Re, Apr. 2025. [Online]. Available: https://www.howdenre.com/sites/howdenre.howdenprod.com/files/2025-04/howdenre_into_the_cyberverse_report_april_2025.pdf

[367] G. Weil, “Closing the AI Accountability Gap: Strict Liability and Punitive Damages for Advanced Artificial Intelligence,” Jan. 13, 2024, *Social Science Research Network, Rochester, NY*: 4694006. doi: 10.2139/ssrn.4694006.

[368] E. Michel-Kerjan and P. A. Raschky, "The effects of government intervention on the market for corporate terrorism insurance," *European Journal of Political Economy*, vol. 27, pp. S122–S132, Dec. 2011, doi: 10.1016/j.ejpoleco.2011.03.006.

[369] G. E. Marchant, B. R. Allenby, and J. R. Herkert, Eds., *The Growing Gap Between Emerging Technologies and Legal-Ethical Oversight: The Pacing Problem*, vol. 7. in The International Library of Ethics, Law and Technology, vol. 7. Dordrecht: Springer Netherlands, 2011. doi: 10.1007/978-94-007-1356-7.

[370] B. Berliner, "Large Risks and Limits of Insurability," *The Geneva Papers on Risk and Insurance*, vol. 10, no. 37, pp. 313–329, 1985, Accessed: Oct. 02, 2025. [Online]. Available: https://www.jstor.org/stable/41950168

[371] T. Baker and P. Siegelman, "Chapter 7: The law and economics of liability insurance: A theoretical and empirical review," 2013. Accessed: Jun. 25, 2025. [Online]. Available: https://www.elgaronline.com/edcollchap/edcoll/9781848441187/9781848441187.00015.xml

[372] R. W. Klein, "A Regulator's Introduction to the Insurance Industry," *NAIC*, 1999, [Online]. Available: https://content.naic.org/sites/default/files/inline-files/prod_serv_marketreg_rii_zb.pdf

[373] "Welcome to the Artificial Intelligence Incident Database." Accessed: May 11, 2026. [Online]. Available: https://incidentdatabase.ai/

[374] "MIT AI Incident Tracker." Accessed: May 11, 2026. [Online]. Available: https://airisk.mit.edu/ai-incident-tracker

[375] "Trust and transparency," OpenAI. Accessed: May 11, 2026. [Online]. Available: https://openai.com/trust-and-transparency/

[376] K. Tangalakis-Lippert and H. Chandonnet, "OpenAI quickly rolled back a new feature that allowed users to make private conversations with ChatGPT searchable," *Business Insider*, Aug. 01, 2025. Accessed: May 11, 2026. [Online]. Available: https://www.businessinsider.com/openai-removes-chatgpt-feature-over-search-engine-privacy-concerns-2025-7

[377] "Anthropic's Transparency Hub." Accessed: May 11, 2026. [Online]. Available: https://www.anthropic.com/transparency/system-trust-reporting

[378] "DAIL – the Database of AI Litigation," Ethical Tech Initiative. [Online]. Available: https://blogs.gwu.edu/law-eti/ai-litigation-database/

[379] Epoch AI, "Data on AI Companies," Epoch AI. Accessed: May 11, 2026. [Online]. Available: https://epoch.ai/data/ai-companies